# Projective Limits of State Spaces
# II. Quantum Formalism

Suzanne Lanéry[1,2] and Thomas Thiemann[1]

[1] Institute for Quantum Gravity, Friedrich-Alexander University Erlangen–Nürnberg, Germany
[2] Mathematics and Theoretical Physics Laboratory, François–Rabelais University of Tours, France

November 11, 2014

## Abstract

In this series of papers, we investigate the projective framework initiated by Jerzy Kijowski [13] and Andrzej Okołów [19, 20], which describes the states of a quantum theory as projective families of density matrices. After discussing the formalism at the classical level in a first paper [15], the present second paper is devoted to the quantum theory. In particular, we inspect in detail how such quantum projective state spaces relate to inductive limit Hilbert spaces and to infinite tensor product constructions. Regarding the quantization of classical projective structures into quantum ones, we extend the results by Okołów [20], that were set up in the context of linear configuration spaces, to configuration spaces given by simply-connected Lie groups, and to holomorphic quantization of complex phase spaces.

# Contents

# 1 Introduction

While finite dimensional symplectic manifolds are comparatively easy to quantize, allowing the formulation of rather systematic procedures (such as geometric quantization [27], of which a brief account is given in appendix A), quantizing infinite dimensional ones (aka. field theories) is substantially more involved: typically, we have to rely on some additional insight, telling us how to break them down into a stack of finite dimensional truncations and how to afterwards reassemble the pieces into a consistent quantum theory. This motivates an approach to quantum field theory introduced in [13, 20], where each such truncation get represented on a 'small' Hilbert space, before sewing these partial quantum theories together into a projective structure. In section 2, we will review this formalism, before inspecting how the quantum state spaces thus obtained compare to those provided by other quantization methods, that are also assembled from 'small' Hilbert spaces, yet sewed in a different manner.

As stressed in [15, section 1], there is a correspondence between projective limits of symplectic manifolds on the classical side and projective limits of state spaces on the quantum side. Accordingly, one can try to formulate a quantization program to turn a classical projective system into a quantum one. In [20] Andrzej Okołów established such a quantization prescription: by identifying appropriate assumptions, he was able to set up what we would call, in the terminology of [15, def. 2.15], a factorizing system of linear configuration spaces, which could then be quantized in a projective form. He subsequently used this construction to obtain the kinematical state space of a certain theory of quantum gravity [19]. In subsection 3.1, we will extend this result to configuration spaces given as simply-connected Lie groups. This is meant as a preparation for a corresponding treatment of loop quantum gravity, but could probably have applications to other gauge field theories as well. Additionally, holomorphic quantization will be discussed in subsection 3.2, following the lines of geometric quantization (note that the quantum projective structure used in [16, subsection 3.2] could be seen as arising from such an holomorphic quantization, although we will arrive at it from a different perspective).

Note that the heuristic picture that was presented at the beginning of [15, section 2] as a justification for the projective formalism becomes a bit more involved when we go over to the quantum theory. In particular, we had justified the directedness of the label set $\mathcal{L}$ by arguing that, given any two experiments, involving observables included respectively in the labels $\eta$ and $\eta'$, we should be able to describe the simultaneous realization of both experiments, hence the need for a label $\eta'' \succcurlyeq \eta, \eta'$. However, this argument obviously does not hold any more in the quantum theory, where complementarity forbids the simultaneous measurement of non-commuting observables. As a way out, we could simply decide to restrict the elementary observables (the ones that are accounted for in $\mathcal{L}$) to a set of mutually compatible observables, but that would force us to hard-code in the theory which observables we intend to actually measure (and to drop those that could *a priori* be measured but will not). This in turn creates severe difficulties, because we need to prescribe how to select in advance the set of those truly measured observables without spoiling predictivity (see eg. the concerns raised in [4]).

In the following, we bypass this discussion by assuming that we get the kinematical quantum state space via the quantization of a nice classical projective limit, so that it will not be a problem, for any finite set of kinematical observables to represent them on a 'simple' Hilbert space, whether these kinematical observables can be simultaneously measured or not. On the other hand it is to be expected that the algebra generated by a finite set of *dynamical* quantum observables will not be easily represented: already on the classical side we had underlined that a finite set of dynamical classical observables may generate an intricate Poisson-algebra (recall the comment before [15, def. 3.21]). Like in the classical formalism we must therefore expect that the exact dynamical observables will have to be approximated by approached ones, which in particular build a tractable algebra.

# 2 Projective limits of quantum state spaces

The crucial point, that goes back to Jerzy Kijowski [13], is that quantum states will be realized as projective families of *density matrices*, and *not* as families of vector states. This is actually a repercussion of the specific viewpoint of this formalism, namely that labels in $\mathcal{L}$ stand for a selection of observables (and not eg. for a selection of states). Indeed, in order to project a state from a more detailed partial quantum theory, represented on an Hilbert space $\mathcal{H}_{\eta'}$, to a coarser one, with Hilbert space $\mathcal{H}_\eta$, we need a map that will retain from a state only the features needed to compute expectation values of the observables on $\mathcal{H}_\eta$: this is what the partial trace on a tensor product factor accomplishes but it can only be defined as a map between density matrices (the partial trace of a pure state can be a mixed state and conversely).

While this has been previously rather seen as a weakness of the construction (see the discussion in [20, section 6.2]), we argue that such a generalized framework will be, in practice, indistinguishable from a more traditional theory on a Hilbert space: given a particular experiment, we should be able to select an $\eta$ such that everything takes place within $\mathcal{H}_\eta$. Moreover it seems advantageous to start with a very large kinematical state space, thus avoiding the inherent arbitrariness of restricting it *a priori* to some particular subspace, and to wait until there is a real need for such a restriction, together with clear requirements on how to perform it (in the light of [15, section 3], this could be because we are forced to consider only the states for which the regularization scheme, needed to implement the dynamics, converges).

In order to clarify the claim made above, that quantum state spaces described as projective limits tend to be 'bigger', we will examine precisely how they compare to those built on inductive limits of Hilbert spaces (theorem 2.9) or on infinite tensor products (theorem 2.11): these are constructions that incorporate ingredients somewhat similar to the projective approach, but differ notably by seeking a global Hilbert space for the quantum theory.

## 2.1 Projective systems of quantum state spaces

In this subsection, we present the projective approach to quantum field theory [13, 20], formulating it in a form as close as possible to the classical formalism set up in [15, section 2]. Indeed, the projective systems of quantum state spaces described here can be seen as the direct equivalents of the factorizing systems we had on the classical side [15, subsection 2.3], replacing Cartesian products of classical phase spaces by tensor products of Hilbert spaces, in accordance to the basic principles of quantum mechanics (in particular, the three-spaces consistency from [15, fig. 2.1] is straightforwardly transformed into its quantum version illustrated in fig. 2.1).

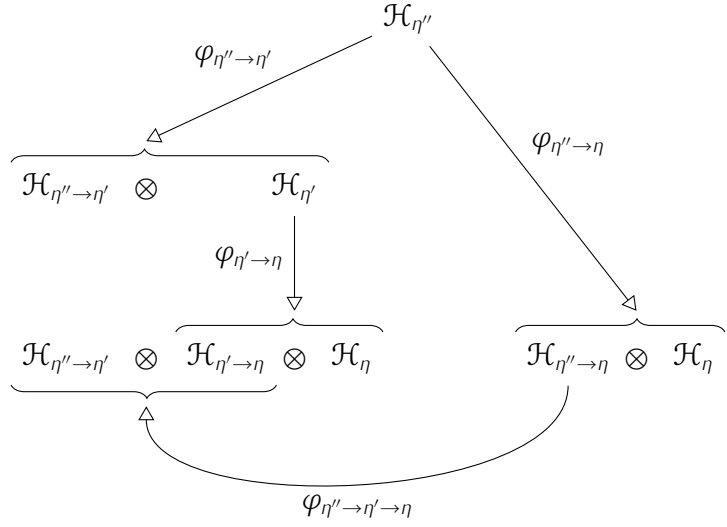

Figure 2.1 – Three-spaces consistency for projective systems of quantum state spaces

While we had convinced ourselves that the factorizing systems are quite generic among the projective systems of classical phase spaces (see the argument laid in [15, prop. 2.10]), their quantum counterparts seem to be even more broadly applicable (as we will show in the case of loop quantum cosmology in forthcoming work, there are systems that are classically not of the factorizing type, yet do admit a quantum formulation within the framework presented below).

**Definition 2.1** A projective system of quantum state spaces is a quintuple:
$$\left(\mathcal{L}, \left(\mathcal{H}_\eta\right)_{\eta\in\mathcal{L}}, \left(\mathcal{H}_{\eta'\to\eta}\right)_{\eta\preccurlyeq\eta'}, \left(\varphi_{\eta'\to\eta}\right)_{\eta\preccurlyeq\eta'}, \left(\varphi_{\eta''\to\eta'\to\eta}\right)_{\eta\preccurlyeq\eta'\preccurlyeq\eta''}\right)$$
where:

1. $\mathcal{L}$ is a preordered, directed set (we denote the pre-order by $\preccurlyeq$, its inverse by $\succcurlyeq$);

2. $\left(\mathcal{H}_\eta\right)_{\eta\in\mathcal{L}}$ is a family of Hilbert spaces indexed by $\mathcal{L}$;

3. $\left(\mathcal{H}_{\eta'\to\eta}\right)_{\eta\preccurlyeq\eta'}$ is a family of Hilbert spaces indexed by $\{\eta,\eta'\in\mathcal{L}\mid\eta\preccurlyeq\eta'\}$, such that $\dim(\mathcal{H}_{\eta\to\eta})=1$ for all $\eta\in\mathcal{L}$;

4. $\left(\varphi_{\eta'\to\eta}\right)_{\eta\preccurlyeq\eta'}$ is a family of isomorphisms of Hilbert spaces $\varphi_{\eta'\to\eta}:\mathcal{H}_{\eta'}\to\mathcal{H}_{\eta'\to\eta}\otimes\mathcal{H}_\eta$ indexed by $\{\eta,\eta'\in\mathcal{L}\mid\eta\preccurlyeq\eta'\}$ such that $\varphi_{\eta\to\eta}$ is trivial (by isomorphism of Hilbert spaces we mean a (bijective) unitary map between Hilbert spaces);

5. $\left(\varphi_{\eta''\to\eta'\to\eta}\right)_{\eta\preccurlyeq\eta'\preccurlyeq\eta''}$ is a family of isomorphisms of Hilbert spaces $\varphi_{\eta''\to\eta'\to\eta}:\mathcal{H}_{\eta''\to\eta}\to\mathcal{H}_{\eta''\to\eta'}\otimes\mathcal{H}_{\eta'\to\eta}$ indexed by $\{\eta,\eta',\eta''\in\mathcal{L}\mid\eta\preccurlyeq\eta'\preccurlyeq\eta''\}$ such that $\varphi_{\eta''\to\eta'\to\eta}$ is trivial whenever two labels among $\eta,\eta',\eta''$ are equal and:

$$\forall \eta \preccurlyeq \eta' \preccurlyeq \eta'' \in \mathcal{L}, \quad (\varphi_{\eta'' \to \eta' \to \eta} \otimes \mathrm{id}_{\mathcal{H}_\eta}) \circ \varphi_{\eta'' \to \eta} = (\mathrm{id}_{\mathcal{H}_{\eta'' \to \eta'}} \otimes \varphi_{\eta' \to \eta}) \circ \varphi_{\eta'' \to \eta'}. \qquad (2.1.1)$$

Whenever possible, we will use the shortened notation $(\mathcal{L}, \mathcal{H}, \varphi)^\otimes$ instead of $\left(\mathcal{L}, (\mathcal{H}_\eta)_{\eta \in \mathcal{L}}, (\mathcal{H}_{\eta' \to \eta})_{\eta \preccurlyeq \eta'}, (\varphi_{\eta' \to \eta})_{\eta \preccurlyeq \eta'}, (\varphi_{\eta'' \to \eta' \to \eta})_{\eta \preccurlyeq \eta' \preccurlyeq \eta''}\right)$.

**Definition 2.2** Let $(\mathcal{L}, \mathcal{H}, \varphi)^\otimes$ be a projective system of quantum state spaces. For $\eta \in \mathcal{L}$, we define $\overline{\mathcal{S}}_\eta$ the space of (self-adjoint) positive semi-definite, traceclass operators on $\mathcal{H}_\eta$ and $\mathcal{S}_\eta$ the space of density matrices:

$$\mathcal{S}_\eta = \left\{ \rho_\eta \in \overline{\mathcal{S}}_\eta \mid \mathrm{Tr}_{\mathcal{H}_\eta} \rho_\eta = 1 \right\}.$$

For $\eta \preccurlyeq \eta' \in \mathcal{L}$, we define:

$$\begin{aligned} \mathrm{Tr}_{\eta' \to \eta} : \overline{\mathcal{S}}_{\eta'} &\to \overline{\mathcal{S}}_\eta \\ \rho_{\eta'} &\mapsto \mathrm{Tr}_{\mathcal{H}_{\eta' \to \eta}} \left( \varphi_{\eta' \to \eta} \circ \rho_{\eta'} \circ \varphi_{\eta' \to \eta}^{-1} \right) \end{aligned}.$$

From eq. (2.1.1), we have:

$$\forall \eta \preccurlyeq \eta' \preccurlyeq \eta'' \in \mathcal{L}, \; \mathrm{Tr}_{\eta' \to \eta} \circ \mathrm{Tr}_{\eta'' \to \eta'} = \mathrm{Tr}_{\eta'' \to \eta}.$$

Hence, $\left(\mathcal{L}, (\overline{\mathcal{S}}_\eta)_{\eta \in \mathcal{L}}, (\mathrm{Tr}_{\eta' \to \eta})_{\eta \preccurlyeq \eta'}\right)$ forms a projective system and we denote its projective limit by $\overline{\mathcal{S}}^\otimes_{(\mathcal{L}, \mathcal{H}, \varphi)}$. The maps $\mathrm{Tr}_{\eta' \to \eta}$ being linear under conical combinations (ie. under addition and multiplication by positive reals), $\overline{\mathcal{S}}^\otimes_{(\mathcal{L}, \mathcal{H}, \varphi)}$ forms a cone (ie. we can equip it with a notion of addition and positive multiplication).

Now, for all $\eta \preccurlyeq \eta' \in \mathcal{L}$, $\mathrm{Tr}_{\eta' \to \eta} \langle \mathcal{S}_{\eta'} \rangle = \mathcal{S}_\eta$, hence $\left(\mathcal{L}, (\mathcal{S}_\eta)_{\eta \in \mathcal{L}}, (\mathrm{Tr}_{\eta' \to \eta})_{\eta \preccurlyeq \eta'}\right)$ also forms a projective system. Accordingly, a state on $(\mathcal{L}, \mathcal{H}, \varphi)^\otimes$ is a family $(\rho_\eta)_{\eta \in \mathcal{L}}$ such that $\forall \eta \in \mathcal{L}$, $\rho_\eta$ is a density matrix over $\mathcal{H}_\eta$ and $\forall \eta \preccurlyeq \eta'$, $\mathrm{Tr}_{\eta' \to \eta} \rho_{\eta'} = \rho_\eta$. We denote the space of states by $\mathcal{S}^\otimes_{(\mathcal{L}, \mathcal{H}, \varphi)}$.

The projective structure conferred to the space of states comes with a natural inductive structure for the observables [20, section 6.2]. Thus, we can define a $C^*$-algebra $\overline{\mathcal{A}}^\otimes_{(\mathcal{L}, \mathcal{H}, \varphi)}$ of bounded operators, as the inductive limit of the algebras $\mathcal{A}_\eta$ that live over each $\mathcal{H}_\eta$. Then, the states defined above can be seen as states on $\overline{\mathcal{A}}^\otimes_{(\mathcal{L}, \mathcal{H}, \varphi)}$ in the sense of [10, part III, def. 2.2.8]. Looking at states over a $C^*$-algebra is in a sense more fundamental than looking at density matrices over a specific representation of this algebra. Indeed, any such representation can be split into cyclic components, and each cyclic component arises from a state over the algebra via the GNS construction. Reciprocally, any state $\rho$ over the algebra defines a corresponding GNS representation, and the states that can be represented as density matrices over this representation form the folium of $\rho$ (see also prop. 2.8 on this point). The irreducible representations are precisely the ones that arise from pure states (states that cannot be written as a non trivial convex superposition of other states) [10, part III, theorem 2.2.17], and Fell's theorem [6] tells us that, whenever a state yields a faithful GNS representation, its folium will be, in a definite sense, dense in the set of all states over the algebra.

Note, however, that the framework presented here differs substantially from the one of Algebraic Quantum Field Theory [10, part III]. In both cases one is looking at states over an inductive limit $C^*$-algebra. But here the building blocks of our algebra of observables will be in practice very *small* algebras: each $\mathcal{A}_\eta$, instead of being meant to include all the operators needed to interpret any arbitrary experiment taking place in some given region of spacetime, should be thought as only containing the operators needed for the description of *finitely* many experiments. More deeply, the purpose of giving the algebra of observables an inductive limit structure is in our case not so much to encode additional physically essential information (eg. the localization of the operators and associated causal structure) but rather to arrive at a description of the space of states as concrete and as convenient as possible (by building it from small representations that are well under control and suitable for calculation). Of course, we could combine both aspects, by decorating the projective structure with this extra information: for example, we could map a region in spacetime to the set of all $\eta$ that can be seen as contained in this region.

**Definition 2.3** We consider the same objects as in def. 2.2. For $\eta \in \mathcal{L}$, we denote by $\mathcal{A}_\eta$ the algebra of bounded operators on $\mathcal{H}_\eta$ and, for $\eta \preccurlyeq \eta' \in \mathcal{L}$, we define:

$$\begin{aligned} \iota_{\eta' \leftarrow \eta} : \mathcal{A}_\eta &\to \mathcal{A}_{\eta'} \\ A_\eta &\mapsto \varphi_{\eta' \to \eta}^{-1} \circ \left( \mathrm{id}_{\mathcal{H}_{\eta' \to \eta}} \otimes A_\eta \right) \circ \varphi_{\eta' \to \eta} \end{aligned}.$$

By definition $\iota_{\eta' \leftarrow \eta}$ is injective and, from eq. (2.1.1), we have:

$$\forall \eta \preccurlyeq \eta' \preccurlyeq \eta'', \, \iota_{\eta'' \leftarrow \eta'} \circ \iota_{\eta' \leftarrow \eta} = \iota_{\eta'' \leftarrow \eta}. \tag{2.3.1}$$

Accordingly, an operator over $\overline{\mathcal{S}}^\otimes_{(\mathcal{L}, \mathcal{H}, \varphi)}$ is an equivalence class in $\bigcup_{\eta \in \mathcal{L}} \mathcal{A}_\eta$ for the equivalence relation defined by:

$$\begin{aligned} \forall \eta, \eta' \in \mathcal{L}, \forall A_\eta \in \mathcal{A}_\eta, \forall A_{\eta'} \in \mathcal{A}_{\eta'}, \\ A_\eta \sim A_{\eta'} \Leftrightarrow \left( \exists \eta'' \in \mathcal{L} \, / \, \eta \preccurlyeq \eta'', \eta' \preccurlyeq \eta'' \, \& \, \iota_{\eta'' \leftarrow \eta} (A_\eta) = \iota_{\eta'' \leftarrow \eta'} (A_{\eta'}) \right) \end{aligned} \tag{2.3.2}$$

The space of operators over $\overline{\mathcal{S}}^\otimes_{(\mathcal{L}, \mathcal{H}, \varphi)}$ will be denoted by $\mathcal{A}^\otimes_{(\mathcal{L}, \mathcal{H}, \varphi)}$. For $A = [A_\eta]_\sim \in \mathcal{A}^\otimes_{(\mathcal{L}, \mathcal{H}, \varphi)}$ and $\rho = (\rho_\eta)_{\eta \in \mathcal{L}} \in \overline{\mathcal{S}}^\otimes_{(\mathcal{L}, \mathcal{H}, \varphi)}$, we can define $\mathrm{Tr} \rho A := \mathrm{Tr}_{\mathcal{H}_\eta} \rho_\eta A_\eta$. The definition of the equivalence relation ensures that this is well-defined.

**Proposition 2.4** For $\eta \preccurlyeq \eta' \in \mathcal{L}$, the map $\iota_{\eta' \leftarrow \eta}$ is an injective $C^*$-algebra morphism (ie. an injective, isometric $*$-algebra morphism). Hence, $\mathcal{A}^\otimes_{(\mathcal{L}, \mathcal{H}, \varphi)}$ can be equipped with a normed $*$-algebra structure as an inductive limit of $C^*$-algebras. And denoting by $\overline{\mathcal{A}}^\otimes_{(\mathcal{L}, \mathcal{H}, \varphi)}$ the completion of $\mathcal{A}^\otimes_{(\mathcal{L}, \mathcal{H}, \varphi)}$ with respect to the operator norm, $\overline{\mathcal{A}}^\otimes_{(\mathcal{L}, \mathcal{H}, \varphi)}$ is a $C^*$-algebra.

Then, for all $\rho \in \mathcal{S}^\otimes_{(\mathcal{L}, \mathcal{H}, \varphi)}$, $\mathrm{Tr}(\rho \, \cdot)$ can be extended by continuity as a state over $\overline{\mathcal{A}}^\otimes_{(\mathcal{L}, \mathcal{H}, \varphi)}$.

**Proof** That $\iota_{\eta' \leftarrow \eta}$ is a $C^*$-algebra morphism for any $\eta \preccurlyeq \eta'$ is ensured by $\varphi_{\eta' \to \eta}$ being a Hilbert space isomorphism and by the properties of the tensor product of operators.

Next, let $\rho \in \mathcal{S}^\otimes_{(\mathcal{L}, \mathcal{H}, \varphi)}$, and let $[A_\eta]_\sim, [B_{\eta'}]_\sim \in \mathcal{A}^\otimes_{(\mathcal{L}, \mathcal{H}, \varphi)}$. For any $\eta'' \succcurlyeq \eta', \eta$ and any $a, b \in \mathbb{C}$, we have:

1. $\mathrm{Tr} \rho (a A + b B) = \mathrm{Tr}_{\mathcal{H}_{\eta''}} \rho_{\eta''} \left( a \iota_{\eta'' \leftarrow \eta}(A_\eta) + b \iota_{\eta'' \leftarrow \eta'}(B_{\eta'}) \right) = a \, \mathrm{Tr} (\rho A) + b \, \mathrm{Tr} (\rho B)$;

2. $\text{Tr}(\rho A) = \text{Tr}_{\mathcal{H}_\eta}(\rho_\eta A_\eta) \leqslant \|A_\eta\| = \|A\|$ ;

3. $\text{Tr}(\rho \mathbb{1}) = \text{Tr}_{\mathcal{H}_\eta}(\rho_\eta \, \text{id}_{\mathcal{H}_\eta}) = 1$ ;

4. $\text{Tr}(\rho A^+ A) = \text{Tr}_{\mathcal{H}_\eta}(\rho_\eta A_\eta^+ A_\eta) \geqslant 0$ .

$\square$

**Proposition 2.5** For $\eta \in \mathcal{L}$, we denote by $\mathcal{O}_\eta$ the algebra of densely defined (possibly unbounded) normal operators on $\mathcal{H}_\eta$. An observable over a projective limit of quantum state spaces $\mathcal{S}^\otimes_{(\mathcal{L},\mathcal{H},\varphi)}$ is defined as an equivalence class in $\bigcup_{\eta \in \mathcal{L}} \mathcal{O}_\eta$ in analogy to def. 2.3.

The space of observables over $\mathcal{S}^\otimes_{(\mathcal{L},\mathcal{H},\varphi)}$ will be denoted by $\mathcal{O}^\otimes_{(\mathcal{L},\mathcal{H},\varphi)}$. For $O = [O_\eta]_\sim \in \mathcal{O}^\otimes_{(\mathcal{L},\mathcal{H},\varphi)}$, we can define the spectrum $\varsigma(O)$ of $O$ as the spectrum $\varsigma(O_\eta)$ of any representative $O_\eta$ of $O$ and, for $W$ a measurable subset of $\varsigma(O)$, we can define $\mathbb{I}_W(O)$ as the equivalence class $[\mathbb{I}_W(O_\eta)]_\sim \in \mathcal{A}^\otimes_{(\mathcal{L},\mathcal{H},\varphi)}$ of the spectral projector $\mathbb{I}_W(O_\eta)$.

Hence, for a state $\rho = (\rho_\eta)_{\eta \in \mathcal{L}} \in \mathcal{S}^\otimes_{(\mathcal{L},\mathcal{H},\varphi)}$, we can define the probability of measuring $O$ in $W$ as $\rho[O \in W] := \text{Tr} \, \rho \, \mathbb{I}_W(O)$.

**Proof** $\varsigma(O_\eta)$ being independent of the choice of a representative $O_\eta$ comes from:

$$\forall \eta' \succcurlyeq \eta, \ \varsigma\left(\varphi^{-1}_{\eta' \to \eta} \circ \left(\text{id}_{\mathcal{H}_{\eta' \to \eta}} \otimes O_\eta\right) \circ \varphi_{\eta' \to \eta}\right) = \varsigma\left(O_\eta\right),$$

where we used [21, theorem VIII.33] together with the fact that $\varphi_{\eta' \to \eta}$ is an Hilbert space isomorphism.

That $O_\eta \sim O_{\eta'}$ implies $\mathbb{I}_W(O_\eta) \sim \mathbb{I}_W(O_{\eta'})$ comes from:

$$\forall \eta' \succcurlyeq \eta, \ \mathbb{I}_W\left(\varphi^{-1}_{\eta' \to \eta} \circ \left(\text{id}_{\mathcal{H}_{\eta' \to \eta}} \otimes O_\eta\right) \circ \varphi_{\eta' \to \eta}\right) = \varphi^{-1}_{\eta' \to \eta} \circ \left(\text{id}_{\mathcal{H}_{\eta' \to \eta}} \otimes \mathbb{I}_W(O_\eta)\right) \circ \varphi_{\eta' \to \eta}.$$

$\square$

## 2.2 Maps between quantum state spaces

In order to investigate the relations between the spaces of quantum states assembled this way and more standard constructions, we will use the same tool as we used to make the connection between classical projective structures and infinite dimensional symplectic manifolds, namely extensions and restrictions of the label set. As in [15, subsection 2.2], the strategy will be to extend the label set by adding to it a greatest element (associated to the 'big' Hilbert space to which we want to make contact), before restricting ourselves to this greatest element alone (thus ending with a trivial projective system that can be identified with the target Hilbert space). A bit unintuitively, the non-trivial switch of state space occurs during the first step: this is because a greatest element forms by definition a cofinal part, which makes the second step innocuous.

**Proposition 2.6** Let $(\mathcal{L}, \mathcal{H}, \varphi)^\otimes$ be a projective system of quantum state spaces and let $\mathcal{L}'$ be a directed subset of $\mathcal{L}$. We define the map:

$$\sigma : \overline{\mathcal{S}}^{\otimes}_{(\mathcal{L},\mathcal{H},\varphi)} \to \overline{\mathcal{S}}^{\otimes}_{(\mathcal{L}',\mathcal{H},\varphi)}$$
$$(\rho_\eta)_{\eta \in \mathcal{L}} \mapsto (\rho_\eta)_{\eta \in \mathcal{L}'} .$$

$\sigma$ is conically linear (ie. compatible with addition and multiplication by positive scalars).

Moreover, we have a map $\alpha : \overline{\mathcal{A}}^{\otimes}_{(\mathcal{L}',\mathcal{H},\varphi)} \to \overline{\mathcal{A}}^{\otimes}_{(\mathcal{L},\mathcal{H},\varphi)}$ such that:

$$\forall \rho \in \overline{\mathcal{S}}^{\otimes}_{(\mathcal{L},\mathcal{H},\varphi)}, \forall A \in \overline{\mathcal{A}}^{\otimes}_{(\mathcal{L}',\mathcal{H},\varphi)}, \ \mathrm{Tr}\,(\rho\,\alpha(A)) = \mathrm{Tr}\,(\sigma(\rho)\,A), \tag{2.6.1}$$

and $\alpha$ is a $C^*$-algebra morphism.

If $\mathcal{L}'$ is cofinal in $\mathcal{L}$, we have in addition that $\sigma$ and $\alpha$ are bijective maps.

**Proof** The proof works in the same way as in the classical case [15, prop. 2.5]. The conical linearity of $\sigma$ and morphism property of $\alpha$ comes from their definition (with $\alpha$ being defined in analogy to the classical case). Eq. (2.6.1) can be first checked for $\mathcal{S}^{\otimes}_{(\mathcal{L},\mathcal{H},\varphi)}$ and $\mathcal{A}_{(\mathcal{L}',\mathcal{H},\varphi)}$ and expanded by continuity and conical linearity. □

**Proposition 2.7** In particular, if $\mathcal{L}$ admits a greatest element $\Lambda$, there exist bijective maps $\sigma : \overline{\mathcal{S}}^{\otimes}_{(\mathcal{L},\mathcal{H},\varphi)} \to \overline{\mathcal{S}}_\Lambda$ ($\overline{\mathcal{S}}_\Lambda$ being the space of (self-adjoint) positive semi-definite, traceclass operators over $\mathcal{H}_\Lambda$) and $\alpha : \mathcal{A}_\Lambda \to \overline{\mathcal{A}}^{\otimes}_{(\mathcal{L},\mathcal{H},\varphi)}$ such that $\forall \rho \in \overline{\mathcal{S}}^{\otimes}_{(\mathcal{L},\mathcal{H},\varphi)}, \forall A \in \mathcal{A}_\Lambda, \ \mathrm{Tr}\,(\rho\,\alpha(A)) = \mathrm{Tr}_\Lambda\,(\sigma(\rho)\,A)$.

**Proof** This is an application of prop. 2.6 for the cofinal part $\mathcal{L}' = \{\Lambda\}$ of $\mathcal{L}$, using the obvious identification of $\overline{\mathcal{S}}^{\otimes}_{(\{\Lambda\},\mathcal{H},\varphi)}$ with $\overline{\mathcal{S}}_\Lambda$ and $\overline{\mathcal{A}}^{\otimes}_{(\{\Lambda\},\mathcal{H},\varphi)}$ with $\mathcal{A}_\Lambda$. □

If we choose a particular state in a projective limit of quantum state spaces, we can use it as a vacuum to construct a corresponding GNS representation of the inductive limit algebra of observables [20, section 6.2], and this representation will naturally inherit a structure of inductive limit Hilbert space (like the Hilbert space used as starting point in LQG, see [1]). We will specialize in the case where the vacuum state we are using projects as a pure state on every $\mathcal{H}_\eta$. Whether there exists such a pure projective state of course depends on the specific projective structure under consideration, however we will be interested in situations where the natural vacuum turns out to be of this type (notably in prop. 3.5 and [16, prop. 3.17]). In this case, the inductive limit Hilbert space is obtained from a collection of 'reference states' $\zeta_{\eta' \to \eta} \in \mathcal{H}_{\eta' \to \eta}$, that allows us to see the tensor product factor $\mathcal{H}_\eta$ as a vector subspace in $\mathcal{H}_{\eta'}$. Any density matrix over such an inductive limit $\mathcal{H}_\zeta$ can then be unambiguously mapped to a state in the projective limit, but the converse typically does not hold, and in fact, we can formulate a handy condition to check if a state on the projective structure has its counterpart as a density matrix on $\mathcal{H}_\zeta$.

**Proposition 2.8** Let $(\mathcal{L}, \mathcal{H}, \varphi)^\otimes$ be a projective system of quantum state spaces and let $\rho = (\rho_\eta)_{\eta \in \mathcal{L}} \in \mathcal{S}^{\otimes}_{(\mathcal{L},\mathcal{H},\varphi)}$. For all $\eta \in \mathcal{L}$, $\rho_\eta$ is a state over $\mathcal{A}_\eta$ and we denote by $\mathcal{H}^{\mathrm{GNS}}_{\rho_\eta}, (\cdot)^{\mathrm{GNS}}$ the GNS representation constructed from this state [10, section III.2.2].

Then, for all $\eta \preccurlyeq \eta' \in \mathcal{L}$, there exists an injective linear map $\tau_{\eta' \leftarrow \eta} : \mathcal{H}^{\mathrm{GNS}}_{\rho_\eta} \to \mathcal{H}^{\mathrm{GNS}}_{\rho_{\eta'}}$. $\tau_{\eta' \leftarrow \eta}$ is isometric onto its image and satisfies:

$$\forall A_\eta \in \mathcal{A}_\eta, \ \tau_{\eta' \leftarrow \eta} \circ A^{\mathrm{GNS}}_\eta = \left(\iota_{\eta' \leftarrow \eta}(A_\eta)\right)^{\mathrm{GNS}} \circ \tau_{\eta' \leftarrow \eta}. \tag{2.8.1}$$

Moreover, for all $\eta \preccurlyeq \eta' \preccurlyeq \eta'' \in \mathcal{L}$, we have $\tau_{\eta'' \leftarrow \eta'} \circ \tau_{\eta' \leftarrow \eta} = \tau_{\eta'' \leftarrow \eta}$, hence we can define an Hilbert space $\mathcal{H}_\rho^{\text{GNS}}$ as (the completion of) the inductive limit of $\left(\mathcal{L}, \left(\mathcal{H}_{\rho_\eta}^{\text{GNS}}\right)_{\eta \in \mathcal{L}}, \left(\tau_{\eta' \leftarrow \eta}\right)_{\eta \preccurlyeq \eta'}\right)$ and $\mathcal{H}_\rho^{\text{GNS}}$ can be identified with the GNS representation of $\overline{\mathcal{A}}_{(\mathcal{L}, \mathcal{H}, \varphi)}^{\otimes}$ arising from the state $\rho$ (prop. 2.4).

If in addition there exists, for all $\eta \in \mathcal{L}$, a vector $\zeta_\eta \in \mathcal{H}_\eta$ such that $\rho_\eta = |\zeta_\eta\rangle\langle\zeta_\eta|$, then $\mathcal{H}_{\rho_\eta}^{\text{GNS}} \approx \mathcal{H}_\eta$ and for all $\eta \preccurlyeq \eta'$ there exists a vector $\zeta_{\eta' \to \eta} \in \mathcal{H}_{\eta' \to \eta}$ such that the map $\tau_{\eta' \leftarrow \eta}$ is given by:

$$\forall \psi \in \mathcal{H}_\eta, \ \tau_{\eta' \leftarrow \eta}(\psi) = \varphi_{\eta' \to \eta}^{-1}\left(\zeta_{\eta' \to \eta} \otimes \psi\right). \tag{2.8.2}$$

**Proof** *Explicit description of $\mathcal{H}_{\rho_\eta}^{\text{GNS}}$.* Let $\eta \in \mathcal{L}$. By applying the spectral theorem to the (self-adjoint) positive semi-definite, traceclass, normalized operator $\rho_\eta$, there exist $N_\eta \in \mathbb{N} \sqcup \{\infty\}$, an orthonormal family $\left(\zeta_\eta^{(k)}\right)_{k \leqslant N_\eta}$ in $\mathcal{H}_\eta$ and a family of *strictly* positive reals $\left(p_\eta^{(k)}\right)_{k \leqslant N_\eta}$ such that:

$$\rho_\eta = \sum_{k \leqslant N_\eta} p_\eta^{(k)} \left|\zeta_\eta^{(k)}\right\rangle \left\langle\zeta_\eta^{(k)}\right|,$$

with $\sum_{k \leqslant N_\eta} p_\eta^{(k)} = 1$.

We define the map $\Psi_{\rho_\eta}$ by:

$$\Psi_{\rho_\eta} : \mathcal{A}_\eta \to \mathcal{H}_\eta^+ \otimes \mathcal{H}_\eta$$
$$A \mapsto \sum_{k \leqslant N_\eta} \sqrt{p_\eta^{(k)}} \left\langle\zeta_\eta^{(k)}\right| \otimes \left|A\zeta_\eta^{(k)}\right\rangle \ ,$$

where $\mathcal{H}_\eta^+$ is the topological dual of $\mathcal{H}_\eta$ equipped with its natural Hilbert space structure. The map $\Psi_{\rho_\eta}$ is well-defined, for we have:

$$\sum_{k \leqslant N_\eta} \left\|\sqrt{p_\eta^{(k)}} \left\langle\zeta_\eta^{(k)}\right| \otimes \left|A\zeta_\eta^{(k)}\right\rangle\right\|^2 = \sum_{k \leqslant N_\eta} p_\eta^{(k)} \left\|\left|A\zeta_\eta^{(k)}\right\rangle\right\|^2 \leqslant \|A\|^2. \tag{2.8.3}$$

Moreover, it has following properties:

1. $\Psi_{\rho_\eta}$ is $\mathbb{C}$-linear;

2. $\forall A, B \in \mathcal{A}_\eta, \ \Psi_{\rho_\eta}(AB) = \left(\text{id}_{\mathcal{H}_\eta^+} \otimes A\right) \Psi_{\rho_\eta}(B)$;

3. $\forall A, B \in \mathcal{A}_\eta, \ \left\langle\Psi_{\rho_\eta}(A), \Psi_{\rho_\eta}(B)\right\rangle_{\mathcal{H}_\eta^+ \otimes \mathcal{H}_\eta} =$

$$= \sum_{k, k' \leqslant N_\eta} \sqrt{p_\eta^{(k)}} \sqrt{p_\eta^{(k')}} \left\langle\zeta_\eta^{(k')} \mid \zeta_\eta^{(k)}\right\rangle \otimes \left\langle A\zeta_\eta^{(k)} \mid B\zeta_\eta^{(k')}\right\rangle$$

$$= \sum_{k \leqslant N_\eta} p_\eta^{(k)} \otimes \left\langle\zeta_\eta^{(k)} \mid A^+ B \zeta_\eta^{(k)}\right\rangle = \text{Tr}_{\mathcal{H}_\eta} \rho_\eta A^+ B \ ;$$

4. $\overline{\Psi_{\rho_\eta}\langle\mathcal{A}_\eta\rangle} = \overline{\text{Vect}\left\{\sqrt{p_\eta^{(k)}} \left\langle\zeta_\eta^{(k)}\right| \otimes |\psi\rangle \mid k \leqslant N_\eta, \ \psi \in \mathcal{H}_\eta\right\}}$

('⊂': by definition of $\Psi_{\rho_\eta}$; '⊃': by considering operators of the form $|\psi\rangle\langle\zeta_\eta^{(k)}|$)

$$= \overline{\text{Vect}\left\{\sqrt{p_\eta^{(k)}}\left\langle\zeta_\eta^{(k)}\right|\,\middle|\,k\leqslant N_\eta\right\}}\otimes\mathcal{H}_\eta =: \mathcal{K}_{\rho_\eta}\otimes\mathcal{H}_\eta.$$

Therefore, we can identify $\mathcal{H}_{\rho_\eta}^{\text{GNS}}$ with $\mathcal{K}_{\rho_\eta}\otimes\mathcal{H}_\eta$ and we have:

$$\forall A\in\mathcal{A}_\eta,\ A^{\text{GNS}} = \text{id}_{\mathcal{K}_{\rho_\eta}}\otimes A.$$

*Definition of the injections $\tau_{\eta'\leftarrow\eta}$.* Let $\eta\preccurlyeq\eta'\in\mathcal{L}$. We define a $\mathbb{C}$-linear map:

$$\tau_{\eta'\leftarrow\eta}:\text{Vect}\left\{\sqrt{p_\eta^{(k)}}\left\langle\zeta_\eta^{(k)}\right|\,\middle|\,k\leqslant N_\eta\right\}\otimes\mathcal{H}_\eta \to \text{Vect}\left\{\sqrt{p_{\eta'}^{(k')}}\left\langle\zeta_{\eta'}^{(k')}\right|\,\middle|\,k'\leqslant N_{\eta'}\right\}\otimes\mathcal{H}_{\eta'}$$

by:

$$\forall k\leqslant N_\eta,\ \forall\psi\in\mathcal{H}_\eta,\ \tau_{\eta'\leftarrow\eta}\left(\sqrt{p_\eta^{(k)}}\langle\zeta_\eta^{(k)}|\otimes|\psi\rangle\right) =$$

$$= \sum_{k'\leqslant N_{\eta'},e}\left\langle\varphi_{\eta'\to\eta}^{-1}\left(e\otimes\zeta_\eta^{(k)}\right)\,\middle|\,\zeta_{\eta'}^{(k')}\right\rangle\sqrt{p_{\eta'}^{(k')}}\left\langle\zeta_{\eta'}^{(k')}\right|\otimes\left|\varphi_{\eta'\to\eta}^{-1}(e\otimes\psi)\right\rangle$$

$$= \sum_e \left\langle\sqrt{\rho_{\eta'}}\circ\varphi_{\eta'\to\eta}^{-1}\left(e\otimes\zeta_\eta^{(k)}\right)\right|\otimes\left|\varphi_{\eta'\to\eta}^{-1}(e\otimes\psi)\right\rangle,$$

where $(e)$ is an orthonormal basis of $\mathcal{H}_{\eta'\to\eta}$ and $\sqrt{\rho_{\eta'}}$ is defined via spectral resolution. We have:

$$\forall k\leqslant N_\eta,\ \forall\psi\in\mathcal{H}_\eta,\ \left\|\tau_{\eta'\leftarrow\eta}\left(\sqrt{p_\eta^{(k)}}\langle\zeta_\eta^{(k)}|\otimes|\psi\rangle\right)\right\|^2 =$$

$$= \sum_{e,e'}\left\langle\sqrt{\rho_{\eta'}}\circ\varphi_{\eta'\to\eta}^{-1}\left(e\otimes\zeta_\eta^{(k)}\right)\,\middle|\,\sqrt{\rho_{\eta'}}\circ\varphi_{\eta'\to\eta}^{-1}\left(e'\otimes\zeta_\eta^{(k)}\right)\right\rangle\left\langle\varphi_{\eta'\to\eta}^{-1}(e'\otimes\psi)\,\middle|\,\varphi_{\eta'\to\eta}^{-1}(e\otimes\psi)\right\rangle$$

$$= \sum_e \left\langle\varphi_{\eta'\to\eta}^{-1}\left(e\otimes\zeta_\eta^{(k)}\right)\,\middle|\,\rho_{\eta'}\circ\varphi_{\eta'\to\eta}^{-1}\left(e\otimes\zeta_\eta^{(k)}\right)\right\rangle\|\psi\|^2$$

$$= \left\langle\zeta_\eta^{(k)}\,\middle|\,\left(\text{Tr}_{\eta'\to\eta}\rho_{\eta'}\right)\zeta_\eta^{(k)}\right\rangle\|\psi\|^2$$

$$= p_\eta^{(k)}\|\psi\|^2 = \left\|\sqrt{p_\eta^{(k)}}\langle\zeta_\eta^{(k)}|\otimes|\psi\rangle\right\|^2.$$

Therefore $\tau_{\eta'\leftarrow\eta}$ is well-defined and can be extended as an injection $\mathcal{H}_{\rho_\eta}^{\text{GNS}}\to\mathcal{H}_{\rho_{\eta'}}^{\text{GNS}}$, which is isometric onto its image. For $A_\eta\in\mathcal{A}_\eta$ we can check directly that eq. (2.8.1) is satisfied, using $A_\eta^{\text{GNS}} = \text{id}_{\mathcal{K}_{\rho_\eta}}\otimes A_\eta$.

Next, we have:

$$\sum_{k\leqslant N_\eta}\text{Tr}_{\mathcal{H}_{\eta'}}\rho_{\eta'}\,\iota_{\eta'\leftarrow\eta}\left(|\zeta_\eta^{(k)}\rangle\langle\zeta_\eta^{(k)}|\right) = \text{Tr}_{\mathcal{H}_\eta}\rho_\eta = 1$$

$$= \sum_{k \leqslant N_\eta, k' \leqslant N_{\eta'}, e} p_{\eta'}^{(k')} \left| \left\langle \zeta_{\eta'}^{(k')} \middle| \varphi_{\eta' \to \eta}^{-1} \left( e \otimes \zeta_\eta^{(k)} \right) \right\rangle \right|^2 ,$$

but since all $p_{\eta'}^{(k')}$ are strictly positive and $\sum_{k' \leqslant N_{\eta'}} p_{\eta'}^{(k')} = 1$, this implies:

$$\forall k' \leqslant N_{\eta'}, \sum_{k \leqslant N_\eta, e} \left| \left\langle \zeta_{\eta'}^{(k')} \middle| \varphi_{\eta' \to \eta}^{-1} \left( e \otimes \zeta_\eta^{(k)} \right) \right\rangle \right|^2 = 1 = \left\| \zeta_{\eta'}^{(k')} \right\|^2 ,$$

and therefore:

$$\forall k' \leqslant N_{\eta'}, \zeta_{\eta'}^{(k')} = \sum_{k \leqslant N_\eta, e} \left\langle \varphi_{\eta' \to \eta}^{-1} \left( e \otimes \zeta_\eta^{(k)} \right) \middle| \zeta_{\eta'}^{(k')} \right\rangle \left| \varphi_{\eta' \to \eta}^{-1} \left( e \otimes \zeta_\eta^{(k)} \right) \right\rangle . \tag{2.8.4}$$

With this we can now prove:

$$\forall A_\eta \in \mathcal{A}_\eta, \tau_{\eta' \leftarrow \eta} \circ \Psi_{\rho_\eta}(A_\eta) =$$

$$= \sum_{k \leqslant N_\eta, k' \leqslant N_{\eta'}, e} \left\langle \varphi_{\eta' \to \eta}^{-1} \left( e \otimes \zeta_\eta^{(k)} \right) \middle| \zeta_{\eta'}^{(k')} \right\rangle \sqrt{p_{\eta'}^{(k')}} \left\langle \zeta_{\eta'}^{(k')} \middle| \otimes \middle| \varphi_{\eta' \to \eta}^{-1} \left( e \otimes A_\eta \zeta_\eta^{(k)} \right) \right\rangle$$

$$= \sum_{k' \leqslant N_{\eta'}} \sqrt{p_{\eta'}^{(k')}} \left\langle \zeta_{\eta'}^{(k')} \middle| \otimes \middle| \iota_{\eta' \leftarrow \eta}(A_\eta) \zeta_{\eta'}^{(k')} \right\rangle$$

$$= \Psi_{\rho_{\eta'}} \circ \iota_{\eta' \leftarrow \eta}(A_\eta) . \tag{2.8.5}$$

Thus, for $\eta \preccurlyeq \eta' \preccurlyeq \eta'' \in \mathcal{L}$, $\tau_{\eta'' \leftarrow \eta'} \circ \tau_{\eta' \leftarrow \eta} = \tau_{\eta'' \leftarrow \eta}$ follows from eq. (2.3.1) together with point 2.8.4 above.

*Inductive limit Hilbert space as GNS representation of $\overline{\mathcal{A}}_{(\mathcal{L}, \mathcal{H}, \varphi)}^\otimes$.* Let $\mathcal{H}_\rho^{\text{GNS}}$ be (the completion of) the inductive limit of $\left( \mathcal{L}, \left( \mathcal{H}_{\rho_\eta}^{\text{GNS}} \right)_{\eta \in \mathcal{L}}, \left( \tau_{\eta' \leftarrow \eta} \right)_{\eta \preccurlyeq \eta'} \right)$. Eq. (2.8.5) ensures that we can consistently assemble the family of maps $\left( \Psi_{\rho_\eta} \right)_{\eta \in \mathcal{L}}$ into a map $\Psi_\rho : \mathcal{A}_{(\mathcal{L}, \mathcal{H}, \varphi)}^\otimes \to \mathcal{H}_\rho^{\text{GNS}}$, and, by eq. (2.8.3), we can extend this map to $\overline{\mathcal{A}}_{(\mathcal{L}, \mathcal{H}, \varphi)}^\otimes$. Now, the properties of the individual $\Psi_{\rho_\eta}$ ensure that $\Psi_\rho$ has the following properties:

5. $\Psi_\rho$ is $\mathbb{C}$-linear;

6. $\forall \eta, \eta' \in \mathcal{L}, \forall [A_\eta]_\sim, [B_{\eta'}]_\sim \in \mathcal{A}_{(\mathcal{L}, \mathcal{H}, \varphi)}^\otimes, \forall \eta'' \succcurlyeq \eta, \eta', \Psi_\rho \left( [A_{\eta''} B_{\eta''}]_\sim \right) = \left[ A_{\eta''}^{\text{GNS}} \Psi_{\rho_{\eta''}}(B_{\eta''}) \right]_\sim$;

7. $\forall A, B \in \mathcal{A}_{(\mathcal{L}, \mathcal{H}, \varphi)}^\otimes, \left\langle \Psi_\rho(A), \Psi_\rho(B) \right\rangle_{\mathcal{H}_\rho^{\text{GNS}}} = \operatorname{Tr} \rho A^+ B$;

8. $\overline{\Psi_\rho \left\langle \overline{\mathcal{A}}_{(\mathcal{L}, \mathcal{H}, \varphi)}^\otimes \right\rangle} = \mathcal{H}_\rho^{\text{GNS}}$.

Therefore, we can identify $\mathcal{H}_\rho^{\text{GNS}}$ with the GNS representation of $\overline{\mathcal{A}}_{(\mathcal{L}, \mathcal{H}, \varphi)}^\otimes$ arising from the state $\rho$.

*Note.* This result could be proved at a more abstract level, by directly using eq. (2.8.5) to *define* $\tau_{\eta' \leftarrow \eta}$. Here we gave the explicit expressions as an added bonus.

*Pure projective state.* We now assume that for all $\eta \in \mathcal{L}$ $N_\eta = 0$ and we define $\forall \eta \in \mathcal{L}, \zeta_\eta := \zeta_\eta^{(0)}$. Thus $\forall \eta \in \mathcal{L}, \mathcal{K}_{\rho_\eta} \approx \mathbb{C}$ and therefore $\mathcal{H}_{\rho_\eta}^{\text{GNS}} \approx \mathcal{H}_\eta$. Then, for $\eta \preccurlyeq \eta'$, eq. (2.8.4) becomes:

$$\zeta_{\eta'} = \sum_{e \text{ BON of } \mathcal{H}_{\eta' \to \eta}} \left\langle \varphi_{\eta' \to \eta}^{-1}\left(e \otimes \zeta_\eta\right) \mid \zeta_{\eta'} \right\rangle \left| \varphi_{\eta' \to \eta}^{-1}\left(e \otimes \zeta_\eta\right) \right\rangle,$$

hence, defining $\zeta_{\eta' \to \eta} := \sum_{e \text{ BON of } \mathcal{H}_{\eta' \to \eta}} \left\langle \varphi_{\eta' \to \eta}^{-1}\left(e \otimes \zeta_\eta\right) \mid \zeta_{\eta'} \right\rangle |e\rangle$, we get $\zeta_{\eta'} = \varphi_{\eta' \to \eta}^{-1}\left(\zeta_{\eta' \to \eta} \otimes \zeta_\eta\right)$.

Inserting into the definition of $\tau_{\eta' \leftarrow \eta}$ and applying the identification $\mathcal{K}_{\rho_\eta} \approx \mathbb{C}$ provides eq. (2.8.2). □

**Theorem 2.9** Let $(\mathcal{L}, \mathcal{H}, \varphi)^\otimes$ be a projective system of quantum state spaces and suppose there exists a family of vectors $\left(\zeta_{\eta' \to \eta}\right)_{\eta \preccurlyeq \eta'}$ such that:

1. $\forall \eta \preccurlyeq \eta', \zeta_{\eta' \to \eta} \in \mathcal{H}_{\eta' \to \eta}$ & $\|\zeta_{\eta' \to \eta}\| = 1$;

2. $\forall \eta \preccurlyeq \eta' \preccurlyeq \eta'', \varphi_{\eta'' \to \eta' \to \eta}(\zeta_{\eta'' \to \eta}) = \zeta_{\eta'' \to \eta'} \otimes \zeta_{\eta' \to \eta}$.

We define an Hilbert space $\mathcal{H}_\zeta$ as (the completion of) the inductive limit of $\left(\mathcal{L}, \left(\mathcal{H}_\eta\right)_{\eta \in \mathcal{L}}, \left(\tau_{\eta' \leftarrow \eta}\right)_{\eta \preccurlyeq \eta'}\right)$, where the injective maps $\tau_{\eta' \leftarrow \eta}$ are defined as:

$$\forall \eta \preccurlyeq \eta' \in \mathcal{L}, \quad \tau_{\eta' \leftarrow \eta} : \begin{array}{rcl} \mathcal{H}_\eta & \to & \mathcal{H}_{\eta'} \\ \psi & \mapsto & \varphi_{\eta' \to \eta}^{-1}\left(\zeta_{\eta' \to \eta} \otimes \psi\right) \end{array}.$$

Then, there exist maps $\sigma : \overline{\mathcal{S}}_\zeta \to \overline{\mathcal{S}}_{(\mathcal{L}, \mathcal{H}, \varphi)}^\otimes$ and $\alpha : \overline{\mathcal{A}}_{(\mathcal{L}, \mathcal{H}, \varphi)}^\otimes \to \mathcal{A}_\zeta$ ($\overline{\mathcal{S}}_\zeta$ being the space of (self-adjoint) positive semi-definite, traceclass operators over $\mathcal{H}_\zeta$ and $\mathcal{A}_\zeta$ the algebra of bounded operators on $\mathcal{H}_\zeta$) such that:

3. $\forall \rho \in \overline{\mathcal{S}}_\zeta, \forall A \in \overline{\mathcal{A}}_{(\mathcal{L}, \mathcal{H}, \varphi)}^\otimes, \text{Tr}_{\mathcal{H}_\zeta}\left(\rho\, \alpha(A)\right) = \text{Tr}\left(\sigma(\rho)\, A\right)$;

4. $\sigma$ is injective;

5. $\sigma \langle \mathcal{S}_\zeta \rangle = \left\{ \left(\rho_\eta\right)_{\eta \in \mathcal{L}} \;\middle|\; \sup_{\eta \in \mathcal{L}} \left( \inf_{\eta' \succcurlyeq \eta} \text{Tr}_{\mathcal{H}_{\eta'}}\left(\rho_{\eta'}\, \Theta_{\eta'|\eta}\right) \right) = 1 \right\}$,

    where $\mathcal{S}_\zeta$ is the space of density matrices over $\mathcal{H}_\zeta$ and:

$$\Theta_{\eta'|\eta} := \varphi_{\eta' \to \eta}^{-1} \circ \left( |\zeta_{\eta' \to \eta}\rangle\langle\zeta_{\eta' \to \eta}| \otimes \text{id}_{\mathcal{H}_\eta} \right) \circ \varphi_{\eta' \to \eta}.$$

We will, in the proof below, rely heavily on the so-called trace norm, which, for a positive traceclass operator is just its trace. The reason why this is the appropriate norm for our purpose is twofold. First, it plays nicely with the partial traces, since the trace norm of a partial trace of $\rho$ is always bounded by the trace norm of $\rho$ itself (it is obviously equal in the case of a positive $\rho$, and the bound follows by decomposing a general $\rho$ into positive and negative parts, or by invoking the next point). Second, it supports the physical interpretation of quantum states, revolving around the evaluation of observable expectation values, since the trace norm of $\rho$ is precisely the norm of the continuous linear functional $A \mapsto \text{Tr}\, \rho A$ defined on the algebra of bounded operators (this can be proven using the polar decomposition of $\rho$ [21, theorem VI.10]). An additional advantage is that the traceclass operators form a Banach space with respect to this norm [22].

**Lemma 2.10** Let $\mathcal{H}$ be an Hilbert space. For any traceclass operator $\rho$ on $\mathcal{H}$ we define its trace norm $\|\rho\|_1$ (aka. Schatten-norm with $p = 1$ [22]) by:

$$\|\rho\|_1 := \operatorname{Tr} \sqrt{\rho^+ \rho}.$$

Let $(\mathcal{J}_\alpha)_\alpha$ be a family of closed vector subspaces of $\mathcal{H}$, forming a directed preordered set under inclusion, and such that:

$$\mathcal{H} = \overline{\bigcup_\alpha \mathcal{J}_\alpha}.$$

Define $\Theta_\alpha$ to be the orthogonal projection on $\mathcal{J}_\alpha$.

The following statements hold:

1. for any (self-adjoint) positive semi-definite, traceclass operator $\rho$ on $\mathcal{H}$, the net $(\Theta_\alpha \rho \Theta_\alpha)_\alpha$ converges in trace norm to $\rho$;

2. if $(\rho_\alpha)_\alpha$ is a net of (self-adjoint) positive semi-definite, traceclass operators on $\mathcal{H}$ such that:

    $$\forall \alpha, \alpha' \mid \mathcal{J}_\alpha \subset \mathcal{J}_{\alpha'}, \rho_\alpha = \Theta_\alpha \rho_{\alpha'} \Theta_\alpha,$$

    and if $\sup_\alpha \operatorname{Tr} \rho_\alpha = l < \infty$, then there exists a (self-adjoint) positive semi-definite, traceclass operator $\rho$ on $\mathcal{H}$ such that $\rho_\alpha = \Theta_\alpha \rho \Theta_\alpha$ and $\operatorname{Tr} \rho = l$.

**Proof** The trace norm is well-defined, since for any traceclass operator $\rho$ on $\mathcal{H}$, $\rho^+ \rho$ is a self-adjoint positive semi-definite operator on $\mathcal{H}$, so its square-root can be defined by spectral resolution, and this square-root is traceclass (by definition of $\rho$ being traceclass).

*Statement 2.10.1.* Let $\rho$ be a (self-adjoint) positive semi-definite, traceclass operator on $\mathcal{H}$. There exist real numbers $p_k \geqslant 0$ ($k \in \mathbb{N}$) and vectors $\psi_k$ in $\mathcal{H}$ with $\|\psi_k\| = 1$ such that:

$$\rho = \sum_k p_k |\psi_k\rangle\langle\psi_k| \quad \& \quad \sum_k p_k = \operatorname{Tr} \rho \geqslant 0.$$

Hence, we have for any $\alpha$:

$$\|\rho - \Theta_\alpha \rho \Theta_\alpha\|_1 \leqslant \sum_k p_k \left\| |\psi_k\rangle\langle\psi_k| - |\Theta_\alpha \psi_k\rangle\langle\Theta_\alpha \psi_k| \right\|_1.$$

Let $\epsilon > 0$ and let $N \in \mathbb{N}$ such that:

$$\sum_{k > N} 2 p_k \leqslant \frac{\epsilon}{2}.$$

Since $\mathcal{H}$ is the completion of the union of the $\mathcal{J}_\alpha$ (which are directed with respect to inclusion subordinate to the labels $\alpha$), there exists, for every $k \leqslant N$, an $\alpha_k$ such that $\|\psi_k - \Theta_{\alpha_k} \psi_k\| \leqslant \frac{\epsilon}{6}$. And since the family $(\mathcal{J}_\alpha)_\alpha$ is directed under inclusion, there exists $\alpha$ such that $\bigcup_{k \leqslant N} \mathcal{J}_{\alpha_k} \subset \mathcal{J}_\alpha$. Let $\alpha'$ such that $\mathcal{J}_{\alpha'} \supset \mathcal{J}_\alpha$. Then, we have:

$$\forall k \leqslant N, \|\psi_k - \Theta_{\alpha'} \psi_k\| \leqslant \frac{\epsilon}{6}.$$

On the other hand, for any $k \leqslant N$, the non-zero eigenvalues of $|\psi_k\rangle\langle\psi_k| - |\Theta_{\alpha'} \psi_k\rangle\langle\Theta_{\alpha'} \psi_k|$ are:

$$\lambda_{\pm} = \frac{\mu^2}{2} \pm \mu \sqrt{1 - \frac{3\mu^2}{4}} \text{ with } \mu := \|\psi_k - \Theta_{\alpha'}\psi_k\|,$$

each one with multiplicity 1. So from $\mu \leqslant \frac{\epsilon}{6}$ and $\mu \leqslant 1$ ($\|\psi_k\| = 1$), we have:

$$\left\| |\psi_k\rangle\langle\psi_k| - |\Theta_{\alpha'}\psi_k\rangle\langle\Theta_{\alpha'}\psi_k| \right\|_1 = |\lambda_+| + |\lambda_-| \leqslant 3\mu \leqslant \frac{\epsilon}{2}.$$

Therefore,

$$\|\rho - \Theta_{\alpha'}\rho\Theta_{\alpha'}\|_1 \leqslant \left( \sum_{k \leqslant N} p_k \frac{\epsilon}{2} \right) + \frac{\epsilon}{2} \leqslant \epsilon.$$

*Statement 2.10.2.* Since the family $(\mathfrak{J}_\alpha)_\alpha$ is directed and each $\mathfrak{J}_\alpha$ is a vector subspace of $\mathcal{H}$, $\mathfrak{J} := \bigcup_\alpha \mathfrak{J}_\alpha$ is a vector subspace of $\mathcal{H}$ and, by hypothesis, $\mathfrak{J}$ is dense in $\mathcal{H}$.

For any $\psi, \psi' \in \mathfrak{J}$, we define:

$$\rho_{\psi,\psi'} := \langle \psi, \rho_\alpha \psi' \rangle \text{ for } \alpha \text{ such that } \psi, \psi' \in \mathfrak{J}_\alpha.$$

$\rho_{\psi,\psi'}$ is well-defined, since there exists $\alpha_\psi$, resp. $\alpha_{\psi'}$, such that $\psi \in \mathfrak{J}_{\alpha_\psi}$, resp. $\psi \in \mathfrak{J}_{\alpha_\psi}$, hence there exists $\alpha$ such that $\psi, \psi' \in \mathfrak{J}_\alpha$; and if $\alpha'$ is an other index such that $\psi, \psi' \in \mathfrak{J}_{\alpha'}$, then there exists $\alpha''$ with $\mathfrak{J}_\alpha, \mathfrak{J}_{\alpha'} \subset \mathfrak{J}_{\alpha''}$, so we have:

$$\langle \psi, \rho_\alpha \psi' \rangle = \langle \Theta_\alpha \psi, \rho_{\alpha''} \Theta_\alpha \psi' \rangle = \langle \psi, \rho_{\alpha''} \psi' \rangle = \langle \Theta_{\alpha'} \psi, \rho_{\alpha''} \Theta_{\alpha'} \psi' \rangle = \langle \psi, \rho_{\alpha'} \psi' \rangle.$$

Moreover, $(\psi, \psi') \mapsto \rho_{\psi,\psi'}$ is a positive semi-definite, sesquilinear form on $\mathfrak{J}$ and:

$$\forall \psi, \psi' \in \mathfrak{J}, \ |\rho_{\psi,\psi'}| \leqslant l \ \|\psi\| \ \|\psi'\|,$$

hence, there exists a positive semi-definite, self-adjoint, bounded operator $\rho$ on $\mathcal{H}$, such that:

$$\forall \psi, \psi' \in \mathfrak{J}, \ \rho_{\psi,\psi'} := \langle \psi, \rho \psi' \rangle.$$

So, for any $\alpha$ and any $\psi, \psi' \in \mathcal{H}$, we have:

$$\langle \psi, \Theta_\alpha \rho \Theta_\alpha \psi' \rangle = \langle \Theta_\alpha \psi, \rho \Theta_\alpha \psi' \rangle = \langle \Theta_\alpha \psi, \rho_\alpha \Theta_\alpha \psi' \rangle = \langle \psi, \rho_\alpha \psi' \rangle,$$

therefore $\rho_\alpha = \Theta_\alpha \rho \Theta_\alpha$.

Now, suppose $\rho$ would not be traceclass. Then, there would exist a finite orthonormal family $\psi_k$, $k \in \{1, \ldots, N\}$ such that:

$$\sum_{k=1}^{N} \langle \psi_k, \rho \psi_k \rangle > l + 1,$$

Next, like in the proof of statement 2.10.1, we can find $\alpha$ satisfying:

$$\forall k \leqslant N, \ \forall \alpha' \ | \ \mathfrak{J}_{\alpha'} \supset \mathfrak{J}_\alpha, \ \|\psi_k - \Theta_{\alpha'}\psi_k\| \leqslant \frac{1}{N(2l+1)}.$$

Hence, using $\|\rho\| \leqslant l$ (where $\|\cdot\|$ is the operator norm):

$$\sum_{k=1}^{N} \langle \psi_k, \rho \psi_k \rangle \leqslant \sum_{k=1}^{N} \left( \langle \Theta_{\alpha'}\psi_k, \rho \Theta_{\alpha'}\psi_k \rangle + \frac{2l+1}{N(2l+1)} \right)$$

$$\leqslant \operatorname{Tr} \rho_{\alpha'} + 1 \leqslant l + 1 \,,$$

which would then be contradictory.

Lastly, $\rho$ being a (self-adjoint) positive semi-definite, traceclass operator on $\mathcal{H}$, the first statement implies that the net $(\rho_\alpha)_\alpha$ converges in trace norm to $\rho$, hence $\lim_\alpha \operatorname{Tr} \rho_\alpha = \operatorname{Tr} \rho$. So $\operatorname{Tr} \rho \leqslant l$ and, for any $\epsilon > 0$, there exists $\alpha_\epsilon$ such that:

$$\forall \alpha' \,/\, \mathcal{J}_{\alpha'} \supset \mathcal{J}_{\alpha_\epsilon},\ \operatorname{Tr} \rho_{\alpha'} \leqslant \operatorname{Tr} \rho + \epsilon \,.$$

Therefore, for any $\alpha$, choosing $\alpha'$ such that $\mathcal{J}_\alpha \cup \mathcal{J}_{\alpha_\epsilon} \subset \mathcal{J}_{\alpha'}$, we have:

$$\operatorname{Tr} \rho_\alpha \leqslant \operatorname{Tr} \rho_{\alpha'} \leqslant \operatorname{Tr} \rho + \epsilon \,.$$

Thus, $l \leqslant \operatorname{Tr} \rho$, hence $\operatorname{Tr} \rho = l$. $\square$

**Proof of theorem 2.9** *Existence of $\sigma$ and $\alpha$ satisfying 2.9.3.* The inductive limit defining $\mathcal{H}_\zeta$ is consistent since for all $\eta \preccurlyeq \eta' \preccurlyeq \eta'' \in \mathcal{L}$ and for all $\psi \in \mathcal{H}_\eta$:

$$\tau_{\eta'' \leftarrow \eta'} \circ \tau_{\eta' \leftarrow \eta}(\psi) = \varphi^{-1}_{\eta'' \to \eta'} \circ \left( \mathrm{id}_{\mathcal{H}_{\eta'' \to \eta'}} \otimes \varphi_{\eta' \to \eta} \right)^{-1} \left( \zeta_{\eta'' \to \eta'} \otimes \left( \zeta_{\eta' \to \eta} \otimes \psi \right) \right)$$

$$= \varphi^{-1}_{\eta'' \to \eta} \circ \left( \varphi_{\eta'' \to \eta' \to \eta} \otimes \mathrm{id}_{\mathcal{H}_\eta} \right)^{-1} \left( \left( \zeta_{\eta'' \to \eta'} \otimes \zeta_{\eta' \to \eta} \right) \otimes \psi \right)$$

$$= \varphi^{-1}_{\eta'' \to \eta} \left( \zeta_{\eta'' \to \eta} \otimes \psi \right) = \tau_{\eta'' \leftarrow \eta}(\psi) \,.$$

Additionally, for all $\eta \in \mathcal{L}$, we call $\tau_{\zeta \leftarrow \eta}$ the injective map $\mathcal{H}_\eta \to \mathcal{H}_\zeta$.

Let $\eta \in \mathcal{L}$. We define an Hilbert space $\mathcal{H}_{\zeta \to \eta}$ as (the completion of) the inductive limit of $\left( \{ \kappa \in \mathcal{L} \mid \kappa \succcurlyeq \eta \}, \left( \mathcal{H}_{\kappa \to \eta} \right)_{\kappa \succcurlyeq \eta}, \left( \tau_{\kappa' \leftarrow \kappa \to \eta} \right)_{\kappa' \succcurlyeq \kappa \succcurlyeq \eta} \right)$, where the injective maps $\tau_{\kappa' \leftarrow \kappa \to \eta}$ are defined as:

$$\forall \kappa' \succcurlyeq \kappa \succcurlyeq \eta, \quad \begin{array}{rcl} \tau_{\kappa' \leftarrow \kappa \to \eta} : \mathcal{H}_{\kappa \to \eta} & \to & \mathcal{H}_{\kappa' \to \eta} \\ \psi & \mapsto & \varphi^{-1}_{\kappa' \to \kappa \to \eta} (\zeta_{\kappa' \to \kappa} \otimes \psi) \end{array} \,.$$

We can prove that $\forall \kappa'' \succcurlyeq \kappa' \succcurlyeq \kappa \succcurlyeq \eta$, $\tau_{\kappa'' \leftarrow \kappa' \to \eta} \circ \tau_{\kappa' \leftarrow \kappa \to \eta} = \tau_{\kappa'' \leftarrow \kappa \to \eta}$ in a way similar to above, using:

$$\left( \mathrm{id}_{\mathcal{H}_{\kappa'' \to \kappa'}} \otimes \varphi_{\kappa' \to \kappa \to \eta} \right) \circ \varphi_{\kappa'' \to \kappa' \to \eta} = \left( \varphi_{\kappa'' \to \kappa' \to \kappa} \otimes \mathrm{id}_{\mathcal{H}_{\kappa \to \eta}} \right) \circ \varphi_{\kappa'' \to \kappa \to \eta} \,, \tag{2.9.1}$$

which can be proved by acting on both sides with $\left( \cdot \otimes \mathrm{id}_{\mathcal{H}_\eta} \right) \circ \varphi_{\kappa'' \to \eta}$ and using repeatedly eq. (2.1.1). Additionally, for all $\kappa \succcurlyeq \eta$, we call $\tau_{\zeta \leftarrow \kappa \to \eta}$ the injective map $\mathcal{H}_{\kappa \to \eta} \to \mathcal{H}_{\zeta \to \eta}$.

Then, we can combine the isomorphisms $\varphi_{\kappa \to \eta} : \mathcal{H}_\kappa \to \mathcal{H}_{\kappa \to \eta} \otimes \mathcal{H}_\eta$ defined for $\kappa \succcurlyeq \eta$ into an isomorphism $\varphi_{\zeta \to \eta} : \mathcal{H}_\zeta \to \mathcal{H}_{\zeta \to \eta} \otimes \mathcal{H}_\eta$, for we have, for all $\kappa' \succcurlyeq \kappa \succcurlyeq \eta$:

$$\varphi_{\kappa' \to \eta} \circ \tau_{\kappa' \leftarrow \kappa} = \left( \tau_{\kappa' \leftarrow \kappa \to \eta} \otimes \mathrm{id}_{\mathcal{H}_\eta} \right) \circ \varphi_{\kappa \to \eta} \,,$$

as can be shown using eq. (2.1.1).

Similarly, we can combine the isomorphisms $\varphi_{\kappa \to \eta' \to \eta} : \mathcal{H}_{\kappa \to \eta} \to \mathcal{H}_{\kappa \to \eta'} \otimes \mathcal{H}_{\eta' \to \eta}$ defined for $\kappa \succcurlyeq \eta' \succcurlyeq \eta$ into an isomorphism $\varphi_{\zeta \to \eta' \to \eta} : \mathcal{H}_{\zeta \to \eta} \to \mathcal{H}_{\zeta \to \eta'} \otimes \mathcal{H}_{\eta' \to \eta}$, for we have, for all $\kappa' \succcurlyeq \kappa \succcurlyeq \eta' \succcurlyeq \eta$:

$$\varphi_{\kappa'\to\eta'\to\eta} \circ \tau_{\kappa'\leftarrow\kappa\to\eta} = \left(\tau_{\kappa'\leftarrow\kappa\to\eta'} \otimes \mathrm{id}_{\mathcal{H}_{\eta'\to\eta}}\right) \circ \varphi_{\kappa\to\eta'\to\eta},$$

as can be shown using eq. (2.9.1).

Moreover, we have (again from eq. (2.1.1)):

$$\left(\varphi_{\zeta\to\eta'\to\eta} \otimes \mathrm{id}_{\mathcal{H}_\eta}\right) \circ \varphi_{\zeta\to\eta} = \left(\mathrm{id}_{\mathcal{H}_{\zeta\to\eta'}} \otimes \varphi_{\eta'\to\eta}\right) \circ \varphi_{\zeta\to\eta'}.$$

Now, if we define $\mathcal{L}_\zeta := \mathcal{L} \sqcup \{\zeta\}$ and extend the preorder on $\mathcal{L}$ to $\mathcal{L}_\zeta$ by requiring $\forall \eta \in \mathcal{L}$, $\eta \prec \zeta$, we can therefore assemble these objects into a projective system of quantum state spaces $\left(\mathcal{L}_\zeta, \mathcal{H}, \varphi\right)^\otimes$.

Using prop. 2.6, we then have maps $\sigma_\searrow : \overline{\mathcal{S}}^\otimes_{(\mathcal{L}\sqcup\{\zeta\},\mathcal{H},\varphi)} \to \overline{\mathcal{S}}^\otimes_{(\mathcal{L},\mathcal{H},\varphi)}$ and $\alpha_\nwarrow : \overline{\mathcal{A}}^\otimes_{(\mathcal{L},\mathcal{H},\varphi)} \to \overline{\mathcal{A}}^\otimes_{(\mathcal{L}\sqcup\{\zeta\},\mathcal{H},\varphi)}$, and, using prop. 2.7, we have maps $\sigma_\zeta^{-1} : \overline{\mathcal{S}}_\zeta \to \overline{\mathcal{S}}^\otimes_{(\mathcal{L}\sqcup\{\zeta\},\mathcal{H},\varphi)}$ and $\alpha_\zeta^{-1} : \overline{\mathcal{A}}^\otimes_{(\mathcal{L}\sqcup\{\zeta\},\mathcal{H},\varphi)} \to \mathcal{A}_\zeta$. Hence, we define:

$$\sigma := \sigma_\searrow \circ \sigma_\zeta^{-1} \quad \& \quad \alpha := \alpha_\zeta^{-1} \circ \alpha_\nwarrow.$$

*Properties of $\sigma$ (2.9.4 and 2.9.5).* For all $\eta \preccurlyeq \eta'$, we define:

$$\Theta_{\eta'|\eta} := \varphi_{\eta'\to\eta}^{-1} \circ \left(|\zeta_{\eta'\to\eta}\rangle\langle\zeta_{\eta'\to\eta}| \otimes \mathrm{id}_{\mathcal{H}_\eta}\right) \circ \varphi_{\eta'\to\eta} = \tau_{\eta'\leftarrow\eta}\,\tau_{\eta'\leftarrow\eta}^+,$$

which is the orthogonal projection on the image of $\tau_{\eta'\leftarrow\eta}$ in $\mathcal{H}_{\eta'}$, and, for all $\eta \in \mathcal{L}$, $\Theta_{\zeta|\eta}$, which is the orthogonal projection on the image of $\tau_{\zeta\leftarrow\eta}$ in $\mathcal{H}_\zeta$ and satisfy:

$$\Theta_{\zeta|\eta} \circ \tau_{\zeta\leftarrow\eta} = \tau_{\zeta\leftarrow\eta} \quad \& \quad \forall \eta' \succcurlyeq \eta,\ \Theta_{\zeta|\eta} \circ \tau_{\zeta\leftarrow\eta'} = \tau_{\zeta\leftarrow\eta'} \circ \Theta_{\eta'|\eta}.$$

We start by deriving a useful identity, for $\eta \preccurlyeq \eta' \in \mathcal{L}$ and $A$ a self-adjoint, traceclass operator on $\mathcal{H}_{\eta'}$:

$$\tau_{\eta'\leftarrow\eta}^+\, A\, \tau_{\eta'\leftarrow\eta} = \mathrm{Tr}_{\mathcal{H}_{\eta'\to\eta}}\left[\left(|\zeta_{\eta'\to\eta}\rangle\langle\zeta_{\eta'\to\eta}| \otimes \mathrm{id}_{\mathcal{H}_\eta}\right)\left(\varphi_{\eta'\to\eta} A\, \varphi_{\eta'\to\eta}^{-1}\right)\left(|\zeta_{\eta'\to\eta}\rangle\langle\zeta_{\eta'\to\eta}| \otimes \mathrm{id}_{\mathcal{H}_\eta}\right)\right]$$

$$= \mathrm{Tr}_{\eta'\to\eta}\left(\Theta_{\eta'|\eta}\, A\, \Theta_{\eta'|\eta}\right). \tag{2.9.2}$$

Let $\rho_\zeta \in \mathcal{S}_\zeta$. For $\kappa \in \mathcal{L}$, we define $\widetilde{\rho}_\kappa := \Theta_{\zeta|\kappa}\, \rho_\zeta\, \Theta_{\zeta|\kappa}$, and $\delta_\kappa := \rho_\zeta - \widetilde{\rho}_\kappa$. $\widetilde{\rho}_\kappa$ is a positive semi-definite, traceclass operator, and, from lemma 2.10.1, $\delta_\kappa$ converges in trace norm to 0.

Moreover, for any $\kappa \in \mathcal{L}$, $\delta_\kappa$ is self-adjoint, so we can write $\delta_\kappa = \delta_\kappa^+ - \delta_\kappa^-$ where $\delta_\kappa^\pm$ are the positive and negative parts of $\delta_\kappa$ (defined by spectral resolution), hence $\left(\sigma(\delta_\kappa^\pm)\right)_\eta$ are positive semi-definite, self-adjoint operators on $\mathcal{H}_\eta$, and from the conical linearity of $\sigma$:

$$\forall \kappa \in \mathcal{L},\ \sigma(\rho_\zeta) = \sigma(\widetilde{\rho}_\kappa) + \sigma(\delta_\kappa^+) - \sigma(\delta_\kappa^-) =: \sigma(\widetilde{\rho}_\kappa) + \sigma(\delta_\kappa),$$

hence:

$$\forall \kappa \in \mathcal{L},\ \forall \eta \in \mathcal{L},\ \left(\sigma(\rho_\zeta)\right)_\eta = \left(\sigma(\widetilde{\rho}_\kappa)\right)_\eta + \left(\sigma(\delta_\kappa)\right)_\eta.$$

Additionally, we have:

$$\mathrm{Tr}_{\mathcal{H}_\eta}\left(\sigma(\delta_\kappa^\pm)\right)_\eta = \mathrm{Tr}\left(\sigma(\delta_\kappa^\pm)\mathbb{1}\right) = \mathrm{Tr}_{\mathcal{H}_\zeta}\left(\delta_\kappa^\pm\, \alpha(\mathbb{1})\right) = \mathrm{Tr}_{\mathcal{H}_\zeta}\left(\delta_\kappa^\pm\, \mathrm{id}_{\mathcal{H}_\zeta}\right) = \mathrm{Tr}_{\mathcal{H}_\zeta}\delta_\kappa^\pm,$$

where $\mathbb{1} \in \overline{\mathcal{A}}^\otimes_{(\mathcal{L},\mathcal{H},\varphi)}$ is the equivalence class of $\mathrm{id}_{\mathcal{H}_\eta}$. So, we get:

$$\forall \kappa \in \mathcal{L}, \; \left\|(\sigma(\delta_\kappa))_\eta\right\|_1 \leqslant \left\|(\sigma(\delta_\kappa^+))_\eta\right\|_1 + \left\|(\sigma(\delta_\kappa^-))_\eta\right\|_1 = \mathrm{Tr}_{\mathcal{H}_\zeta} \, \delta_\kappa^+ + \mathrm{Tr}_{\mathcal{H}_\zeta} \, \delta_\kappa^- = \|\delta_\kappa\|_1,$$

therefore the net $\left((\sigma(\widetilde{\rho}_\kappa))_\eta\right)_{\kappa \in \mathcal{L}}$ converges in trace norm to $(\sigma(\rho_\zeta))_\eta$.

Now, for $\eta' \succcurlyeq \kappa \in \mathcal{L}$, we have:

$$(\sigma(\widetilde{\rho}_\kappa))_{\eta'} = \mathrm{Tr}_{\zeta \to \eta'} \, \widetilde{\rho}_\kappa = \mathrm{Tr}_{\zeta \to \eta'} \, \Theta_{\zeta|\eta'} \, \widetilde{\rho}_\kappa \, \Theta_{\zeta|\eta'}$$

$$= \tau^+_{\zeta \leftarrow \eta'} \, \widetilde{\rho}_\kappa \, \tau_{\zeta \leftarrow \eta'}$$

$$= \tau_{\eta' \leftarrow \kappa} \, \tau^+_{\zeta \leftarrow \kappa} \, \rho_\zeta \, \tau_{\zeta \leftarrow \kappa} \, \tau^+_{\eta' \leftarrow \kappa}$$

$$= \varphi^{-1}_{\eta' \to \kappa} \left( |\zeta_{\eta' \to \kappa}\rangle\langle\zeta_{\eta' \to \kappa}| \otimes \left(\tau^+_{\zeta \leftarrow \kappa} \, \rho_\zeta \, \tau_{\zeta \leftarrow \kappa}\right) \right) \varphi_{\eta' \to \kappa}, \quad (2.9.3)$$

and, for $\eta' \succcurlyeq \kappa' \succcurlyeq \kappa \in \mathcal{L}$:

$$\Theta_{\eta'|\kappa} \, (\sigma(\widetilde{\rho}_{\kappa'}))_{\eta'} \, \Theta_{\eta'|\kappa} = \tau_{\eta' \leftarrow \kappa} \, \tau^+_{\kappa' \leftarrow \kappa} \, \tau^+_{\zeta \leftarrow \kappa'} \, \rho_\zeta \, \tau_{\zeta \leftarrow \kappa'} \, \tau_{\kappa' \leftarrow \kappa} \, \tau^+_{\eta' \leftarrow \kappa}$$

$$= (\sigma(\widetilde{\rho}_\kappa))_{\eta'}.$$

Hence, for all $\eta, \kappa \in \mathcal{L}$, and for all $\eta' \in \mathcal{L}$ such that $\eta' \succcurlyeq \eta$ and $\eta' \succcurlyeq \kappa$, we have:

$$\mathrm{Tr}_{\eta' \to \eta} \, \Theta_{\eta'|\kappa} \, \left(\sigma(\widetilde{\rho}_{\eta'})\right)_{\eta'} \, \Theta_{\eta'|\kappa} = (\sigma(\widetilde{\rho}_\kappa))_\eta. \quad (2.9.4)$$

On the other hand:

$$\left\| \left[ \mathrm{Tr}_{\eta' \to \eta} \, \Theta_{\eta'|\kappa} \, \left(\sigma(\rho_\zeta)\right)_{\eta'} \, \Theta_{\eta'|\kappa} \right] - (\sigma(\widetilde{\rho}_\kappa))_\eta \right\|_1$$

$$= \left\| \left[ \mathrm{Tr}_{\eta' \to \eta} \, \Theta_{\eta'|\kappa} \, \left(\sigma(\rho_\zeta)\right)_{\eta'} \, \Theta_{\eta'|\kappa} \right] - \left[ \mathrm{Tr}_{\eta' \to \eta} \, \Theta_{\eta'|\kappa} \, \left(\sigma(\widetilde{\rho}_{\eta'})\right)_{\eta'} \, \Theta_{\eta'|\kappa} \right] \right\|_1$$

$$\leqslant \left\| \Theta_{\eta'|\kappa} \, \left(\sigma(\rho_\zeta) - \sigma(\widetilde{\rho}_{\eta'})\right)_{\eta'} \, \Theta_{\eta'|\kappa} \right\|_1$$

$$= \left\| \Theta_{\eta'|\kappa} \, \left(\sigma(\delta_{\eta'})\right)_{\eta'} \, \Theta_{\eta'|\kappa} \right\|_1 \leqslant \|\delta_{\eta'}\|_1,$$

as can be shown by decomposing the self-adjoint operator $\delta_{\eta'}$ in positive and negative parts. Therefore, we have:

$$\lim_{\eta' \succcurlyeq \kappa, \eta} \mathrm{Tr}_{\eta' \to \eta} \left( \Theta_{\eta'|\kappa} \, \left(\sigma(\rho_\zeta)\right)_{\eta'} \, \Theta_{\eta'|\kappa} \right) = (\sigma(\widetilde{\rho}_\kappa))_\eta, \quad (2.9.5)$$

where the limit is taken in the trace norm. And we can now take the net limit on $\kappa$:

$$\lim_{\kappa \in \mathcal{L}} \lim_{\eta' \succcurlyeq \kappa, \eta} \mathrm{Tr}_{\eta' \to \eta} \left( \Theta_{\eta'|\kappa} \, \left(\sigma(\rho_\zeta)\right)_{\eta'} \, \Theta_{\eta'|\kappa} \right) = (\sigma(\rho_\zeta))_\eta. \quad (2.9.6)$$

Now, for $\rho_\zeta \neq \rho'_\zeta \in \mathcal{S}_\zeta$, there should exist $\kappa \in \mathcal{L}$ such that:

$$\left(\tau^+_{\zeta \leftarrow \kappa} \circ \rho_\zeta \circ \tau_{\zeta \leftarrow \kappa}\right) \neq \left(\tau^+_{\zeta \leftarrow \kappa} \circ \rho'_\zeta \circ \tau_{\zeta \leftarrow \kappa}\right),$$

which, from eq. (2.9.3), implies:

$$\forall \eta \succcurlyeq \kappa, \; (\sigma(\widetilde{\rho}_\kappa))_\eta \neq (\sigma(\widetilde{\rho}'_\kappa))_\eta,$$

but, using eq. (2.9.5), $(\sigma(\widetilde{\rho}_\kappa))_\eta$ can be computed from $\sigma(\rho_\zeta)$, hence $\sigma(\rho_\zeta) \neq \sigma(\rho'_\zeta)$. Therefore, $\sigma|_{S_\zeta}$ is injective, so, from the conical linearity of $\sigma$, $\sigma$ is injective.

Then, for $\rho_\zeta \in S_\zeta$, we have (from eq. (2.9.6)):

$$\lim_{\eta \in \mathcal{L}} \lim_{\eta' \succcurlyeq \eta} \mathrm{Tr}_{\mathcal{H}_{\eta'}} \left( \left(\sigma(\rho_\zeta)\right)_{\eta'} \Theta_{\eta'|\eta} \right) = \mathrm{Tr}_{\mathcal{H}_\zeta} \rho_\zeta = 1,$$

and, since the net $\left( \mathrm{Tr}_{\mathcal{H}_{\eta'}} \left( \left(\sigma(\rho_\zeta)\right)_{\eta'} \Theta_{\eta'|\eta} \right) \right)_{\eta' \succcurlyeq \eta}$ is decreasing, while the net:

$$\left( \lim_{\eta' \succcurlyeq \eta} \mathrm{Tr}_{\mathcal{H}_{\eta'}} \left( \left(\sigma(\rho_\zeta)\right)_{\eta'} \Theta_{\eta'|\eta} \right) \right)_{\eta \in \mathcal{L}} = \left( \mathrm{Tr}_{\mathcal{H}_\zeta} \left( \widetilde{\rho}_\eta \right) \right)_{\eta \in \mathcal{L}},$$

is increasing, the limits are given respectively by the infimum and by the supremum, so:

$$\sigma(\rho_\zeta) \in \left\{ (\rho_\eta)_{\eta \in \mathcal{L}} \in S^\otimes_{(\mathcal{L},\mathcal{H},\varphi)} \;\middle|\; \sup_{\eta \in \mathcal{L}} \left( \inf_{\eta' \succcurlyeq \eta} \mathrm{Tr}_{\mathcal{H}_{\eta'}} \left( \rho_{\eta'} \Theta_{\eta'|\eta} \right) \right) = 1 \right\}.$$

To prove that this condition indeed characterizes $\sigma \langle S_\zeta \rangle$, we now consider $(\rho_\eta)_{\eta \in \mathcal{L}} \in S^\otimes_{(\mathcal{L},\mathcal{H},\varphi)}$ such that:

$$\sup_{\eta \in \mathcal{L}} \left( \inf_{\eta' \succcurlyeq \eta} \mathrm{Tr}_{\mathcal{H}_{\eta'}} \left( \rho_{\eta'} \Theta_{\eta'|\eta} \right) \right) = 1.$$

Let $\eta \in \mathcal{L}$. We have $0 \leqslant \inf_{\eta' \succcurlyeq \eta} \mathrm{Tr}_{\mathcal{H}_{\eta'}} \left( \rho_{\eta'} \Theta_{\eta'|\eta} \right) = \mu_\eta \leqslant 1$. We consider the net $(\check{\rho}_{\eta'|\eta})_{\eta' \succcurlyeq \eta}$, where $\check{\rho}_{\eta'|\eta}$ is a positive semi-definite, traceclass operator on $\mathcal{H}_\eta$ defined by:

$$\check{\rho}_{\eta'|\eta} := \mathrm{Tr}_{\eta' \to \eta} \left( \Theta_{\eta'|\eta} \rho_{\eta'} \Theta_{\eta'|\eta} \right).$$

For $\eta'' \succcurlyeq \eta' \succcurlyeq \eta \in \mathcal{L}$, we have:

$$\check{\rho}_{\eta'|\eta} - \check{\rho}_{\eta''|\eta} = \mathrm{Tr}_{\eta'' \to \eta} \left( \left( \iota_{\eta'' \leftarrow \eta'}(\Theta_{\eta'|\eta}) - \Theta_{\eta''|\eta} \right) \rho_{\eta''} \left( \iota_{\eta'' \leftarrow \eta'}(\Theta_{\eta'|\eta}) - \Theta_{\eta''|\eta} \right) \right),$$

and $\iota_{\eta'' \leftarrow \eta'}(\Theta_{\eta'|\eta}) - \Theta_{\eta''|\eta} = \varphi^{-1}_{\eta'' \to \eta'} \circ \left[ \left( \mathrm{id}_{\mathcal{H}_{\eta'' \to \eta'}} - |\zeta_{\eta'' \to \eta'}\rangle\langle\zeta_{\eta'' \to \eta'}| \right) \otimes \Theta_{\eta'|\eta} \right] \circ \varphi_{\eta'' \to \eta'}$,

hence $\check{\rho}_{\eta'|\eta} - \check{\rho}_{\eta''|\eta}$ is also a positive semi-definite, traceclass operator on $\mathcal{H}_\eta$. Its trace is $\mathrm{Tr}_{\mathcal{H}_{\eta'}} \left( \rho_{\eta'} \Theta_{\eta'|\eta} \right) - \mathrm{Tr}_{\mathcal{H}_{\eta''}} \left( \rho_{\eta''} \Theta_{\eta''|\eta} \right) = \mathrm{Tr}_{\mathcal{H}_\eta} \check{\rho}_{\eta'|\eta} - \mathrm{Tr}_{\mathcal{H}_\eta} \check{\rho}_{\eta''|\eta}$, therefore $\left( \mathrm{Tr}_{\mathcal{H}_\eta} \check{\rho}_{\eta'|\eta} \right)_{\eta' \succcurlyeq \eta}$ is decreasing and converges to $\mu_\eta$.

Thus, $(\check{\rho}_{\eta'|\eta})_{\eta' \succcurlyeq \eta}$ is a Cauchy net and, since the traceclass operators form a Banach space with respect to the trace norm [22], it converges in trace norm to a positive semi-definite, traceclass operator $\check{\rho}_\eta$ on $\mathcal{H}_\eta$, with $\mathrm{Tr}_{\mathcal{H}_\eta} \check{\rho}_\eta = \mu_\eta$.

Moreover, for $\kappa \preccurlyeq \kappa' \in \mathcal{L}$, we have:

$$\tau^+_{\kappa' \leftarrow \kappa} \check{\rho}_{\kappa'} \tau_{\kappa' \leftarrow \kappa} = \lim_{\eta \succcurlyeq \kappa'} \tau^+_{\kappa' \leftarrow \kappa} \check{\rho}_{\eta|\kappa'} \tau_{\kappa' \leftarrow \kappa}$$

$$= \lim_{\eta \succcurlyeq \kappa'} \check{\rho}_{\eta|\kappa} = \check{\rho}_\kappa \quad \text{(using eq. (2.9.2))}.$$

Hence, since $\sup_{\kappa \in \mathcal{L}} \mathrm{Tr}_{\mathcal{H}_\kappa} \check{\rho}_\kappa = 1$, there exists, from lemma 2.10.2, an operator $\check{\rho}_\zeta \in S_\zeta$ satisfying:

$$\forall \kappa \in \mathcal{L},\ \tau^+_{\zeta \leftarrow \kappa}\ \check{\rho}_\zeta\ \tau_{\zeta \leftarrow \kappa} = \check{\rho}_\kappa\,.$$

Therefore, we have:

$$\forall \eta, \kappa \in \mathcal{L},\ \forall \eta' \succcurlyeq \eta, \kappa,\ \left(\sigma\left(\widetilde{(\check{\rho}_\zeta)}_\kappa\right)\right)_\eta = \mathrm{Tr}_{\eta' \to \eta}\, \Theta_{\eta'|\kappa}\, \check{\rho}_{\eta'}\, \Theta_{\eta'|\kappa} \quad \text{(using eqs. (2.9.3) and (2.9.4))},$$

hence, applying for $\eta' = \kappa \succcurlyeq \eta \in \mathcal{L}$:

$$\left(\sigma\left(\check{\rho}_\zeta\right)\right)_\eta = \lim_{\kappa \succcurlyeq \eta} \mathrm{Tr}_{\kappa \to \eta}\, \check{\rho}_\kappa$$

$$= \lim_{\kappa \succcurlyeq \eta}\, \lim_{\kappa' \succcurlyeq \kappa} \mathrm{Tr}_{\kappa \to \eta}\, \check{\rho}_{\kappa'|\kappa}$$

On the other hand, we can show as above that for any $\kappa \preccurlyeq \kappa'$, $\rho_\kappa - \check{\rho}_{\kappa'|\kappa}$ is a positive semi-definite, traceclass operator on $\mathcal{H}_\kappa$, with trace smaller than $1 - \mu_\kappa$. Thus, $\rho = \sigma\left(\check{\rho}_\zeta\right)$. $\square$

Finally, we also consider the case of infinite tensor products [25, 24]. Given a family of Hilbert spaces $(\mathcal{J}_\lambda)_{\lambda \in \mathcal{F}}$, we can build its infinite tensor product (ITP) $\mathcal{H}_\mathcal{F}$, which will in general be a non-separable Hilbert space. On the other hand, we can also build a projective system of quantum state spaces where the 'small' Hilbert spaces are given by tensor products of finitely many $\mathcal{J}_\lambda$. In this case, we still can map density matrices on $\mathcal{H}_\mathcal{F}$ to states in the projective limit, but this mapping will no longer be injective, because we can define considerably more observables over the infinite tensor product, and we can use them to distinguish between states that are indistinguishable if we solely use the algebra of observables defined over the projective system. However, if we believe that these latter observables (which can be sensible only to correlations between finitely many $\mathcal{J}_\lambda$) are the only *experimentally measurable* ones, additional distinctions between states might be objectionable. Interestingly, the ITP $\mathcal{H}_\mathcal{F}$, while being a really huge Hilbert space, still fails (except in absolutely degenerate cases) to reproduce the full state space of the projective system: this can be traced back to the fact that the latter allows to model states that are patently more 'statistical' than any state realizable on $\mathcal{H}_\mathcal{F}$. Also, grouping the tensor product factors $\mathcal{J}_\lambda$ into finite tensor products *before* performing the ITP construction generically gives rise to inequivalent Hilbert spaces (ie. ITP's are not associative, see [25, section 4.2]), while such a grouping does not affect the projective state space (as a consequence of prop. 2.6).

**Theorem 2.11** Let $(\mathcal{J}_\lambda)_{\lambda \in \mathcal{F}}$ be a family of Hilbert spaces and define:

1. $\mathcal{L} := \{\Lambda \subset \mathcal{F} \mid \#\Lambda < \infty\}$ equipped with the preorder $\subset$;
2. $\forall \Lambda \in \mathcal{L},\ \mathcal{H}_\Lambda := \bigotimes_{\lambda \in \Lambda} \mathcal{J}_\lambda$;
3. $\forall \Lambda \subset \Lambda' \in \mathcal{L},\ \mathcal{H}_{\Lambda' \to \Lambda} := \mathcal{H}_{\Lambda' \setminus \Lambda}$ with $\varphi_{\Lambda' \to \Lambda}$ the natural identification $\mathcal{H}_{\Lambda'} \to \mathcal{H}_{\Lambda' \setminus \Lambda} \otimes \mathcal{H}_\Lambda$.

Then, we can complete these elements into a projective system of quantum state spaces $(\mathcal{L}, \mathcal{H}, \varphi)^\otimes$.

Let $\mathcal{H}_\mathcal{F}$ be infinite tensor product of $(\mathcal{J}_\lambda)_{\lambda \in \mathcal{F}}$. There exist maps $\sigma : \overline{\mathcal{S}}_\mathcal{F} \to \overline{\mathcal{S}}^\otimes_{(\mathcal{L}, \mathcal{H}, \varphi)}$ and $\alpha : \overline{\mathcal{A}}^\otimes_{(\mathcal{L}, \mathcal{H}, \varphi)} \to \mathcal{A}_\mathcal{F}$ such that:

4. $\forall \rho \in \overline{\mathcal{S}}_\mathcal{F},\ \forall A \in \overline{\mathcal{A}}^\otimes_{(\mathcal{L}, \mathcal{H}, \varphi)},\ \mathrm{Tr}_{\mathcal{H}_\mathcal{F}}(\rho\, \alpha(A)) = \mathrm{Tr}(\sigma(\rho)\, A)$;
5. if $\{\lambda \in \mathcal{F} \mid \dim \mathcal{J}_\lambda > 1\}$ is infinite, $\sigma$ is neither injective nor surjective.

**Proof** Clearly, $(\mathcal{L}, \subset)$ is a directed set and, defining, for any $\Lambda \subset \Lambda' \subset \Lambda'' \in \mathcal{L}$, $\varphi_{\Lambda'' \to \Lambda' \to \Lambda}$ as the natural identification $\mathcal{H}_{\Lambda'' \setminus \Lambda} \to \mathcal{H}_{\Lambda'' \setminus \Lambda'} \otimes \mathcal{H}_{\Lambda' \setminus \Lambda}$, we obtain a projective system of quantum state spaces $(\mathcal{L}, \mathcal{H}, \varphi)^{\otimes}$.

*Existance of $\sigma$ and $\alpha$ satisfying 2.11.4.* The ITP $\mathcal{H}_{\mathcal{F}}$ arising from $(\mathcal{J}_\lambda)_{\lambda \in \mathcal{F}}$ can be written as [25, chapter 4]:

$$\mathcal{H}_{\mathcal{F}} = \overline{\bigoplus_{[f]} \mathcal{H}_{[f]}},$$

where the $[f]$ are equivalence classes in $\left\{ (f_\lambda)_{\lambda \in \mathcal{F}} \in (\mathcal{J}_\lambda)_{\lambda \in \mathcal{F}} \;\middle|\; \sum_{\lambda \in \mathcal{F}} \left| \|f_\lambda\|_{\mathcal{J}_\lambda} - 1 \right| \text{ converges} \right\}$ for the equivalence relation:

$$(f_\lambda)_{\lambda \in \mathcal{F}} \simeq (g_\lambda)_{\lambda \in \mathcal{F}} \Leftrightarrow \sum_{\lambda \in \mathcal{F}} \left| \langle f_\lambda, g_\lambda \rangle_{\mathcal{J}_\lambda} - 1 \right| \text{ converges },$$

and the Hilbert space $\mathcal{H}_{[f]}$ is (the completion of) the inductive limit of $\left( \mathcal{L}, \left( \bigotimes_{\lambda \in \Lambda} \mathcal{J}_\lambda \right)_{\Lambda \in \mathcal{L}}, \left( \tau^f_{\Lambda' \leftarrow \Lambda} \right)_{\Lambda \subset \Lambda'} \right)$, the inductive maps $\tau^f_{\Lambda' \leftarrow \Lambda}$ being defined as:

$$\forall \Lambda \subset \Lambda', \quad \begin{array}{c} \tau^f_{\Lambda' \leftarrow \Lambda} : \bigotimes_{\lambda \in \Lambda} \mathcal{J}_\lambda \to \bigotimes_{\lambda \in \Lambda'} \mathcal{J}_\lambda \\ \psi \mapsto \left( \bigotimes_{\lambda \in \Lambda' \setminus \Lambda} f_\lambda \right) \otimes \psi \end{array},$$

for $f$ some representative of $[f]$ such that $\forall \lambda \in \mathcal{F}$, $\|f_\lambda\|_{\mathcal{J}_\lambda} = 1$.

Now, we can identify $\mathcal{H}_{[f]}$ with the Hilbert space $\mathcal{H}_{\zeta^f}$ constructed as in theorem 2.9 for the family $\left( \zeta^f_{\Lambda' \to \Lambda} \right)_{\Lambda \subset \Lambda'}$ given by:

$$\forall \Lambda, \zeta^f_{\Lambda \to \Lambda} = 1 \quad \& \quad \forall \Lambda \subsetneq \Lambda', \zeta^f_{\Lambda' \to \Lambda} := \bigotimes_{\lambda \in \Lambda' \setminus \Lambda} f_\lambda.$$

Hence, as in the proof of theorem 2.9, we can construct for all $\Lambda \in \mathcal{L}$ an Hilbert space $\mathcal{H}_{[f] \to \Lambda}$ and an Hilbert space isomorphism $\varphi_{[f] \to \Lambda} : \mathcal{H}_{[f]} \to \mathcal{H}_{[f] \to \Lambda} \otimes \mathcal{H}_\Lambda$, and for all $\Lambda \subset \Lambda'$, we can construct an Hilbert space isomorphism $\varphi_{[f] \to \Lambda' \to \Lambda} : \mathcal{H}_{[f] \to \Lambda} \to \mathcal{H}_{[f] \to \Lambda'} \otimes \mathcal{H}_{\Lambda' \to \Lambda}$, satisfying:

$$\left( \varphi_{[f] \to \Lambda' \to \Lambda} \otimes \mathrm{id}_{\mathcal{H}_\Lambda} \right) \circ \varphi_{[f] \to \Lambda} = \left( \mathrm{id}_{\mathcal{H}_{[f] \to \Lambda'}} \otimes \varphi_{\Lambda' \to \Lambda} \right) \circ \varphi_{[f] \to \Lambda'}.$$

Now, we define:

$$\forall \Lambda \in \mathcal{L}, \quad \mathcal{H}_{\mathcal{F} \to \Lambda} = \overline{\bigoplus_{[f]} \mathcal{H}_{[f] \to \Lambda}}, \; \varphi_{\mathcal{F} \to \Lambda} = \overline{\bigoplus_{[f]} \varphi_{[f] \to \Lambda}} \quad \& \quad \forall \Lambda \subset \Lambda', \; \varphi_{\mathcal{F} \to \Lambda' \to \Lambda} = \overline{\bigoplus_{[f]} \varphi_{[f] \to \Lambda' \to \Lambda}}.$$

Note that $\mathcal{H}_{\mathcal{F} \to \Lambda}$ can also be identified as the ITP of $(\mathcal{J}_\lambda)_{\lambda \in \mathcal{F} \setminus \Lambda}$.

Defining $\overline{\mathcal{L}} := \mathcal{L} \cup \{\mathcal{F}\}$ and extending the preorder on $\mathcal{L}$ to $\overline{\mathcal{L}}$ by $\forall \Lambda \in \mathcal{L}, \Lambda \subset \mathcal{F}$, we thus have a projective system of quantum state spaces $\left( \overline{\mathcal{L}}, \mathcal{H}, \varphi \right)^{\otimes}$.

As in the proof of theorem 2.9, we can then define $\sigma$ and $\alpha$ by first using prop. 2.7 to go from $\mathcal{H}_{\mathcal{F}}$ to $\left( \overline{\mathcal{L}}, \mathcal{H}, \varphi \right)^{\otimes}$ and then using prop. 2.6 to go from $\left( \overline{\mathcal{L}}, \mathcal{H}, \varphi \right)^{\otimes}$ to $(\mathcal{L}, \mathcal{H}, \varphi)^{\otimes}$.

*Properties of $\sigma$ (2.11.5).* Let $\rho_{\mathcal{F}} \in \overline{\mathcal{S}}_{\mathcal{F}}$. For $\Lambda \in \mathcal{L}$, we have:

$$\left(\sigma(\rho_{\mathcal{F}})\right)_\Lambda = \mathrm{Tr}_{\mathcal{F}\to\Lambda}\,\rho_{\mathcal{F}} = \mathrm{Tr}_{\mathcal{H}_{\mathcal{F}\to\Lambda}}\left(\varphi_{\mathcal{F}\to\Lambda}\circ\rho_{\mathcal{F}}\circ\varphi_{\mathcal{F}\to\Lambda}^{-1}\right)$$

$$= \sum_{[f]} \mathrm{Tr}_{\mathcal{H}_{[f]\to\Lambda}}\left(\varphi_{[f]\to\Lambda}\circ\Pi_{[f]}\circ\rho_{\mathcal{F}}\circ\Pi_{[f]}^{+}\circ\varphi_{[f]\to\Lambda}^{-1}\right),$$

where $\Pi_{[f]} : \mathcal{H}_{\mathcal{F}} \to \mathcal{H}_{[f]}$ is the orthogonal projection on $\mathcal{H}_{[f]}$.

Thus, $\sigma$ does not see the correlations between different $\mathcal{H}_{[f]}$ that might be contained in $\rho_{\mathcal{F}}$, and therefore $\sigma$ cannot be injective if there exist more than one equivalence class (as will be the case if $\{\lambda\in\mathcal{F} \mid \dim\mathcal{J}_\lambda > 1\}$ is infinite, see below). Indeed, if $\psi$, $\psi'$ are two normalized states in $\mathcal{H}_{\mathcal{F}}$ with $\psi\in\mathcal{H}_{[f]}$, $\psi'\in\mathcal{H}_{[g]}$ and $[f]\neq[g]$, then:

$$\sigma\left(\frac{1}{2}|\psi\rangle\langle\psi| + \frac{1}{2}|\psi'\rangle\langle\psi'|\right) = \sigma\left(\left|\frac{\psi+\psi'}{\sqrt{2}}\right\rangle\left\langle\frac{\psi+\psi'}{\sqrt{2}}\right|\right).$$

On the other hand, let $\rho = (\rho_\Lambda)_{\Lambda\in\mathcal{L}} \in \overline{\mathcal{S}}^{\otimes}_{(\mathcal{L},\mathcal{H},\varphi)}$ and suppose there exists $\rho_{\mathcal{F}} \in \overline{\mathcal{S}}_{\mathcal{F}}$ such that $\rho = \sigma(\rho_{\mathcal{F}})$. If $\mathrm{Tr}_{\mathcal{H}_{\mathcal{F}}}\rho_{\mathcal{F}} = 0$, then $\rho_{\mathcal{F}} = 0$, hence $\rho = 0$. Therefore, if $\rho\neq 0$, we have $\mathrm{Tr}_{\mathcal{H}_{\mathcal{F}}}\rho_{\mathcal{F}} > 0$, and therefore there should exist at least one $[f]$ such that $\mathrm{Tr}_{\mathcal{H}_{[f]}}\left(\Pi_{[f]}\rho_{\mathcal{F}}\Pi_{[f]}^{+}\right) > 0$. Using theorem 2.9, we then have:

$$\sup_{\Lambda\in\mathcal{L}}\left(\inf_{\Lambda'\supset\Lambda}\mathrm{Tr}_{\mathcal{H}_{\Lambda'}}\left(\rho_{\Lambda'}\Theta^{f}_{\Lambda'|\Lambda}\right)\right) = \mathrm{Tr}_{\mathcal{H}_{[f]}}\left(\Pi_{[f]}\rho_{\mathcal{F}}\Pi_{[f]}^{+}\right) > 0,$$

with $\Theta^{f}_{\Lambda'|\Lambda} = \varphi_{\Lambda'\to\Lambda}^{-1} \circ \left(\left|\zeta^{f}_{\Lambda'\to\Lambda}\right\rangle\left\langle\zeta^{f}_{\Lambda'\to\Lambda}\right| \otimes \mathrm{id}_{\mathcal{H}_\Lambda}\right) \circ \varphi_{\Lambda'\to\Lambda}$.

Now, we suppose that we have a infinite part $\Gamma\subset\mathcal{F}$ such that $\forall\gamma\in\Gamma$, $\dim\mathcal{J}_\gamma > 1$, and, for all $\gamma\in\Gamma$, we choose $g_\gamma^1$ and $g_\gamma^2$ two normalized vectors in $\mathcal{J}_\gamma$ that are orthogonal with each other. We define:

$$\forall\Lambda\in\mathcal{L}/\Lambda\subset\Gamma,\ \forall\left(\epsilon_\gamma\right)_{\gamma\in\Lambda}\in\{0,1\}^\Lambda,\ g_\Lambda^{(\epsilon)} := \bigotimes_{\gamma\in\Lambda} g_\gamma^{\epsilon_\gamma} \in \mathcal{H}_\Lambda.$$

Then, we choose some $(g_\lambda)_{\lambda\in\mathcal{F}\setminus\Gamma}$, with $\forall\lambda\in\mathcal{F}\setminus\Gamma$, $g_\lambda\in\mathcal{J}_\lambda$ and $\|g_\lambda\|_{\mathcal{J}_\lambda} = 1$. We define, for any $\Lambda\in\mathcal{L}$:

$$\forall\left(\epsilon_\gamma\right)_{\gamma\in\Lambda\cap\Gamma}\in\{0,1\}^{\Lambda\cap\Gamma},\ g_\Lambda^{(\epsilon)} := \left(\bigotimes_{\lambda\in\Lambda\setminus\Gamma} g_\lambda\right) \otimes g_{\Lambda\cap\Gamma}^{(\epsilon_\gamma)}$$

and $\rho_\Lambda := \dfrac{1}{2^{\#(\Lambda\cap\Gamma)}} \sum_{(\epsilon)} \left|g_\Lambda^{(\epsilon)}\right\rangle\left\langle g_\Lambda^{(\epsilon)}\right|.$

We can check that $(\rho_\Lambda)_{\Lambda\in\mathcal{L}} \in \overline{\mathcal{S}}^{\otimes}_{(\mathcal{L},\mathcal{H},\varphi)}$ and we have, for any $f$ and any $\Lambda\subset\Lambda'$:

$$\mathrm{Tr}_{\mathcal{H}_{\Lambda'}}\left(\rho_{\Lambda'}\Theta^{f}_{\Lambda'|\Lambda}\right) = \prod_{\gamma\in(\Lambda'\setminus\Lambda)\cap\Gamma}\mathrm{Tr}_{\mathcal{J}_\gamma}\left(\tfrac{1}{2}\left(|g_\gamma^1\rangle\langle g_\gamma^1| + |g_\gamma^2\rangle\langle g_\gamma^2|\right)|f_\gamma\rangle\langle f_\gamma|\right) \prod_{\lambda\in(\Lambda'\setminus\Lambda)\setminus\Gamma}\mathrm{Tr}_{\mathcal{J}_\lambda}\left(|g_\lambda\rangle\langle g_\lambda|\,|f_\lambda\rangle\langle f_\lambda|\right)$$

$$= \prod_{\gamma\in(\Lambda'\setminus\Lambda)\cap\Gamma}\tfrac{1}{2}\left(|\langle g_\gamma^1, f_\gamma\rangle|^2 + |\langle g_\gamma^2, f_\gamma\rangle|^2\right) \prod_{\lambda\in(\Lambda'\setminus\Lambda)\setminus\Gamma}\mathrm{Tr}_{\mathcal{J}_\lambda}\left(|\langle g_\lambda, f_\lambda\rangle|^2\right).$$

Thus, we get for any $f$ and any $\Lambda\subset\Lambda'$ the bound:

$$\operatorname{Tr}_{\mathcal{H}_{\Lambda'}} \left( \rho_{\Lambda'} \Theta^f_{\Lambda'|\Lambda} \right) \leqslant \frac{2^{\#(\Lambda \cap \Gamma)}}{2^{\#(\Lambda' \cap \Gamma)}}.$$

Hence,

$$\forall \Lambda \in \mathcal{L}, \ \inf_{\Lambda' \supset \Lambda} \operatorname{Tr}_{\mathcal{H}_{\Lambda'}} \left( \rho_{\Lambda'} \Theta^f_{\Lambda'|\Lambda} \right) = 0 \quad \text{and} \quad \sup_{\Lambda \in \mathcal{L}} \left( \inf_{\Lambda' \supset \Lambda} \operatorname{Tr}_{\mathcal{H}_{\Lambda'}} \left( \rho_{\Lambda'} \Theta^f_{\Lambda'|\Lambda} \right) \right) = 0,$$

so $(\rho_\Lambda)_{\Lambda \in \mathcal{L}} \notin \operatorname{Im} \sigma$, and therefore $\sigma$ is not surjective. □

# 3 Quantization in special cases

The motivation of [15, section 2] was to pave the way for a better understanding of how a quantum projective structure as described in the previous section can be constructed starting from a classical field theory. The procedure we have in mind here, is, given an infinite dimensional symplectic manifold, to first build its rendering by a system of finite dimensional manifolds (the partial theories, that encapsulate insights from a careful analysis of how measurements are done experimentally), and then quantize this projective system (with the aim of getting a quantum theory assembled from 'small' Hilbert spaces, on which calculations should be workable).

In this section, we will consider two basic, yet fairly generic cases, namely position and holomorphic representations, assuming that we have a factorizing system on the classical side (see [15, subsection 2.3]). In both cases, the key prerequisite is that the polarizations, that endow each symplectic manifold $\mathcal{M}_\eta$ with the additional structure needed for quantization (the choice of configuration variables or the complex structure, respectively), should be compatible, in an appropriate sense, with the projections defining the projective system.

In this section, all manifolds are assumed to be smooth finite dimensional manifolds and all maps between them are assumed to be smooth.

## 3.1 Position representation

The starting point for position quantization will be a projective limit of classical phase spaces arising from a factorizing system of configuration spaces as described in [15, prop. 2.16]. Then, there is only one additional ingredient required, namely we need to find a family of measures on these configuration spaces that are intertwined by the factorization maps. With this, constructing the projective system of quantum state spaces is a straightforward generalization of [20, subsections 3.4.3 to 3.4.5], since an $L_2$-space over a Cartesian product of measure spaces has a natural tensor product factorization.

Surely, given an arbitrary projective system of phase spaces, it will in general not be possible to rewrite it as arising from a factorizing system of configuration spaces. However, we consider in theorem 3.2 an important case where we indeed get such a factorizing form automatically, namely when each individual phase space can be identified with the cotangent bundle on a simply-connected Lie group (assuming some appropriate compatibility conditions between the projections and these identifications). The idea is that the group structure, together with the favorable topology, fills exactly the gap between the local result from [15, prop. 2.10] and the global factorization we want to have. Also, using Haar measures, we can easily build a family of measures for this factorizing system.

Note that this result in particular covers the situation considered in [20] (looking at $\mathbb{R}^n$ as an additive Lie group), while answering the question raised in this reference, as to whether the construction can be generalized to non-Abelian gauge theories. To make the relation clearer between the objects in [20] and the ones we are using here, let us look in more detail at the assumptions of theorem 3.2. That each 'small' phase space $\mathcal{M}_\eta$ is a cotangent bundle on a simply-connected Lie group, equipped with its canonical symplectic structure, is a weaker version of assumptions 2, 3b and 4 in [20]. The most crucial assumption is that we start from a projective system of phase spaces: on the one hand, the compatibility of the projections with the symplectic structures provides the seeds of the desired factorizations, on the other hand its three-spaces consistency condition will turn into the corresponding condition for the quantum projective system (eq. (2.1.1)). This is ensured in [20] by assumptions 3a and 6. Finally, the condition 3.2.2, corresponding to the rest of assumption 6 in [20], ensures the compatibility of the projection maps both with the configuration polarizations (so that the factorization of the phase spaces will descend to a factorization of the configuration spaces) and with the group structure (otherwise we would not be able to really make use of this structure). Note that, thanks to the compatibility of the projections with the symplectic structures, simply assuming that the map, besides acting independently on the position and momentum variables, is linear in the momentum variables is sufficient to ensure a full compatibility with the group structure, as expressed by eqs. (3.2.1) and (3.2.2).

**Definition 3.1** A factorizing system of measured manifolds is a factorizing system of smooth, finite dimensional, manifolds $(\mathcal{L}, \mathcal{C}, \varphi)^\times$ [15, def. 2.15] such that:

1. for all $\eta \in \mathcal{L}$, $\mathcal{C}_\eta$ is equipped with a smooth measure $\mu_\eta$ (def. A.12);

2. for all $\eta \prec \eta' \in \mathcal{L}$, $\mathcal{C}_{\eta' \to \eta}$ is equipped with a smooth measure $\mu_{\eta' \to \eta}$ (on $\mathcal{C}_{\eta \to \eta}$ we will use the counting measure);

3. for all $\eta \prec \eta' \in \mathcal{L}$, $\varphi_{\eta' \to \eta}$ is volume-preserving, and for all $\eta \prec \eta' \prec \eta'' \in \mathcal{L}$, $\varphi_{\eta'' \to \eta' \to \eta}$ is volume-preserving; in other words, we require:

$$\forall \eta \prec \eta', \quad \varphi_{\eta' \to \eta}^{-1,*} \mu_{\eta'} = \mu_{\eta' \to \eta} \times \mu_\eta,$$

and $\quad \forall \eta \prec \eta' \prec \eta'', \quad \varphi_{\eta'' \to \eta' \to \eta}^{-1,*} \mu_{\eta'' \to \eta} = \mu_{\eta'' \to \eta'} \times \mu_{\eta' \to \eta}.$

**Theorem 3.2** Let $(\mathcal{L}, \mathcal{M}, \pi)^\downarrow$ be a projective system of phase spaces such that:

1. $\forall \eta \in \mathcal{L}$, $\mathcal{M}_\eta = T^*(\mathcal{C}_\eta)$ where $\mathcal{C}_\eta$ is a *simply-connected* Lie group; by relying on left translations, we thus have an identification $L_\eta : \mathcal{M}_\eta \to \mathcal{C}_\eta \times \text{Lie}^*(\mathcal{C}_\eta)$;

2. $\forall \eta \preccurlyeq \eta'$, $\pi_{\eta' \to \eta} = L_\eta^{-1} \circ (\rho_{\eta' \to \eta} \times \lambda_{\eta' \to \eta}) \circ L_{\eta'}$ where $\rho_{\eta' \to \eta}$ is a map $\mathcal{C}_{\eta'} \to \mathcal{C}_\eta$ and $\lambda_{\eta' \to \eta}$ is a *linear* map $\text{Lie}^*(\mathcal{C}_{\eta'}) \to \text{Lie}^*(\mathcal{C}_\eta)$.

Then, there exists a factorizing system of measured manifolds $(\mathcal{L}, (\mathcal{C}, \mu), \varphi)^\times$ such that $(\mathcal{L}, \mathcal{M}, \pi)^\downarrow$ arises from $(\mathcal{L}, \mathcal{C}, \varphi)^\times$ (in the sense of [15, prop. 2.16]).

**Proof** *Conditions on $\rho_{\eta' \to \eta}$ and $\lambda_{\eta' \to \eta}$.* Let $\eta \in \mathcal{L}$ and $x \in \mathcal{C}_\eta$. There exists an open neighborhood $U$ of 0 in $\mathrm{Lie}(\mathcal{C}_\eta)$ such that the map:

$$\begin{aligned} \Psi_x : U &\to \mathcal{C}_\eta \\ X &\mapsto x \cdot \exp(X) \end{aligned},$$

is a diffeomorphism onto its image, hence it provides a local coordinate system around $x$ in $\mathcal{C}_\eta$. We can lift it to a local trivialization of the cotangent bundle $\mathcal{M}_\eta = T^*(\mathcal{C}_\eta)$:

$$\begin{aligned} \widetilde{\Psi}_x : U \times \mathrm{Lie}^*(\mathcal{C}_\eta) &\to \mathcal{M}_\eta \\ X, \ell &\mapsto x \cdot \exp(X), \ell \circ [T_X \Psi_x]^{-1} \end{aligned}.$$

Using [15, eq. (2.16.1)], we then get, for all $\ell \in \mathrm{Lie}^*(\mathcal{C}_\eta)$ and for all $(u,v), (u',v') \in \mathrm{Lie}(\mathcal{C}_\eta) \times \mathrm{Lie}^*(\mathcal{C}_\eta)$:

$$\Omega_{\mathcal{M}_\eta, \widetilde{\Psi}_x(0,\ell)} \left( \left[ T_{(0,\ell)} \widetilde{\Psi}_x \right] (u,v), \left[ T_{(0,\ell)} \widetilde{\Psi}_x \right] (u',v') \right) = v'(u) - v(u').$$

Next, we have:

$$\begin{aligned} L_\eta \circ \widetilde{\Psi}_x : U \times \mathrm{Lie}^*(\mathcal{C}_\eta) &\to \mathcal{C}_\eta \times \mathrm{Lie}^*(\mathcal{C}_\eta) \\ X, \ell &\mapsto x \cdot \exp(X), \ell \circ [T_X \Psi_x]^{-1} \circ [T_1(x \cdot \exp(X) \cdot \cdot)] = \ell \circ \frac{\mathrm{ad}_X}{\mathrm{id}_{\mathrm{Lie}(\mathcal{C}_\eta)} - e^{-\mathrm{ad}_X}} \end{aligned},$$

where we have used:

$$[T_{\exp(X)}(\exp(-X) \cdot \cdot)] \circ [T_X \exp] = \frac{\mathrm{id}_{\mathrm{Lie}(\mathcal{C}_\eta)} - e^{-\mathrm{ad}_X}}{\mathrm{ad}_X},$$

as follows from the Baker–Campbell–Hausdorff formula.

Therefore, for all $x, \ell \in \mathcal{C}_\eta \times \mathrm{Lie}^*(\mathcal{C}_\eta)$ and for all $(u,v), (u',v') \in T_x(\mathcal{C}_\eta) \times \mathrm{Lie}^*(\mathcal{C}_\eta)$, the symplectic form on $\mathcal{M}_\eta$ is given by:

$$\left( L_\eta^{-1,*} \Omega_{\mathcal{M}_\eta} \right)_{(x,\ell)} \left( (u,v), (u',v') \right) = v' \circ L_{\eta,x}^{-1}(u) - v \circ L_{\eta,x}^{-1}(u') + \ell \left( \left[ L_{\eta,x}^{-1}(u), L_{\eta,x}^{-1}(u') \right]_{\mathrm{Lie}(\mathcal{C}_\eta)} \right),$$

where $L_{\eta,x} := [T_1(x \cdot \cdot)] : \mathrm{Lie}(\mathcal{C}_\eta) \to T_x(\mathcal{C}_\eta)$. This allows us to express the map $\underline{\cdot} : T^*(\mathcal{M}_\eta) \to T(\mathcal{M}_\eta)$ (defined from the symplectic structure as in [15, def. 2.1]) and we have for any $p, k \in T_x^*(\mathcal{C}_\eta) \times \mathrm{Lie}(\mathcal{C}_\eta)$:

$$\underline{(p,k)} \circ \left[ T_{L_\eta^{-1}(x,\ell)} L_\eta \right] = \left[ T_{(x,\ell)} L_\eta^{-1} \right] \left( L_{\eta,x}(k), \ell \left( [k, \cdot ]_{\mathrm{Lie}(\mathcal{C}_\eta)} \right) - p \circ L_{\eta,x} \right).$$

Let $\eta \preccurlyeq \eta'$. We can now formulate the conditions on $\rho_{\eta' \to \eta}$ and $\lambda_{\eta' \to \eta}$ for $\pi_{\eta' \to \eta}$ to be compatible with the symplectic structures as:

$$[T_{x'} \rho_{\eta' \to \eta}] \circ L_{\eta',x'} \circ \lambda_{\eta' \to \eta}^* = L_{\eta, \rho_{\eta' \to \eta}(x')}, \tag{3.2.1}$$

and

$$\left[ \lambda_{\eta' \to \eta}^*(\cdot), \lambda_{\eta' \to \eta}^*(\cdot) \right]_{\mathrm{Lie}(\mathcal{C}_{\eta'})} = \lambda_{\eta' \to \eta}^* \left( [\cdot, \cdot]_{\mathrm{Lie}(\mathcal{C}_\eta)} \right). \tag{3.2.2}$$

$\mathcal{C}_\eta$ *as a Lie subgroup of $\mathcal{C}_{\eta'}$.* $\pi_{\eta' \to \eta}$ being surjective, so is $\lambda_{\eta' \to \eta}$, thus $\lambda_{\eta' \to \eta}^* : \mathrm{Lie}(\mathcal{C}_\eta) \to \mathrm{Lie}(\mathcal{C}_{\eta'})$ is injective, and, from eq. (3.2.2), it is a Lie algebra morphism. Therefore, $\lambda_{\eta' \to \eta}^* \langle \mathrm{Lie}(\mathcal{C}_\eta) \rangle$ is a Lie subalgebra of $\mathrm{Lie}(\mathcal{C}_{\eta'})$ so there exists a unique connected Lie subgroup $\widetilde{\mathcal{C}}_\eta$ of $\mathcal{C}_{\eta'}$ such that $T_1\left(\widetilde{\mathcal{C}}_\eta\right) = \lambda^*_{\eta' \to \eta} \langle \text{Lie}(\mathcal{C}_\eta) \rangle$ [26, theorem 3.19]. $\widetilde{\mathcal{C}}_\eta$ is an immersed submanifold in $\mathcal{C}_{\eta'}$ and its tangent space at $x' \in \widetilde{\mathcal{C}}_\eta$ is given by:

$$T_{x'}\left(\widetilde{\mathcal{C}}_\eta\right) = L_{\eta', x'} \circ \lambda^*_{\eta' \to \eta} \langle \text{Lie}(\mathcal{C}_\eta) \rangle.$$

Let $x'_1, x'_2 \in \mathcal{C}_{\eta'}$ and define:

$$\kappa_{x'_1, x'_2} : \mathcal{C}_{\eta'} \to \mathcal{C}_\eta$$
$$x' \mapsto \rho_{\eta' \to \eta}(x'_1 \cdot x') \cdot \rho_{\eta' \to \eta}(x'_2 \cdot x')^{-1}.$$

For any $k \in \text{Lie}(\mathcal{C}_\eta)$, we have:

$$\left[T_1 \kappa_{x'_1, x'_2}\right] \circ \lambda^*_{\eta' \to \eta}(k) =$$

$$= \left[T_{\rho_{\eta' \to \eta}(x'_1)} \left(\cdot \cdot \rho_{\eta' \to \eta}(x'_2)^{-1}\right)\right] \circ \left[T_{x'_1} \rho_{\eta' \to \eta}\right] \circ L_{\eta', x'_1} \circ \lambda^*_{\eta' \to \eta}(k) +$$
$$+ \left[T_{\rho_{\eta' \to \eta}(x'_2)} \left(\rho_{\eta' \to \eta}(x'_1) \cdot (\cdot)^{-1}\right)\right] \circ \left[T_{x'_2} \rho_{\eta' \to \eta}\right] \circ L_{\eta', x'_2} \circ \lambda^*_{\eta' \to \eta}(k)$$

$$= \left[T_{\rho_{\eta' \to \eta}(x'_1)} \left(\cdot \cdot \rho_{\eta' \to \eta}(x'_2)^{-1}\right)\right] \circ L_{\eta, \rho_{\eta' \to \eta}(x'_1)}(k) + \left[T_{\rho_{\eta' \to \eta}(x'_2)} \left(\rho_{\eta' \to \eta}(x'_1) \cdot (\cdot)^{-1}\right)\right] \circ L_{\eta, \rho_{\eta' \to \eta}(x'_2)}(k)$$

(using eq. (3.2.1))

$$= \left[T_1 \left(x \mapsto \rho_{\eta' \to \eta}(x'_1) \cdot x \cdot x^{-1} \cdot \rho_{\eta' \to \eta}(x'_2)^{-1}\right)\right](k) = 0.$$

With this, we get, for any $x' \in \widetilde{\mathcal{C}}_\eta$:

$$\left[T_{x'} \kappa_{x'_1, x'_2}\right] \left\langle T_{x'}\left(\widetilde{\mathcal{C}}_\eta\right)\right\rangle = \left[T_{x'} \kappa_{x'_1, x'_2}\right] \circ L_{\eta', x'} \circ \lambda^*_{\eta' \to \eta} \langle \text{Lie}\left(\mathcal{C}_\eta\right)\rangle$$

$$= \left[T_1 \kappa_{x'_1 \cdot x', x'_2 \cdot x'}\right] \circ \lambda^*_{\eta' \to \eta} \langle \text{Lie}\left(\mathcal{C}_\eta\right)\rangle = \{0\},$$

thus, $\widetilde{\mathcal{C}}_\eta$ being connected by definition, $\kappa_{x'_1, x'_2}$ is constant on $\widetilde{\mathcal{C}}_\eta$ for any $x'_1, x'_2 \in \mathcal{C}_{\eta'}$. In particular, applying with $x'_2 = 1$ gives:

$$\forall x'_o \in \mathcal{C}_{\eta'}, \forall x' \in \widetilde{\mathcal{C}}_\eta, \widetilde{\rho}_{\eta' \to \eta}(x'_o \cdot x') = \widetilde{\rho}_{\eta' \to \eta}(x'_o) \cdot \widetilde{\rho}_{\eta' \to \eta}(x'), \tag{3.2.3}$$

where $\forall x' \in \mathcal{C}_{\eta'}, \widetilde{\rho}_{\eta' \to \eta}(x') := \rho_{\eta' \to \eta}(1)^{-1} \cdot \rho_{\eta' \to \eta}(x')$.

Therefore, $\widetilde{\rho}_{\eta' \to \eta}|_{\widetilde{\mathcal{C}}_\eta \to \mathcal{C}_\eta}$ is a smooth group homomorphism, and, moreover, its derivative at $1$ is a Lie algebra isomorphism, for we have using eq. (3.2.1):

$$\left[T_1 \widetilde{\rho}_{\eta' \to \eta}\right] \circ \lambda^*_{\eta' \to \eta} = L^{-1}_{\eta, \rho_{\eta' \to \eta}(1)} \circ \left[T_1 \rho_{\eta' \to \eta}\right] \circ \lambda^*_{\eta' \to \eta} = \text{id}_{\text{Lie}(\mathcal{C}_\eta)}.$$

Hence, $\widetilde{\rho}_{\eta' \to \eta}|_{\widetilde{\mathcal{C}}_\eta \to \mathcal{C}_\eta}$ is a Lie group isomorphism, for $\widetilde{\mathcal{C}}_\eta$ is connected and $\mathcal{C}_\eta$ is simply-connected [26, prop. 3.26]. We will denote by $\Lambda_{\eta' \leftarrow \eta} : \mathcal{C}_\eta \to \widetilde{\mathcal{C}}_\eta$ its inverse.

*Factorizing system.* We define $\mathcal{C}_{\eta' \to \eta} := \widetilde{\rho}^{-1}_{\eta' \to \eta} \langle 1 \rangle$. $\widetilde{\rho}_{\eta' \to \eta}$ has surjective derivative at each point, so $\mathcal{C}_{\eta' \to \eta}$ is a smooth manifold as level set of a constant rank map [17, theorem 5.22]. Next, we define the map $\varphi_{\eta' \to \eta}$ by:

$$\varphi_{\eta' \to \eta} : \mathcal{C}_{\eta'} \to \mathcal{C}_{\eta' \to \eta} \times \mathcal{C}_{\eta}$$
$$x' \mapsto \left( x' \cdot \left( \Lambda_{\eta' \leftarrow \eta} \circ \widetilde{\rho}_{\eta' \to \eta}(x') \right)^{-1}, \rho_{\eta' \to \eta}(x') \right).$$

$\varphi_{\eta' \to \eta}$ is well-defined for, using eq. (3.2.3), we have for all $x' \in \mathcal{C}_{\eta'}$:

$$\widetilde{\rho}_{\eta' \to \eta} \left( x' \cdot \Lambda_{\eta' \leftarrow \eta} \left( \widetilde{\rho}_{\eta' \to \eta}(x')^{-1} \right) \right) = \widetilde{\rho}_{\eta' \to \eta}(x') \cdot \widetilde{\rho}_{\eta' \to \eta}(x')^{-1} = \mathbf{1}.$$

To prove that $\varphi_{\eta' \to \eta}$ is a bijective map, we define a map $\widetilde{\varphi}_{\eta' \to \eta}$ by:

$$\widetilde{\varphi}_{\eta' \to \eta} : \mathcal{C}_{\eta' \to \eta} \times \mathcal{C}_{\eta} \to \mathcal{C}_{\eta'}$$
$$y, x \mapsto y \cdot \sigma^{-1} \cdot \Lambda_{\eta' \leftarrow \eta}(x),$$

where $\sigma := \Lambda_{\eta' \leftarrow \eta} \left( \rho_{\eta' \to \eta}(\mathbf{1}) \right)$. Using again eq. (3.2.3), we can check that $\varphi_{\eta' \to \eta} \circ \widetilde{\varphi}_{\eta' \to \eta} = \mathrm{id}_{\mathcal{C}_{\eta' \to \eta} \times \mathcal{C}_{\eta}}$ and $\widetilde{\varphi}_{\eta' \to \eta} \circ \varphi_{\eta' \to \eta} = \mathrm{id}_{\mathcal{C}_{\eta'}}$. Since both $\varphi_{\eta' \to \eta}$ and $\widetilde{\varphi}_{\eta' \to \eta}$ are smooth, $\varphi_{\eta' \to \eta}$ is a diffeomorphism.

From eq. (3.2.1), we have:

$$\forall x \in \mathcal{C}_{\eta}, \; [T_x \Lambda_{\eta' \leftarrow \eta}] = L_{\eta', \Lambda_{\eta' \leftarrow \eta}(x)} \circ \lambda^*_{\eta' \to \eta} \circ L^{-1}_{\eta, x}.$$

Thus, for any $y, x \in \mathcal{C}_{\eta' \to \eta} \times \mathcal{C}_{\eta}$, the derivative of $\varphi^{-1}_{\eta' \to \eta}$ satisfies:

$$\forall u \in T_x(\mathcal{C}_{\eta}), \; \left[ T_{y,x} \varphi^{-1}_{\eta' \to \eta} \right](0, u) = L_{\eta', \varphi^{-1}_{\eta' \to \eta}(y,x)} \circ \lambda^*_{\eta' \to \eta} \circ L^{-1}_{\eta, x}(u). \tag{3.2.4}$$

So we get, for any $x', p' \in T^*(\mathcal{C}_{\eta'})$:

$$\pi_{\eta' \to \eta}(x', p') = \left( \rho_{\eta' \to \eta}(x'), p' \circ L_{\eta', x'} \circ \lambda^*_{\eta' \to \eta} \circ L^{-1}_{\eta, \rho_{\eta' \to \eta}(x')} \right)$$
$$= \left( s_{\eta' \to \eta} \circ \varphi_{\eta' \to \eta}(x'), p' \circ [T_{x'} \varphi_{\eta' \to \eta}]^{-1}(0, \cdot) \right),$$

where $s_{\eta' \to \eta} : \mathcal{C}_{\eta' \to \eta} \times \mathcal{C}_{\eta} \to \mathcal{C}_{\eta}$ is the projection on the second Cartesian factor. We can now use [15, prop. 2.17] to build from these objects a factorizing system $(\mathcal{L}, \mathcal{C}, \varphi)^\times$ that gives rise to $(\mathcal{L}, \mathcal{M}, \pi)^\downarrow$.

*Volume forms.* For any $\eta \in \mathcal{L}$, we choose a non-zero $d_\eta$-form $\widetilde{\omega}_\eta$ on $\mathrm{Lie}(\mathcal{C}_\eta)$ (with $d_\eta := \dim \mathcal{C}_\eta$) and we define a right-invariant volume form $\omega_\eta$ on $\mathcal{C}_\eta$ by:

$$\forall x \in \mathcal{C}_\eta, \; \forall u_1, \ldots, u_{d_\eta} \in T_x(\mathcal{C}_\eta), \; \omega_\eta(u_1, \ldots, u_{d_\eta}) := \widetilde{\omega}_\eta \left( R^{-1}_{\eta, x} u_1, \ldots, R^{-1}_{\eta, x} u_{d_\eta} \right),$$

where $R_{\eta, x} := T_\mathbf{1}(\cdot \cdot x)$. We call $\mu_\eta$ the smooth measure arising from the volume form $\omega_\eta$.

Let $\eta \preccurlyeq \eta'$, $y \in \mathcal{C}_{\eta' \to \eta}$ and $w_1, \ldots, w_{d_{\eta' \to \eta}} \in T_y(\mathcal{C}_{\eta' \to \eta}) = \mathrm{Ker}\left[ T_y \rho_{\eta' \to \eta} \right]$ (with $d_{\eta' \to \eta} := \dim \mathcal{C}_{\eta' \to \eta} = d_{\eta'} - d_\eta$). The map:

$$\alpha : \mathrm{Lie}(\mathcal{C}_\eta)^{d_\eta} \to \mathbb{R}$$
$$u_1, \ldots, u_{d_\eta} \mapsto \widetilde{\omega}_{\eta'} \left( R^{-1}_{\eta', y}(w_1), \ldots, R^{-1}_{\eta', y}(w_{d_{\eta' \to \eta}}), A_{\eta' \leftarrow \eta, y}(u_1), \ldots, A_{\eta' \leftarrow \eta, y}(u_{d_\eta}) \right),$$

where $A_{\eta' \leftarrow \eta, y} := \mathrm{Ad}_y \circ \lambda^*_{\eta' \to \eta} \circ \mathrm{Ad}^{-1}_{\rho_{\eta' \to \eta}(\mathbf{1})}$, is a $d_\eta$-form on $\mathrm{Lie}(\mathcal{C}_\eta)$, so there exists $\omega_{\eta' \to \eta, y}(w_1, \ldots, w_{d_{\eta' \to \eta}}) \in \mathbb{R}$ such that:

$$\alpha(u_1, \ldots, u_{d_\eta}) = \omega_{\eta' \to \eta, y}(w_1, \ldots, w_{d_{\eta' \to \eta}}) \widetilde{\omega}_\eta(u_1, \ldots, u_{d_\eta}).$$

Now, using the expression for $\varphi^{-1}_{\eta' \to \eta}$ given above, we have, for any $y, x \in \mathcal{C}_{\eta' \to \eta} \times \mathcal{C}_\eta$:

$$\forall w \in T_y(\mathcal{C}_{\eta' \to \eta}), \ R^{-1}_{\eta', \varphi^{-1}_{\eta' \to \eta}(y,x)} \circ \left[ T_{y,x} \varphi^{-1}_{\eta' \to \eta} \right](w, 0) = R^{-1}_{\eta', y}(w),$$

and, from eq. (3.2.4), we also have:

$$\forall u \in T_x(\mathcal{C}_\eta), \ R^{-1}_{\eta', \varphi^{-1}_{\eta' \to \eta}(y,x)} \circ \left[ T_{y,x} \varphi^{-1}_{\eta' \to \eta} \right](0, u) = A_{\eta' \leftarrow \eta, y} \circ R^{-1}_{\eta, x}(u),$$

where we have used that $\lambda^*_{\eta' \to \eta} = T_1 \Lambda_{\eta' \leftarrow \eta}$ with $\Lambda_{\eta' \leftarrow \eta}$ a group homomorphism. With this, we can check that $\varphi^{-1,*}_{\eta' \to \eta} \omega_{\eta'} = \omega_{\eta' \to \eta} \wedge \omega_\eta$. In particular, this implies that $\omega_{\eta' \to \eta}$ is a smooth volume form on $\mathcal{C}_{\eta' \to \eta}$. Thus, defining $\mu_{\eta' \to \eta}$ to be the corresponding smooth measure, we get $\varphi^{-1,*}_{\eta' \to \eta} \mu_{\eta'} = \mu_{\eta' \to \eta} \times \mu_\eta$.

Finally, for any $\eta \preccurlyeq \eta' \preccurlyeq \eta''$, $\varphi_{\eta'' \to \eta'}$, $\varphi_{\eta' \to \eta}$ and $\varphi_{\eta'' \to \eta}$ are volume-preserving, hence so is $\varphi_{\eta'' \to \eta' \to \eta} \times \mathrm{id}_{\mathcal{C}_\eta}$ (using [15, eq. (2.11.1)]) and therefore $\varphi_{\eta'' \to \eta' \to \eta}$ itself. □

**Proposition 3.3** Let $(\mathcal{L}, \mathcal{C}, \varphi)^\times$ be a factorizing system of measured manifolds. We define:

1. for $\eta \in \mathcal{L}$, $\mathcal{H}_\eta := L_2(\mathcal{C}_\eta, d\mu_\eta)$;
2. for $\eta \prec \eta' \in \mathcal{L}$, $\mathcal{H}_{\eta' \to \eta} := L_2(\mathcal{C}_{\eta' \to \eta}, d\mu_{\eta' \to \eta})$, and:

$$\begin{aligned} \Phi_{\eta' \to \eta} : \mathcal{H}_{\eta'} &\to \mathcal{H}_{\eta' \to \eta} \otimes \mathcal{H}_\eta \\ \psi &\mapsto \psi \circ \varphi^{-1}_{\eta' \to \eta} \end{aligned},$$

with the natural identification of $L_2(\mathcal{C}_{\eta' \to \eta}, d\mu_{\eta' \to \eta}) \otimes L_2(\mathcal{C}_\eta, d\mu_\eta)$ with $L_2(\mathcal{C}_{\eta' \to \eta} \times \mathcal{C}_\eta, d\mu_{\eta' \to \eta} \times d\mu_\eta)$.

Then, we can complete these objects to build a projective system of quantum state spaces $(\mathcal{L}, \mathcal{H}, \Phi)^\otimes$.

**Proof** We define:

3. for $\eta \prec \eta' \prec \eta'' \in \mathcal{L}$,

$$\begin{aligned} \Phi_{\eta'' \to \eta' \to \eta} : \mathcal{H}_{\eta'' \to \eta} &\to \mathcal{H}_{\eta'' \to \eta'} \otimes \mathcal{H}_{\eta' \to \eta} \\ \psi &\mapsto \psi \circ \varphi^{-1}_{\eta'' \to \eta' \to \eta} \end{aligned},$$

with the natural identification of $L_2(\mathcal{C}_{\eta'' \to \eta'}, d\mu_{\eta'' \to \eta'}) \otimes L_2(\mathcal{C}_{\eta' \to \eta}, d\mu_{\eta' \to \eta})$ with $L_2(\mathcal{C}_{\eta'' \to \eta'} \times \mathcal{C}_{\eta' \to \eta}, d\mu_{\eta'' \to \eta'} \times d\mu_{\eta' \to \eta})$;

4. for $\eta \in \mathcal{L}$, $\mathcal{H}_{\eta \to \eta} := \mathbb{C}$ and we define $\Phi_{\eta \to \eta}$ to be the natural isomorphic identification between $\mathcal{H}_\eta$ and $\mathbb{C} \otimes \mathcal{H}_\eta$;

5. for $\eta \preccurlyeq \eta' \in \mathcal{L}$, we define $\Phi_{\eta' \to \eta \to \eta}$ (resp. $\Phi_{\eta' \to \eta' \to \eta}$) to be the natural isomorphic identification between $\mathcal{H}_{\eta' \to \eta}$ and $\mathcal{H}_{\eta' \to \eta} \otimes \mathbb{C}$ (resp. $\mathbb{C} \otimes \mathcal{H}_{\eta' \to \eta}$).

That $\Phi_{\eta' \to \eta}$ for $\eta \prec \eta'$ defines an Hilbert space isomorphism comes from the volume-preserving property of $\varphi_{\eta' \to \eta}$ and from the fact that $L_2(\mathcal{C}, d\mu) \otimes L_2(\mathcal{C}', d\mu')$ can be unitarily identified with $L_2(\mathcal{C} \times \mathcal{C}', d\mu \times d\mu')$ (thanks to Fubini's theorem). Similarly, $\Phi_{\eta'' \to \eta' \to \eta}$ for $\eta \prec \eta' \prec \eta''$ is an Hilbert space isomorphism.

We now just need to check the three-spaces consistency condition eq. (2.1.1). We consider $\eta \prec \eta' \prec \eta''$ (since the condition is trivially satisfied whenever two labels are equal):

$$\forall \psi \in \mathcal{H}_\eta, \ \left( \Phi_{\eta'' \to \eta' \to \eta} \otimes \mathrm{id}_{\mathcal{H}_\eta} \right) \circ \Phi_{\eta'' \to \eta}(\psi) =$$

$$= \left( \psi \circ \varphi^{-1}_{\eta'' \to \eta} \right) \circ \left( \varphi^{-1}_{\eta'' \to \eta' \to \eta} \otimes \mathrm{id}_{\mathcal{C}_\eta} \right)$$

$$= \left(\psi \circ \varphi_{\eta'' \to \eta'}^{-1}\right) \circ \left(\mathrm{id}_{\mathcal{C}_{\eta'' \to \eta'}} \otimes \varphi_{\eta' \to \eta}^{-1}\right) \text{ (using [15, eq. (2.11.1)])}$$

$$= \left(\mathrm{id}_{\mathcal{H}_{\eta'' \to \eta'}} \otimes \Phi_{\eta' \to \eta}\right) \circ \Phi_{\eta'' \to \eta'}(\psi).$$

□

To argue that the quantum projective system composed above actually provides a quantization of the classical one (as specified by the factorizing system of configuration spaces we started from), we need to say how classical observables on the latter are turned to quantum observables on the former. For this, we import the prescriptions of geometric quantization (summarized in appendix A.3, especially in prop. A.14, and rewritten here more explicitly using the benefit of working in a phase space given as a cotangent bundle). Thus, for each $\eta$, we can formulate the quantization condition (the choice of preferred configuration variables is tied to a selection of which observables can be directly quantized) as well as the definition of the quantized observables. The key statement is that the compatibility conditions imposed on the family of measures is sufficient to ensure that these prescriptions, supplied separately for each $\eta$, will fit readily into a coherent picture.

**Proposition 3.4** We consider the same objects as in prop. 3.3. Let $(\mathcal{L}, \mathcal{M}, \widetilde{\varphi})^\times$ be the factorizing system of phase spaces constructed from $(\mathcal{L}, \mathcal{C}, \varphi)^\times$ as in [15, prop. 2.16] and let $f = [f_\eta]_\sim \in \mathcal{O}^\times_{(\mathcal{L}, \mathcal{M}, \widetilde{\varphi})}$ [15, prop. 2.13].

If there exists a representative $f_\eta$ of $f$ such that:

$$\exists \overline{X}_{f_\eta} \in \mathcal{T}^\infty(\mathcal{C}_\eta) / \forall (x, p) \in \mathcal{M}_\eta, \; [T_{(x,p)} \gamma_\eta]\left(X_{f_\eta, (x,p)}\right) = \overline{X}_{f_\eta, x}, \tag{3.4.1}$$

where $\mathcal{T}^\infty(\mathcal{C}_\eta)$ is the space of smooth vector fields on $\mathcal{C}_\eta$, and $\gamma_\eta : \mathcal{M}_\eta = T^*(\mathcal{C}_\eta) \to \mathcal{C}_\eta$ is the bundle projection, then this condition is satisfied by all representatives of $f$. Accordingly, we define:

$$\mathcal{O}^\times_{(\mathcal{L}, \mathcal{C}, \varphi)} := \left\{ f \in \mathcal{O}^\times_{(\mathcal{L}, \mathcal{M}, \widetilde{\varphi})} \;\middle|\; \exists f_\eta \in f \text{ satisfying eq. (3.4.1)} \right\}.$$

For $f_\eta \in f \in \mathcal{O}^\times_{(\mathcal{L}, \mathcal{C}, \varphi)}$, we can define $\widehat{f}_\eta^{\mu_\eta}$ as a densely defined operator on $\mathcal{H}_\eta$ (with dense domain $\mathcal{D}_\eta \subset \mathcal{H}_\eta$) by:

$$\forall \psi \in \mathcal{D}_\eta, \forall x \in \mathcal{C}_\eta, \; \widehat{f}_\eta^{\mu_\eta}(\psi)(x) := \psi(x) f_\eta(x, 0) + i \, [d\psi]_x\left(\overline{X}_{f_\eta, x}\right) + \frac{i}{2} \psi(x) \left(\mathrm{div}_{\mu_\eta} \overline{X}_{f_\eta}\right)(x),$$

where $\mathrm{div}_{\mu_\eta} \overline{X}_{f_\eta}$ is defined by $\mathfrak{L}_{\overline{X}_{f_\eta}} \mu_\eta = \left(\mathrm{div}_{\mu_\eta} \overline{X}_{f_\eta}\right) \mu_\eta$ (def. A.12).

Then, the application:

$$\widehat{\cdot}^\mu : \mathcal{O}^\times_{(\mathcal{L}, \mathcal{C}, \varphi)} \to \mathcal{O}^\otimes_{(\mathcal{L}, \mathcal{H}, \Phi)}$$
$$[f_\eta]_\sim \mapsto \left[\widehat{f}_\eta^{\mu_\eta}\right]_\sim,$$

is well-defined ($\mathcal{O}^\otimes_{(\mathcal{L}, \mathcal{H}, \Phi)}$ has been defined in prop. 2.5).

**Proof** *Quantization condition.* For $\eta \preccurlyeq \eta' \in \mathcal{L}$, we define $\pi_{\eta' \to \eta} : \mathcal{C}_{\eta'} \to \mathcal{C}_\eta$, $\lambda_{\eta' \to \eta} : \mathcal{C}_{\eta'} \to \mathcal{C}_{\eta' \to \eta}$, and $\widetilde{\pi}_{\eta' \to \eta} : \mathcal{M}_{\eta'} \to \mathcal{M}_\eta$, $\widetilde{\lambda}_{\eta' \to \eta} : \mathcal{M}_{\eta'} \to \mathcal{M}_{\eta' \to \eta}$, such that:

$$\forall x' \in \mathcal{C}_{\eta'}, \; \varphi_{\eta' \to \eta}(x') = \left( \lambda_{\eta' \to \eta}(x'), \; \pi_{\eta' \to \eta}(x') \right)$$

$$\& \quad \forall x', p' \in \mathcal{M}_{\eta'}, \; \widetilde{\varphi}_{\eta' \to \eta}(x', p') = \left( \widetilde{\lambda}_{\eta' \to \eta}(x', p'), \; \widetilde{\pi}_{\eta' \to \eta}(x', p') \right).$$

From [15, prop. 2.16], we then have:

$$\forall \eta \preccurlyeq \eta' \in \mathcal{L}, \; \gamma_\eta \circ \widetilde{\pi}_{\eta' \to \eta} = \pi_{\eta' \to \eta} \circ \gamma_{\eta'} \quad \& \quad \gamma_{\eta' \to \eta} \circ \widetilde{\lambda}_{\eta' \to \eta} = \lambda_{\eta' \to \eta} \circ \gamma_{\eta'}.$$

Let $f_\eta \in C^\infty(\mathcal{M}_\eta, \mathbb{R})$ satisfying eq. (3.4.1) and let $\eta' \succcurlyeq \eta$. Using the previous identity, we have:

$$\forall x', p' \in \mathcal{M}_{\eta'}, \; [T_{x'} \, \pi_{\eta' \to \eta}] \circ [T_{x', p'} \, \gamma_{\eta'}] \left( X_{f_\eta \circ \widetilde{\pi}_{\eta' \to \eta}, (x', p')} \right) =$$

$$= \left[ T_{\widetilde{\pi}_{\eta' \to \eta}(x', p')} \, \gamma_\eta \right] \circ [T_{x', p'} \, \widetilde{\pi}_{\eta' \to \eta}] \left( \underline{\left[ \widetilde{\pi}^*_{\eta' \to \eta} \, df_\eta \right]_{x', p'}} \right)$$

$$= \left[ T_{\widetilde{\pi}_{\eta' \to \eta}(x', p')} \, \gamma_\eta \right] \left( \underline{[df_\eta]_{\widetilde{\pi}_{\eta' \to \eta}(x', p')}} \right) \; \text{(using [15, eq. (2.1.1)])}$$

$$= \left[ T_{\widetilde{\pi}_{\eta' \to \eta}(x', p')} \, \gamma_\eta \right] \left( X_{f_\eta, \widetilde{\pi}_{\eta' \to \eta}(x', p')} \right) = \overline{X}_{f_\eta, \pi_{\eta' \to \eta}(x')},$$

and:

$$\forall x', p' \in \mathcal{M}_{\eta'}, \; [T_{x'} \, \lambda_{\eta' \to \eta}] \circ [T_{x', p'} \, \gamma_{\eta'}] \left( X_{f_\eta \circ \widetilde{\pi}_{\eta' \to \eta}, (x', p')} \right) =$$

$$= \left[ T_{\widetilde{\lambda}_{\eta' \to \eta}(x', p')} \, \gamma_{\eta' \to \eta} \right] \circ [T_{x', p'} \, \widetilde{\lambda}_{\eta' \to \eta}] \left( \underline{\left[ \widetilde{\pi}^*_{\eta' \to \eta} \, df_\eta \right]_{x', p'}} \right)$$

$$= 0,$$

since $\underline{\left[ \widetilde{\pi}^*_{\eta' \to \eta} \, df_\eta \right]_{x', p'}} \in \left( \operatorname{Ker} [T_{x', p'} \, \widetilde{\pi}_{\eta' \to \eta}] \right)^\perp = \operatorname{Ker} \left[ T_{x', p'} \, \widetilde{\lambda}_{\eta' \to \eta} \right].$

Therefore, $[T_{x', p'} \, \gamma_{\eta'}] \left( X_{f_\eta \circ \widetilde{\pi}_{\eta' \to \eta}, (x', p')} \right)$ only depends on $x'$. If we now define:

$$\forall x' \in \mathcal{M}_{\eta'}, \; \overline{X}_{f_\eta \circ \widetilde{\pi}_{\eta' \to \eta}, x'} := [T_{x', 0} \, \gamma_{\eta'}] \left( X_{f_\eta \circ \widetilde{\pi}_{\eta' \to \eta}, (x', 0)} \right),$$

we have $\overline{X}_{f_\eta \circ \widetilde{\pi}_{\eta' \to \eta}} \in \mathcal{T}^\infty(\mathcal{C}_{\eta'})$ and:

$$\forall x', p' \in \mathcal{M}_{\eta'}, \; [T_{x', p'} \, \gamma_{\eta'}] \left( X_{f_\eta \circ \widetilde{\pi}_{\eta' \to \eta}, (x', p')} \right) = \overline{X}_{f_\eta \circ \widetilde{\pi}_{\eta' \to \eta}, x'}.$$

Thus, $f_\eta \circ \widetilde{\pi}_{\eta' \to \eta}$ also fulfills eq. (3.4.1) and we moreover have:

$$\forall x' \in \mathcal{C}_{\eta'}, \; [T_{x'} \, \varphi_{\eta' \to \eta}] \left( \overline{X}_{f_\eta \circ \widetilde{\pi}_{\eta' \to \eta}, x'} \right) = \left( 0, \; \overline{X}_{f_\eta, \pi_{\eta' \to \eta}(x')} \right). \tag{3.4.2}$$

On the other hand, for $f_\eta \in C^\infty(\mathcal{M}_\eta, \mathbb{R})$, if there exists $\eta' \succcurlyeq \eta$ such that $f_\eta \circ \widetilde{\pi}_{\eta' \to \eta}$ satisfy eq. (3.4.1), then, in the same way as above:

$$\forall x', p' \in \mathcal{M}^{\eta'}, \; \left[ T_{\widetilde{\pi}_{\eta' \to \eta}(x', p')} \, \gamma_\eta \right] \left( X_{f_\eta, \widetilde{\pi}_{\eta' \to \eta}(x', p')} \right) = [T_{x'} \, \pi_{\eta' \to \eta}] \left( \overline{X}_{f_\eta \circ \widetilde{\pi}_{\eta' \to \eta}, x'} \right).$$

Therefore, since $\widetilde{\pi}_{\eta' \to \eta}$ is surjective, $f_\eta$ also satisfy eq. (3.4.1), and again we have eq. (3.4.2).

*Quantized observable.* Let $f_\eta \in f \in \mathcal{O}^\times_{(\mathcal{L},\mathcal{C},\varphi)}$. We start by deriving an identity for $\mathrm{div}_{\mu_\eta} \overline{X}_{f_\eta}$:

$$\left(\mathrm{div}_{\mu_{\eta'}} \overline{X}_{f_\eta \circ \tilde{\pi}_{\eta'\to\eta}}\right) \mu_{\eta'} = \mathfrak{L}_{\overline{X}_{f_\eta \circ \tilde{\pi}_{\eta'\to\eta}}} \mu_{\eta'}$$

$$= \varphi^*_{\eta'\to\eta} \left[ \mathfrak{L}_{\varphi^{-1,*}_{\eta'\to\eta}\left(\overline{X}_{f_\eta \circ \tilde{\pi}_{\eta'\to\eta}}\right)} \varphi^{-1,*}_{\eta'\to\eta} \mu_{\eta'} \right]$$

$$= \varphi^*_{\eta'\to\eta} \left[ \mathfrak{L}_{\varphi^{-1,*}_{\eta'\to\eta}\left(\overline{X}_{f_\eta \circ \tilde{\pi}_{\eta'\to\eta}}\right)} \mu_{\eta'\to\eta} \times \mu_\eta \right] \quad \text{(using def. 3.1.3)}$$

$$= \varphi^*_{\eta'\to\eta} \left[ \mathfrak{L}_{\left(0, \overline{X}_{f_\eta}\right)} \mu_{\eta'\to\eta} \times \mu_\eta \right] \quad \text{(using eq. (3.4.2))}$$

$$= \varphi^*_{\eta'\to\eta} \left[ \mu_{\eta'\to\eta} \times \left(\mathfrak{L}_{\overline{X}_{f_\eta}} \mu_\eta\right) \right]$$

$$= \left[ \left(\mathrm{div}_{\mu_\eta} \overline{X}_{f_\eta}\right) \circ \pi_{\eta'\to\eta} \right] \mu_{\eta'} .$$

Hence, it follows:

$$\mathrm{div}_{\mu_{\eta'}} \overline{X}_{f_\eta \circ \tilde{\pi}_{\eta'\to\eta}} = \left(\mathrm{div}_{\mu_\eta} \overline{X}_{f_\eta}\right) \circ \pi_{\eta'\to\eta}. \tag{3.4.3}$$

Now, let $\psi \in \Phi^{-1}_{\eta'\to\eta} \langle \mathcal{H}_{\eta'\to\eta} \otimes \mathcal{D}_\eta \rangle$ (where $\mathcal{D}_\eta$ is the domain of $\widehat{f}^{\mu_\eta}_\eta$ and the $\otimes$ is to be understood as a tensor product of vector spaces, that is without any completion in contrast to a tensor product of Hilbert spaces). Then, we have:

$$\forall y, x \in \mathcal{C}_{\eta'\to\eta} \times \mathcal{C}_\eta, \quad \left(\mathrm{id}_{\mathcal{H}_{\eta'\to\eta}} \otimes \widehat{f}^{\mu_\eta}_\eta\right) \circ \Phi_{\eta'\to\eta}(\psi)(y,x) =$$

$$= \psi \circ \varphi^{-1}_{\eta'\to\eta}(y,x)\, f_\eta(x, 0) + i\, [\partial_x \psi \circ \varphi^{-1}_{\eta'\to\eta}]_{(y,x)}\left(\overline{X}_{f_\eta, x}\right) + \frac{i}{2} \psi \circ \varphi^{-1}_{\eta'\to\eta}(y,x) \left(\mathrm{div}_{\mu_\eta} \overline{X}_{f_\eta}\right)(x)$$

$$= \psi \circ \varphi^{-1}_{\eta'\to\eta}(y,x)\, f_\eta \circ \tilde{\pi}_{\eta'\to\eta}(\varphi^{-1}_{\eta'\to\eta}(y,x), 0) + i\, [d\psi]_{\varphi^{-1}_{\eta'\to\eta}(y,x)} \left(\overline{X}_{f_\eta \circ \tilde{\pi}_{\eta'\to\eta}, \varphi^{-1}_{\eta'\to\eta}(y,x)}\right) +$$

$$+ \frac{i}{2} \psi \circ \varphi^{-1}_{\eta'\to\eta}(y,x) \left(\mathrm{div}_{\mu_{\eta'}} \overline{X}_{f_\eta \circ \tilde{\pi}_{\eta'\to\eta}}\right) \circ \varphi^{-1}_{\eta'\to\eta}(y,x) \quad \text{(using eqs. (3.4.2) and (3.4.3))}$$

$$= \left(\Phi_{\eta'\to\eta} \circ \widehat{f_\eta \circ \tilde{\pi}_{\eta'\to\eta}}^{\mu_{\eta'}}\right)(\psi)(y,x) .$$

Therefore, $\forall f_\eta, f_{\eta'} \in f \in \mathcal{O}^\times_{(\mathcal{L},\mathcal{C},\varphi)},\ \widehat{f}^{\mu_\eta}_\eta \sim \widehat{f}^{\mu_{\eta'}}_{\eta'}$. $\square$

We close this subsection with an application of theorem 2.9: under the additional hypothesis that the measures are normalized to unity, we can construct an inductive limit of Hilbert spaces from the $\mathcal{H}_\eta$, whose space of states is naturally embedded in the one of the projective structure developed above. As long as all $\mathcal{C}_\eta$ have finite volume (hence in particular if they are compact), it is always possible to consistently normalize the measures to unity. Note however that, depending on the projective structure under consideration, it may not be possible to equip all $\mathcal{C}_\eta$ with normalizable measures fulfilling the factorization requirement def. 3.1.3 (see eg. the models considered in [18, 19], in particular the discussion in [18, section 1.1] ).

**Proposition 3.5** We consider the same objects as in prop. 3.3, and we now additionally assume:

$$\forall \eta \in \mathcal{L}, \ \mu_\eta(\mathcal{C}_\eta) = 1.$$

Then, we can also construct an Hilbert space $\mathcal{H}_\oplus$ as (the completion of) the inductive limit of $\left(\mathcal{L}, \left(\mathcal{H}_\eta\right)_{\eta \in \mathcal{L}}, \left(\tau_{\eta' \leftarrow \eta}\right)_{\eta \preccurlyeq \eta'}\right)$, where the injective maps $\tau_{\eta' \leftarrow \eta}$ are defined as:

$$\forall \eta \preccurlyeq \eta' \in \mathcal{L}, \quad \begin{array}{c} \tau_{\eta' \leftarrow \eta} : \mathcal{H}_\eta \to \mathcal{H}_{\eta'} \\ \psi \mapsto \psi \circ s_{\eta' \to \eta} \circ \varphi_{\eta' \to \eta} \end{array}, \text{ with } \begin{array}{c} s_{\eta' \to \eta} : \mathcal{C}_{\eta' \to \eta} \times \mathcal{C}_\eta \to \mathcal{C}_\eta \\ (x, y) \mapsto y \end{array}.$$

There exist maps $\sigma : \overline{\mathcal{S}}_\oplus \to \overline{\mathcal{S}}^\otimes_{(\mathcal{L}, \mathcal{H}, \Phi)}$ and $\alpha : \overline{\mathcal{A}}^\otimes_{(\mathcal{L}, \mathcal{H}, \Phi)} \to \mathcal{A}_\oplus$ ($\overline{\mathcal{S}}_\oplus$ being the space of (self-adjoint) positive semi-definite, traceclass operators over $\mathcal{H}_\oplus$ and $\mathcal{A}_\oplus$ the algebra of bounded operators on $\mathcal{H}_\oplus$) such that:

1. $\forall \rho \in \overline{\mathcal{S}}_\oplus, \forall A \in \overline{\mathcal{A}}^\otimes_{(\mathcal{L}, \mathcal{H}, \Phi)}, \ \text{Tr}_{\mathcal{H}_\oplus}(\rho \, \alpha(A)) = \text{Tr}(\sigma(\rho) \, A)$;

2. $\sigma$ is injective;

3. $\sigma \langle \mathcal{S}_\oplus \rangle = \left\{ (\rho_\eta)_{\eta \in \mathcal{L}} \ \middle| \ \sup_{\eta \in \mathcal{L}} \inf_{\eta' \succcurlyeq \eta} \int_{\mathcal{C}_{\eta' \to \eta} \times \mathcal{C}_{\eta' \to \eta}} dx dx' \int_{\mathcal{C}_\eta} dy \ \rho_{\eta'}\left(\varphi^{-1}_{\eta' \to \eta}(x, y); \varphi^{-1}_{\eta' \to \eta}(x', y)\right) = 1 \right\}$

where $\mathcal{S}_\oplus$ is the space of density matrices over $\mathcal{H}_\oplus$ and $\rho_\eta(\cdot\,;\,\cdot)$ is the integral kernel of $\rho_\eta$.

**Proof** This is an application of theorem 2.9, where for $\eta \preccurlyeq \eta' \in \mathcal{L}$, we define:

$$\zeta_{\eta' \to \eta} \equiv 1 \in \mathcal{H}_{\eta' \to \eta}.$$

We have $\forall \eta \preccurlyeq \eta', \ \|\zeta_{\eta' \to \eta}\| = 1$, since $\mu_{\eta' \to \eta}(\mathcal{C}_{\eta' \to \eta}) = \mu_{\eta'}(\mathcal{C}_{\eta'}) / \mu_\eta(\mathcal{C}_\eta) = 1$, and $\forall \eta \preccurlyeq \eta' \preccurlyeq \eta''$, $\Phi_{\eta'' \to \eta' \to \eta}(\zeta_{\eta'' \to \eta'}) \equiv 1 \equiv \zeta_{\eta'' \to \eta'} \otimes \zeta_{\eta' \to \eta}$. □

Finally, note that, as far as the construction of the quantum projective state space and observables thereof is concerned, we can actually dispense from having a factorizing system of measures. Indeed, if we just have families $\left(\mu_\eta\right)_{\eta \in \mathcal{L}}$ and $\left(\mu_{\eta' \to \eta}\right)_{\eta \preccurlyeq \eta'}$ of smooth measures, which do *not* satisfy the compatibility conditions from def. 3.1.3, we can rely on the canonical identification introduced in prop. A.15 to relate the position representation built on the measure $\mu_{\eta'}$ with the one built on the measure $\varphi^*_{\eta' \to \eta}(\mu_{\eta' \to \eta} \times \mu_\eta)$. Provided this conversion is incorporated in the definition of the quantum projective structure, one can check that the three-spaces consistency condition still holds. In contrast, the consistency of the measures is essential for the inductive construction of prop. 3.5, where it ensures the compatibility of the reference states $\zeta_{\eta' \to \eta}$.

## 3.2 Holomorphic representation

We now turn to the holomorphic representation. In order to get the scalar product right, we cannot spare, when doing holomorphic quantization, a formulation using a prequantum bundle $\mathcal{B}_\eta$ (see [27, section 8] and appendix A, def. A.2) constructed over $\mathcal{M}_\eta$ (this differs from the previous subsection, for when dealing with configuration representation, the relevant part of the bundle structure is flat and, as a result, the prequantum bundle only needs to be taken into account if a unified context for describing various representations is demanded). Therefore, we begin by examining how to arrange prequantum bundles built on the $\mathcal{M}_\eta$ into a form of factorizing structure suitable for quantization. More precisely, we are looking for a way to connect the $\mathcal{B}_\eta$ bundles that will provide the required tensor product factorizations of the corresponding $L_2$-spaces of bundle cross-sections (the prequantum Hilbert spaces, see [27, section 8] and def. A.3).

To address this question, we go back to the basics underlying the tensor product decomposition of the $L_2$-space of complex-valued functions over a Cartesian product $\mathcal{A} \times \mathcal{B}$: there, the tensor product of a function on $\mathcal{A}$ with a function on $\mathcal{B}$ is obtained as their pointwise product. Accordingly, what we need in the hermitian line bundle case is an operation to make the 'product' of a point in the bundle above $\mathcal{A}$ with a point in the one above $\mathcal{B}$, and we want this operation to be valued in the bundle we happen to have above $\mathcal{A} \times \mathcal{B}$.

**Definition 3.6** Let $\mathcal{M}$, $\mathcal{N}'$ and $\mathcal{N}$ be three smooth, finite dimensional, manifolds, and let $\varphi : \mathcal{M} \to \mathcal{N}' \times \mathcal{N}$ be a diffeomorphism. Let $\mathcal{B}_\mathcal{M}$, $\mathcal{B}_{\mathcal{N}'}$ and $\mathcal{B}_\mathcal{N}$ be hermitian line bundles (def. A.1), respectively with base $\mathcal{M}$, $\mathcal{N}'$ and $\mathcal{N}$. We call a smooth map $\zeta : \mathcal{B}_{\mathcal{N}'} \times \mathcal{B}_\mathcal{N} \to \mathcal{B}_\mathcal{M}$ a factorization of $\mathcal{B}_\mathcal{M}$ compatible with $\varphi$ iff:

1. $\varphi \circ \Pi_{\mathcal{B}_\mathcal{M}} \circ \zeta = \Pi_{\mathcal{B}_{\mathcal{N}'}} \times \Pi_{\mathcal{B}_\mathcal{N}}$, where $\Pi_{\mathcal{B}_\mathcal{M}}$, $\Pi_{\mathcal{B}_{\mathcal{N}'}}$ and $\Pi_{\mathcal{B}_\mathcal{N}}$ are the bundles projections of $\mathcal{B}_\mathcal{M}$, $\mathcal{B}_{\mathcal{N}'}$ and $\mathcal{B}_\mathcal{N}$ respectively;

2. $\forall z' \in \mathcal{B}_{\mathcal{N}'}, \forall z \in \mathcal{B}_\mathcal{N}, \quad |\zeta(z', z)| = |z'|\,|z|$;

3. $\forall z' \in \mathcal{B}_{\mathcal{N}'}, \forall z \in \mathcal{B}_\mathcal{N}, \forall \lambda', \lambda \in \mathbb{C}, \quad \zeta(\lambda' z', \lambda z) = \lambda' \lambda\, \zeta(z', z)$.

**Proposition 3.7** We consider the same objects as in def. 3.6. Moreover, we assume that $\mathcal{N}'$ and $\mathcal{N}$ are equipped with smooth measures $\mu_{\mathcal{N}'}$ and $\mu_\mathcal{N}$, and we equip $\mathcal{M}$ with the smooth measure $\mu_\mathcal{M} := \varphi^*(\mu_{\mathcal{N}'} \times \mu_\mathcal{N})$. Then, there exists a unique Hilbert space isomorphism:

$$\Phi_\zeta : L_2\left(\mathcal{M} \to \mathcal{B}_\mathcal{M}, d\mu_\mathcal{M}\right) \to L_2\left(\mathcal{N}' \to \mathcal{B}_{\mathcal{N}'}, d\mu_{\mathcal{N}'}\right) \otimes L_2\left(\mathcal{N} \to \mathcal{B}_\mathcal{N}, d\mu_\mathcal{N}\right),$$

such that:

$$\forall s' \in L_2\left(\mathcal{N}' \to \mathcal{B}_{\mathcal{N}'}, d\mu_{\mathcal{N}'}\right), \forall s \in L_2\left(\mathcal{N} \to \mathcal{B}_\mathcal{N}, d\mu_\mathcal{N}\right), \quad \Phi_\zeta\left(\widetilde{\zeta}(s', s)\right) = s' \otimes s,$$

where $\forall x', x \in \mathcal{N}' \times \mathcal{N}, \widetilde{\zeta}(s', s) \circ \varphi^{-1}(x', x) := \zeta\left(s'(x'), s(x)\right)$.

**Proof** We define $\mathcal{H}_\mathcal{M} := L_2(\mathcal{M} \to \mathcal{B}_\mathcal{M}, d\mu_\mathcal{M})$, and similarly $\mathcal{H}_{\mathcal{N}'}$ and $\mathcal{H}_\mathcal{N}$.

We first want to prove that $\mathrm{Vect}\left\{\widetilde{\zeta}(s', s) \,\middle|\, s' \in \mathcal{H}_{\mathcal{N}'} \,\&\, s \in \mathcal{H}_\mathcal{N}\right\}$ is dense in $\mathcal{H}_\mathcal{M}$. It is well-defined as a vector subspace of $\mathcal{H}_\mathcal{M}$ for $\forall s' \in \mathcal{H}_\mathcal{M}, \forall s \in \mathcal{H}_\mathcal{N}, \widetilde{\zeta}(s', s)$ is a cross-section of $\mathcal{B}_\mathcal{M}$ (from def. 3.6.1) and $\left\|\widetilde{\zeta}(s', s)\right\|_\mathcal{M} = \|s'\|_{\mathcal{N}'} \|s\|_\mathcal{N}$ (from def. 3.6.2 and Fubini's theorem), hence $\widetilde{\zeta}(s', s) \in \mathcal{H}_\mathcal{M}$.

The cross-sections with compact support are dense in $\mathcal{H}_\mathcal{M}$, and, by partition of the unity, they are linear combinations of cross-sections with compact support of the form $W := \varphi^{-1}\langle V' \times V \rangle$, where $V'$ is a trivialization patch for $\mathcal{B}_{\mathcal{N}'}$ and $V$ a trivialization patch for $\mathcal{B}_{\mathcal{N}}$. Given a non-zero cross-section $s$, resp. $s'$, of $\mathcal{B}_{\mathcal{N}'}|_{V'}$, resp. $\mathcal{B}_{\mathcal{N}}|_V$, we define a non-zero cross-section $s''$ of $\mathcal{B}_{\mathcal{M}}|_W$ by:

$$\forall x', x \in V' \times V, \; s'' \circ \varphi(x', x) := \zeta\left(s'(x'), s(x)\right).$$

Thus, using the trivialization defined by $s'$, resp. $s$, $s''$, to identify the vector subspace of $\mathcal{H}_{\mathcal{N}'}$, resp. $\mathcal{H}_{\mathcal{N}}$, $\mathcal{H}_{\mathcal{M}}$, of cross-sections with compact support in $V'$, resp. $V$, $W$, with $L_2\left(V', d\mu_{\mathcal{N}'}|_{V'}\right)$, resp. $L_2\left(V, d\mu_{\mathcal{N}}|_V\right)$, $L_2\left(W, d\mu_{\mathcal{N}}|_W\right)$, the restriction $\widetilde{\zeta}_W$ of $\widetilde{\zeta}$ to these vector subspaces is given by:

$$\widetilde{\zeta}_W(f' s', f s) = \left(f' \otimes f\right) \circ \left(\varphi|_{W \to V' \times V}\right) s'' \text{ (from def. 3.6.3).}$$

Hence, its image is dense in $L_2\left(W, d\mu_{\mathcal{N}}|_W\right) = L_2\left(V', d\mu_{\mathcal{N}'}|_{V'}\right) \otimes L_2\left(W, d\mu_{\mathcal{N}}|_W\right)$ (which equality follows from def. 3.6.2 and Fubini's theorem).

Now, the application:

$$\widetilde{\zeta} \;:\; \mathcal{H}_{\mathcal{N}'} \times \mathcal{H}_{\mathcal{N}} \;\to\; \mathcal{H}_{\mathcal{M}}$$
$$(s', s) \;\mapsto\; \widetilde{\zeta}(s', s)$$

is a bilinear map (from def. 3.6.3) and satisfies (from def. 3.6.2 and Fubini's theorem):

$$\forall s', t' \in \mathcal{H}_{\mathcal{N}'}, \forall s, t \in \mathcal{H}_{\mathcal{N}}, \;\; \left\langle \widetilde{\zeta}(s', s), \widetilde{\zeta}(t', t) \right\rangle_{\mathcal{M}} = \langle s', t' \rangle_{\mathcal{N}'} \langle s, t \rangle_{\mathcal{N}}.$$

Hence, there exists a unique Hilbert space isomorphism $\Phi_\zeta^{-1} : \mathcal{H}_{\mathcal{N}'} \otimes \mathcal{H}_{\mathcal{N}} \to \overline{\text{Vect Im}\widetilde{\zeta}} = \mathcal{H}_{\mathcal{M}}$, such that $\forall s' \in \mathcal{H}_{\mathcal{N}'}, \forall s \in \mathcal{H}_{\mathcal{N}}, \;\; \Phi_\zeta^{-1}(s' \otimes s) = \widetilde{\zeta}(s', s)$. $\square$

With this, we can now present the announced factorizing structure for prequantum bundles. As usual, we need to require an appropriate 'three-spaces consistency' that will support the corresponding consistency of the projective limits we are ultimately interested in (fig. 3.1 looks slightly different from what we had for factorizing system of phase spaces in [15, fig. 2.2], because we are forced to define the maps $\zeta$ in the direction opposite to our standard convention for factorizing maps). Note that we also have a compatibility condition involving the connection of the prequantum bundles, that will come into play when (pre-)quantizing observables and expressing the holomorphic condition.

**Definition 3.8** Let $(\mathcal{L}, \mathcal{M}, \varphi)^\times$ be a factorizing system of finite dimensional phase spaces [15, def. 2.12]. A factorizing system of prequantum bundles for $(\mathcal{L}, \mathcal{M}, \varphi)^\times$ is a quadruple:

$$\left(\left(\mathcal{B}_\eta, \nabla_\eta\right)_{\eta \in \mathcal{L}}, \left(\mathcal{B}_{\eta' \to \eta}, \nabla_{\eta' \to \eta}\right)_{\eta \preccurlyeq \eta'}, \left(\zeta_{\eta' \to \eta}\right)_{\eta \preccurlyeq \eta'}, \left(\zeta_{\eta'' \to \eta' \to \eta}\right)_{\eta \preccurlyeq \eta' \preccurlyeq \eta''}\right)$$

such that:
1. $\forall \eta \in \mathcal{L}$, $\left(\mathcal{B}_\eta, \nabla_\eta\right)$ is a prequantum bundle for $\mathcal{M}_\eta$ (def. A.2);

2. $\forall \eta \preccurlyeq \eta' \in \mathcal{L}$, $\left(\mathcal{B}_{\eta' \to \eta}, \nabla_{\eta' \to \eta}\right)$ is a prequantum bundle for $\mathcal{M}_{\eta' \to \eta}$ (except for the case $\eta = \eta'$: $\mathcal{M}_{\eta \to \eta}$ has only one element and $\mathcal{B}_{\eta \to \eta} = \mathbb{C}$);

3. $\forall \eta \preccurlyeq \eta' \in \mathcal{L}$, $\zeta_{\eta' \to \eta} : \mathcal{B}_{\eta' \to \eta} \times \mathcal{B}_\eta \to \mathcal{B}_{\eta'}$ is a factorization of $\mathcal{B}_{\eta'}$ compatible with $\varphi_{\eta' \to \eta} : \mathcal{M}_{\eta'} \to$

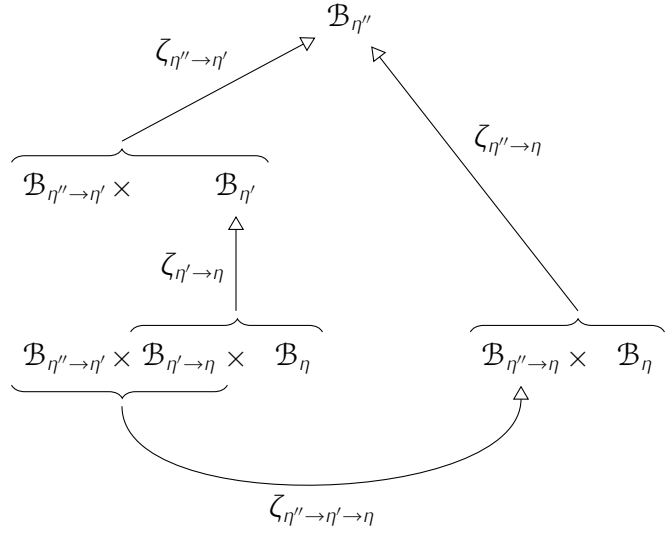

Figure 3.1 – Three-spaces consistency for factorizing systems of prequantum bundles

$\mathcal{M}_{\eta'\to\eta} \times \mathcal{M}_\eta$;

4. $\forall \eta \preccurlyeq \eta' \in \mathcal{L}$, $\forall y \in \mathcal{B}_{\eta'\to\eta}$, $\forall z \in \mathcal{B}_\eta$,

$$\left[T_{y,z}\,\zeta_{\eta'\to\eta}\right]\left\langle \mathrm{Hor}_y(\mathcal{B}_{\eta'\to\eta}, \nabla_{\eta'\to\eta}) \times \mathrm{Hor}_z(\mathcal{B}_\eta, \nabla_\eta)\right\rangle = \mathrm{Hor}_{\zeta_{\eta'\to\eta}(y,z)}(\mathcal{B}_{\eta'}, \nabla_{\eta'}), \tag{3.8.1}$$

where $\mathrm{Hor}_z(\mathcal{B}_\eta, \nabla_\eta)$ is defined as the $\nabla_\eta$-horizontal subspace of $T_z(\mathcal{B}_\eta)$ for $z \in \mathcal{B}_\eta$, and $\mathrm{Hor}_y(\mathcal{B}_{\eta'\to\eta}, \nabla_{\eta'\to\eta})$ is defined similarly for $y \in \mathcal{B}_{\eta'\to\eta}$;

5. $\forall \eta \preccurlyeq \eta' \preccurlyeq \eta'' \in \mathcal{L}$, $\zeta_{\eta''\to\eta'\to\eta} : \mathcal{B}_{\eta''\to\eta'} \times \mathcal{B}_{\eta'\to\eta} \to \mathcal{B}_{\eta''\to\eta}$ is a smooth map such that:

$$\zeta_{\eta''\to\eta} \circ (\zeta_{\eta''\to\eta'\to\eta} \times \mathrm{id}_{\mathcal{B}_\eta}) = \zeta_{\eta''\to\eta'} \circ (\mathrm{id}_{\mathcal{B}_{\eta''\to\eta'}} \times \zeta_{\eta'\to\eta}). \tag{3.8.2}$$

Def. 3.8 seems to require a lot, so it is reassuring that, at least in the topologically trivial case, we can construct such a structure for any factorizing system of phase spaces satisfying nothing but the quantization rule [27, section 8.3], which is anyhow mandatory to ensure the existence of prequantum bundles for the $\mathcal{M}_\eta$.

**Theorem 3.9** Let $(\mathcal{L}, \mathcal{M}, \varphi)^\times$ be a factorizing system of finite dimensional phase spaces such that:
1. $\forall \eta \in \mathcal{L}$, $\mathcal{M}_\eta$ is simply-connected;

2. $\forall \eta \in \mathcal{L}$, $\forall S$ a closed oriented 2-surface in $\mathcal{M}_\eta$, $\int_S \Omega_\eta \in 2\pi\,\mathbb{Z}$, where $\Omega_\eta$ is the symplectic structure of $\mathcal{M}_\eta$.

Then there exists a factorizing system of prequantum bundles for $(\mathcal{L}, \mathcal{M}, \varphi)^\times$.

**Proof** For $\eta \preccurlyeq \eta' \in \mathcal{L}$, we have that $\mathcal{M}_{\eta'\to\eta}$ is simply-connected, otherwise $\mathcal{M}_{\eta'} \approx \mathcal{M}_{\eta'\to\eta} \times \mathcal{M}_\eta$ would not be simply-connected.

Besides, for any oriented 2-surface $S_{\eta'}$ in $\mathcal{M}_{\eta'}$, we have:

$$\int_{S_{\eta'}} \Omega_{\eta'} = \int_{\varphi_{\eta' \to \eta} \circ S_{\eta'}} \varphi_{\eta' \to \eta}^{-1,*} \Omega_{\eta'}$$

$$= \int_{\varphi_{\eta' \to \eta} \circ S_{\eta'}} \Omega_{\eta' \to \eta} \times \Omega_\eta = \int_{S_{\eta' \to \eta}} \Omega_{\eta' \to \eta} + \int_{S_\eta} \Omega_\eta, \quad (3.9.1)$$

where $S_{\eta' \to \eta}$, resp. $S_\eta$, is the projection on $\mathcal{M}_{\eta' \to \eta}$, resp. $\mathcal{M}_\eta$, of $\varphi_{\eta' \to \eta} \circ S_{\eta'}$ (which is an oriented 2-surface in $\mathcal{M}_{\eta' \to \eta} \times \mathcal{M}_\eta$). In particular, if $S_{\eta' \to \eta}$ is a closed oriented 2-surface in $\mathcal{M}_{\eta' \to \eta}$, applying eq. (3.9.1) to $S_{\eta'} := \varphi_{\eta' \to \eta}^{-1} \circ (S_{\eta' \to \eta} \times \{x_\eta\})$, where $x_\eta$ is any point in $\mathcal{M}_\eta$, gives:

$$\int_{S_{\eta' \to \eta}} \Omega_{\eta' \to \eta} = \int_{S_{\eta'}} \Omega_{\eta'} \in 2\pi \mathbb{Z}.$$

Let $(x_\eta^o)_{\eta \in \mathcal{L}} \in \mathcal{S}_{(\mathcal{L}, \mathcal{M}, \varphi)}^\times$ and let $\eta \in \mathcal{L}$. The construction in [27, section 8.3] tells us that, thanks to the conditions 3.9.1 and 3.9.2, we can construct a prequantum bundle $(\mathcal{B}_\eta, \nabla_\eta)$ for $\mathcal{M}_\eta$ in such a way that $\mathcal{B}_\eta$ can be identified with the equivalence classes in:

$$\{(x_\eta, z_\eta, \gamma_\eta) \mid x_\eta \in \mathcal{M}_\eta, z_\eta \in \mathbb{C}, \gamma_\eta \text{ is a piecewise smooth path from } x_\eta^o \text{ to } x_\eta\},$$

for the equivalence relation:

$$((x_\eta, z_\eta, \gamma_\eta) \simeq (x_\eta', z_\eta', \gamma_\eta')) \iff \begin{cases} x_\eta = x_\eta' \\ z_\eta' = z_\eta \exp\left(-i \int_{\Sigma_\eta(\gamma_\eta, \gamma_\eta')} \Omega_\eta\right) \end{cases},$$

where $\Sigma_\eta(\gamma_\eta, \gamma_\eta')$ is any oriented 2-surface in $\mathcal{M}_\eta$ such that $\partial \Sigma_\eta(\gamma_\eta, \gamma_\eta') = \gamma_\eta'^{-1} \cdot \gamma_\eta$. Moreover, the $\nabla_\eta$-parallel transport along some path $\gamma_\eta'$ in $\mathcal{M}_\eta$ is then given by:

$$P_{\gamma_\eta'}^{\nabla_\eta}\left([\gamma_\eta'(0), z_\eta, \gamma_\eta]_\simeq\right) = [\gamma_\eta'(1), z_\eta, \gamma_\eta' \cdot \gamma_\eta]_\simeq.$$

Since we proved above that, for all $\eta \preccurlyeq \eta'$, $\mathcal{M}_{\eta' \to \eta}$ also fulfills these conditions, we can make the same construction to obtain a prequantum bundle $(\mathcal{B}_{\eta' \to \eta}, \nabla_{\eta' \to \eta})$, using as origin the point $x_{\eta' \to \eta}^o \in \mathcal{M}_{\eta' \to \eta}$, defined by $\varphi_{\eta' \to \eta}(x_{\eta'}^o) = (x_{\eta' \to \eta}^o, x_\eta^o)$.

Now, for $\eta \preccurlyeq \eta' \in \mathcal{L}$, we define $\zeta_{\eta' \to \eta} : \mathcal{B}_{\eta' \to \eta} \times \mathcal{B}_\eta \to \mathcal{B}_{\eta'}$ by:

$$\zeta_{\eta' \to \eta}\left([x_{\eta' \to \eta}, z_{\eta' \to \eta}, \gamma_{\eta' \to \eta}]_\simeq, [x_\eta, z_\eta, \gamma_\eta]_\simeq\right) := \left[\varphi_{\eta' \to \eta}^{-1}(x_{\eta' \to \eta}, x_\eta), z_{\eta' \to \eta} z_\eta, \varphi_{\eta' \to \eta}^{-1}(\gamma_{\eta' \to \eta}, \gamma_\eta)\right]_\simeq.$$

This is a well-defined map, for we have, using eq. (3.9.1):

$$\exp\left(-i \int_{\Sigma_{\eta'}\left(\varphi_{\eta' \to \eta}^{-1}(\gamma_{\eta' \to \eta}, \gamma_\eta), \varphi_{\eta' \to \eta}^{-1}(\gamma_{\eta' \to \eta}', \gamma_\eta')\right)} \Omega_{\eta'}\right) =$$

$$= \exp\left(-i \int_{\Sigma_{\eta' \to \eta}(\gamma_{\eta' \to \eta}, \gamma_{\eta' \to \eta}')} \Omega_{\eta' \to \eta}\right) \exp\left(-i \int_{\Sigma_\eta(\gamma_\eta, \gamma_\eta')} \Omega_\eta\right).$$

Moreover, we can check that it fulfills defs. 3.6.1 to 3.6.3.

Let $x_{\eta'} \in \mathcal{M}_{\eta'}$ and $(x_{\eta' \to \eta}, x_\eta) = \varphi_{\eta' \to \eta}(x_{\eta'})$. Let $\gamma_\eta$ and $\gamma_{\eta' \to \eta}$ be piecewise smooth paths from $x_\eta^o$ to $x_\eta$, and $x_{\eta' \to \eta}^o$ to $x_{\eta' \to \eta}$, respectively. We can choose local coordinates around $x_{\eta' \to \eta}$ in $\mathcal{M}_{\eta' \to \eta}$ and around $x_\eta$ in $\mathfrak{M}_\eta$. Hence, we have diffeomorphisms $\psi_{\eta'\to\eta} : [-1, 1]^{d_{\eta'\to\eta}} \to U_{\eta'\to\eta}$, resp. $\psi_\eta : [-1, 1]^{d_\eta} \to U_\eta$, where $d_{\eta'\to\eta} := \dim \mathfrak{M}_{\eta'\to\eta}$, resp. $d_\eta := \dim \mathfrak{M}_\eta$, and $U_{\eta'\to\eta}$, resp. $U_\eta$, is an open neighborhood of $x_{\eta'\to\eta}$ in $\mathfrak{M}_{\eta'\to\eta}$, resp. of $x_\eta$ in $\mathfrak{M}_\eta$. This provides us local trivializations of the bundles $\mathcal{B}_{\eta'}$, $\mathcal{B}_{\eta'\to\eta}$ and $\mathcal{B}_\eta$, by:

$$\forall r_{\eta'\to\eta} \in [-1, 1]^{d_{\eta'\to\eta}}, \forall z_{\eta'\to\eta} \in \mathbb{C},$$

$$\Psi_{\eta'\to\eta}(r_{\eta'\to\eta}, z_{\eta'\to\eta}) = \left[\psi_{\eta'\to\eta}(r_{\eta'\to\eta}), z_{\eta'\to\eta}, \chi_{r_{\eta'\to\eta}} \cdot \gamma_{\eta'\to\eta}\right]_\simeq,$$

$$\forall r_\eta \in [-1, 1]^{d_\eta}, \forall z_\eta \in \mathbb{C},$$

$$\Psi_\eta(r_\eta, z_\eta) = \left[\psi_\eta(r_\eta), z_\eta, \chi_{r_\eta} \cdot \gamma_\eta\right]_\simeq,$$

$$\forall r_{\eta'\to\eta}, r_\eta \in [-1, 1]^{d_{\eta'\to\eta}} \times [-1, 1]^{d_\eta}, \forall z_{\eta'} \in \mathbb{C},$$

$$\Psi_{\eta'}(r_{\eta'\to\eta}, r_\eta, z_{\eta'}) = \left[\varphi_{\eta'\to\eta}^{-1}\left(\psi_{\eta'\to\eta}(r_{\eta'\to\eta}), \psi_\eta(r_\eta)\right), z_{\eta'}, \varphi_{\eta'\to\eta}^{-1}\left(\chi_{r_{\eta'\to\eta}} \cdot \gamma_{\eta'\to\eta}, \chi_{r_\eta} \cdot \gamma_\eta\right)\right]_\simeq,$$

where $\chi_{r_{\eta'\to\eta}} : \tau \mapsto \psi_{\eta'\to\eta}(\tau r_{\eta'\to\eta})$ and $\chi_{r_\eta} : \tau \mapsto \psi_\eta(\tau r_\eta)$.

And we have:

$$\forall r_{\eta'\to\eta}, r_\eta \in [-1, 1]^{d_{\eta'\to\eta}} \times [-1, 1]^{d_\eta}, \forall z_{\eta'\to\eta}, z_\eta \in \mathbb{C},$$

$$\Psi_{\eta'}^{-1} \circ \zeta\left(\Psi_{\eta'\to\eta}(r_{\eta'\to\eta}, z_{\eta'\to\eta}), \Psi_\eta(r_\eta, z_\eta)\right) = (r_{\eta'\to\eta}, r_\eta, z_{\eta'\to\eta} z_\eta),$$

Therefore, $\zeta_{\eta'\to\eta}$ is smooth.

Then, for $\gamma'_{\eta'\to\eta}$ a path in $\mathfrak{M}_{\eta'\to\eta}$, and $\gamma'_\eta$ a path in $\mathfrak{M}_\eta$, we have:

$$P^{\nabla_{\eta'}}_{\varphi_{\eta'\to\eta}^{-1}(\gamma'_{\eta'\to\eta}, \gamma'_\eta)} \circ \zeta_{\eta'\to\eta}\left(\left[\gamma'_{\eta'\to\eta}(0), z_{\eta'\to\eta}, \gamma_{\eta'\to\eta}\right]_\simeq, \left[\gamma'_\eta(0), z_\eta, \gamma_\eta\right]_\simeq\right) =$$

$$= \left[\varphi_{\eta'\to\eta}^{-1}\left(\gamma'_{\eta'\to\eta}(1), \gamma'_\eta(1)\right), z_{\eta'\to\eta} z_\eta, \varphi_{\eta'\to\eta}^{-1}\left(\gamma'_{\eta'\to\eta} \cdot \gamma_{\eta'\to\eta}, \gamma'_\eta \cdot \gamma_\eta\right)\right]_\simeq$$

$$= \zeta_{\eta'\to\eta}\left(P^{\nabla_{\eta'\to\eta}}_{\gamma'_{\eta'\to\eta}}\left(\left[\gamma'_{\eta'\to\eta}(0), z_{\eta'\to\eta}, \gamma_{\eta'\to\eta}\right]_\simeq\right), P^{\nabla_\eta}_{\gamma'_\eta}\left(\left[\gamma'_\eta(0), z_\eta, \gamma_\eta\right]_\simeq\right)\right),$$

hence $P^{\nabla_{\eta'}}_{\varphi_{\eta'\to\eta}^{-1}(\gamma'_{\eta'\to\eta}, \gamma'_\eta)} \circ \zeta_{\eta'\to\eta} = \zeta_{\eta'\to\eta} \circ \left(P^{\nabla_{\eta'\to\eta}}_{\gamma'_{\eta'\to\eta}}, P^{\nabla_\eta}_{\gamma'_\eta}\right)$. Therefore, eq. (3.8.1) is fulfilled.

Lastly, for $\eta \preccurlyeq \eta' \preccurlyeq \eta'' \in \mathcal{L}$, we can in a similar way define a map $\zeta_{\eta''\to\eta'\to\eta} : \mathcal{B}_{\eta''\to\eta'} \times \mathcal{B}_{\eta'\to\eta}$ and we have:

$$\zeta_{\eta''\to\eta} \circ (\zeta_{\eta''\to\eta'\to\eta} \times \mathrm{id}_{\mathcal{B}_\eta})\left(\left[x_{\eta''\to\eta'}, z_{\eta''\to\eta'}, \gamma_{\eta''\to\eta'}\right]_\simeq, \left[x_{\eta'\to\eta}, z_{\eta'\to\eta}, \gamma_{\eta'\to\eta}\right]_\simeq, \left[x_\eta, z_\eta, \gamma_\eta\right]_\simeq\right) =$$

$$= \left[\varphi_{\eta''\to\eta}^{-1}\left(\varphi_{\eta''\to\eta'\to\eta}^{-1}(x_{\eta''\to\eta'}, x_{\eta'\to\eta}), x_\eta\right), z_{\eta''\to\eta'} z_{\eta'\to\eta} z_\eta, \varphi_{\eta''\to\eta}^{-1}\left(\varphi_{\eta''\to\eta'\to\eta}^{-1}(\gamma_{\eta''\to\eta'}, \gamma_{\eta'\to\eta}), \gamma_\eta\right)\right]_\simeq$$

$$= \left[\varphi_{\eta''\to\eta'}^{-1}\left(x_{\eta''\to\eta'}, \varphi_{\eta'\to\eta}^{-1}(x_{\eta'\to\eta}, x_\eta)\right), z_{\eta''\to\eta'} z_{\eta'\to\eta} z_\eta, \varphi_{\eta''\to\eta'}^{-1}\left(\gamma_{\eta''\to\eta'}, \varphi_{\eta'\to\eta}^{-1}(\gamma_{\eta'\to\eta}, \gamma_\eta)\right)\right]_\simeq$$

(using [15, eq. (2.11.1)])

$$= \zeta_{\eta''\to\eta'} \circ \left(\mathrm{id}_{\mathcal{B}_{\eta''\to\eta'}} \times \zeta_{\eta'\to\eta}\right)\left(\left[x_{\eta''\to\eta'}, z_{\eta''\to\eta'}, \gamma_{\eta''\to\eta'}\right]_\simeq, \left[x_{\eta'\to\eta}, z_{\eta'\to\eta}, \gamma_{\eta'\to\eta}\right]_\simeq, \left[x_\eta, z_\eta, \gamma_\eta\right]_\simeq\right),$$

therefore eq. (3.8.2) holds. $\square$

The last ingredient we need in order to perform prequantization are measures on the $\mathcal{M}_\eta$ and $\mathcal{M}_{\eta'\to\eta}$, and they should be compatible, like we asked when setting up the configuration representation. But this is in fact something we can get automatically and in a very straightforward way from the structure $(\mathcal{L}, \mathcal{M}, \varphi)^\times$, since a symplectic form gives us a natural volume form and the compatibility of the symplectic forms is enough to ensure the compatibility of their associated volume form.

**Proposition 3.10** Let $(\mathcal{L}, \mathcal{M}, \varphi)^\times$ be a factorizing system of finite dimensional phase spaces. We define:

1. for $\eta \in \mathcal{L}$, the volume form $\omega_\eta := \dfrac{1}{(d_\eta/2)!} \Omega_\eta \wedge \ldots \wedge \Omega_\eta = \dfrac{1}{(d_\eta/2)!} \Omega_\eta^{\wedge d_\eta/2}$ on $\mathcal{M}_\eta$ (where $d_\eta = \dim \mathcal{M}_\eta$) and the corresponding smooth measure $\mu_\eta$ on $\mathcal{M}_\eta$;

2. for $\eta \prec \eta' \in \mathcal{L}$, the volume form $\omega_{\eta'\to\eta} := \dfrac{1}{(d_{\eta'\to\eta}/2)!} \Omega_{\eta'\to\eta} \wedge \ldots \wedge \Omega_{\eta'\to\eta} = \dfrac{1}{(d_{\eta'\to\eta}/2)!} \Omega_{\eta'\to\eta}^{\wedge d_{\eta'\to\eta}/2}$ on $\mathcal{M}_{\eta'\to\eta}$ (where $d_{\eta'\to\eta} = \dim \mathcal{M}_{\eta'\to\eta}$) and the corresponding smooth measure $\mu_{\eta'\to\eta}$ on $\mathcal{M}_{\eta'\to\eta}$.

Then, this equips $(\mathcal{L}, \mathcal{M}, \varphi)^\times$ with a structure of factorizing system of measured manifolds (def. 3.1).

**Proof** That $\omega_\eta$, resp. $\omega_{\eta'\to\eta}$, is a nowhere-vanishing top-dimensional form on $\mathcal{M}_\eta$, resp. $\mathcal{M}_{\eta'\to\eta}$, can be checked in local Darboux coordinates.

What is left to prove is the compatibility of these definitions of the volume forms with the maps $\varphi_{\eta'\to\eta}$ and $\varphi_{\eta''\to\eta'\to\eta}$ (def. 3.1.3). For $\eta \prec \eta'$, we have:

$$\varphi_{\eta'\to\eta}^{-1,*} \omega_{\eta'} = \frac{1}{(d_{\eta'}/2)!} \left( \varphi_{\eta'\to\eta}^{-1,*} \Omega_{\eta'} \right)^{\wedge d_{\eta'}/2}$$

$$= \frac{1}{(d_{\eta'}/2)!} \left( \Omega_{\eta'\to\eta} \times \Omega_\eta \right)^{\wedge d_{\eta'}/2} \quad \text{(since } \varphi_{\eta'\to\eta} \text{ is a symplectomorphism)}$$

$$= \frac{1}{(d_{\eta'\to\eta}/2)! \, (d_\eta/2)!} \left( \Omega_{\eta'\to\eta} \right)^{\wedge d_{\eta'\to\eta}/2} \wedge \left( \Omega_\eta \right)^{\wedge d_\eta/2}$$

$$= \omega_{\eta'\to\eta} \wedge \omega_\eta ,$$

hence $\varphi_{\eta'\to\eta}^{-1,*} \mu_{\eta'} = \mu_{\eta'\to\eta} \times \mu_\eta$, and similarly, for $\eta \prec \eta' \prec \eta''$:

$$\varphi_{\eta''\to\eta'\to\eta}^{-1,*} \mu_{\eta''\to\eta} = \mu_{\eta''\to\eta'} \times \mu_{\eta'\to\eta} .$$

$\square$

On the grounds of the preliminaries developed so far, the prequantization of a factorizing system of prequantum bundles is actually very similar to what we did for the position quantization, and, again, the link connecting the classical structure and the (pre-)quantum one is demonstrated by exposing the correspondence between observables.

**Proposition 3.11** Let $(\mathcal{L}, \mathcal{M}, \varphi)^\times$ be a factorizing system of finite dimensional phase spaces, equipped with a structure of factorizing system of measured manifolds according to prop. 3.10, and let $\left( (\mathcal{B}_\eta, \nabla_\eta)_{\eta \in \mathcal{L}}, (\mathcal{B}_{\eta' \to \eta}, \nabla_{\eta' \to \eta})_{\eta \preccurlyeq \eta'}, (\zeta_{\eta' \to \eta})_{\eta \preccurlyeq \eta'}, (\zeta_{\eta'' \to \eta' \to \eta})_{\eta \preccurlyeq \eta' \preccurlyeq \eta''} \right)$ be a factorizing system of prequantum bundles for $(\mathcal{L}, \mathcal{M}, \varphi)^\times$.

We define:
1. for $\eta \in \mathcal{L}$, $\mathcal{H}_\eta^{\text{preQ}} := L_2(\mathcal{M}_\eta \to \mathcal{B}_\eta, d\mu_\eta)$;
2. for $\eta \prec \eta' \in \mathcal{L}$, $\mathcal{H}_{\eta' \to \eta}^{\text{preQ}} := L_2(\mathcal{M}_{\eta' \to \eta} \to \mathcal{B}_{\eta' \to \eta}, d\mu_{\eta' \to \eta})$, and $\Phi_{\eta' \to \eta}^{\text{preQ}} := \Phi_{\zeta_{\eta' \to \eta}} : \mathcal{H}_{\eta'}^{\text{preQ}} \to \mathcal{H}_{\eta' \to \eta}^{\text{preQ}} \otimes \mathcal{H}_\eta^{\text{preQ}}$.

Then, we can complete these elements into a projective system of quantum state spaces $(\mathcal{L}, \mathcal{H}^{\text{preQ}}, \Phi^{\text{preQ}})^\otimes$.

**Proof** The proof works like the one for prop. 3.3, using prop. 3.7 and:

$$\forall \eta \prec \eta' \prec \eta'', \forall s'' \in \mathcal{H}_{\eta'' \to \eta'}^{\text{preQ}}, \forall s' \in \mathcal{H}_{\eta' \to \eta}^{\text{preQ}}, \forall s \in \mathcal{H}_\eta^{\text{preQ}},$$

$$\Phi_{\zeta_{\eta'' \to \eta'}}^{-1} \circ \left( \text{id}_{\mathcal{H}_{\eta'' \to \eta'}^{\text{preQ}}} \otimes \Phi_{\zeta_{\eta' \to \eta}}^{-1} \right) (s'' \otimes s' \otimes s) = \widetilde{\zeta}_{\eta'' \to \eta'} \left( s'', \widetilde{\zeta}_{\eta' \to \eta}(s', s) \right)$$

$$= \widetilde{\zeta}_{\eta'' \to \eta} \left( \widetilde{\zeta}_{\eta'' \to \eta' \to \eta}(s'', s'), s \right) \text{ (from eq. (3.8.2) and [15, eq. (2.11.1)])}$$

$$= \Phi_{\zeta_{\eta'' \to \eta}}^{-1} \circ \left( \Phi_{\zeta_{\eta'' \to \eta' \to \eta}}^{-1} \otimes \text{id}_{\mathcal{H}_\eta^{\text{preQ}}} \right) (s'' \otimes s' \otimes s).$$

$\square$

**Proposition 3.12** We consider the same objects as in prop. 3.11. For $f_\eta \in C^\infty(\mathcal{M}_\eta, \mathbb{R})$, we define the prequantization $\widehat{f}_\eta$ of $f_\eta$ as a densely defined operator on $\mathcal{H}_\eta^{\text{preQ}}$ (def. A.3).

Let $f_\eta \in C^\infty(\mathcal{M}_\eta, \mathbb{R})$ and $f_{\eta'} \in C^\infty(\mathcal{M}_{\eta'}, \mathbb{R})$, such that $f_\eta \sim f_{\eta'}$ (the equivalence relation is defined in [15, eq. (2.4.1)], where we use $\pi_{\eta' \to \eta}$ from [15, eq. (2.13.1)]). Then, $\widehat{f}_\eta \sim \widehat{f}_{\eta'}$ (with the equivalence relation defined in eq. (2.3.2)).

Hence, a classical observable $f = [f_\eta]_\sim \in \mathcal{O}_{(\mathcal{L},\mathcal{M},\varphi)}^\times$ [15, prop. 2.13] defines a prequantum observable $\widehat{f} := \left[\widehat{f}_\eta\right]_\sim \in \mathcal{O}_{(\mathcal{L},\mathcal{H}^{\text{preQ}},\Phi^{\text{preQ}})}^\otimes$ (prop. 2.5).

**Proof** Let $f_\eta \in C^\infty(\mathcal{M}_\eta, \mathbb{R})$ and let $\eta' \succcurlyeq \eta$. Let $s' \in \Phi_{\eta' \to \eta}^{\text{preQ},-1} \left\langle \mathcal{H}_{\eta' \to \eta}^{\text{preQ}} \otimes \mathcal{D}_\eta^{\text{preQ}} \right\rangle$ (where $\mathcal{D}_\eta^{\text{preQ}} \subset \mathcal{H}_\eta^{\text{preQ}}$ is the dense domain of $\widehat{f}_\eta$ and the $\otimes$ is to be understood as a tensor product of vector spaces). We define:

$$\Phi_{\eta' \to \eta}^{\text{preQ}}(s') =: \sum_\alpha t^\alpha \otimes s^\alpha, \text{ with } \forall \alpha, s^\alpha \in \mathcal{D}_\eta^{\text{preQ}}.$$

Then, we have:

$$\forall y, x \in \mathcal{M}_{\eta' \to \eta} \times \mathcal{M}_\eta, \left[ \Phi_{\eta' \to \eta}^{\text{preQ},-1} \circ \left( \text{id}_{\mathcal{H}_{\eta' \to \eta}^{\text{preQ}}} \otimes \widehat{f}_\eta \right) \circ \Phi_{\eta' \to \eta}^{\text{preQ}}(s') \right] \circ \varphi_{\eta' \to \eta}^{-1}(y, x) =$$

$$= \sum_\alpha \zeta_{\eta' \to \eta} \left[ t^\alpha(y), \left( f_\eta(x) s^\alpha(x) + i \nabla_{\eta, X_{f_\eta}} s^\alpha(x) \right) \right]$$

$$= \sum_\alpha f_\eta \circ \pi_{\eta' \to \eta} \circ \varphi^{-1}_{\eta' \to \eta}(y, x) \left[ \widetilde{\zeta}_{\eta' \to \eta} (t^\alpha, s^\alpha) \right] \circ \varphi^{-1}_{\eta' \to \eta}(y, x) +$$

$$+ i \sum_\alpha \left[ \nabla_{\eta', T\varphi^{-1}_{\eta' \to \eta}(0, X_{f_\eta})} \widetilde{\zeta}_{\eta' \to \eta} (t^\alpha, s^\alpha) \right] \circ \varphi^{-1}_{\eta' \to \eta}(y, x) \quad \text{(using eq. (3.8.1) and def. 3.6.3)}$$

$$= \left[ \left( f_\eta \circ \pi_{\eta' \to \eta} \right) s' + i \nabla_{\eta', X_{f_\eta \circ \pi_{\eta' \to \eta}}} s' \right] \circ \varphi^{-1}_{\eta' \to \eta}(y, x)$$

$$= \left[ \widehat{f_\eta \circ \pi_{\eta' \to \eta}} (s') \right] \circ \varphi^{-1}_{\eta' \to \eta}(y, x).$$

Therefore, we have $\forall f_\eta \in C^\infty(\mathcal{M}_\eta, \mathbb{R}), \forall \eta' \succcurlyeq \eta, \widehat{f_\eta} \sim \widehat{f_\eta \circ \pi_{\eta' \to \eta}}$. Hence, $\forall f_\eta \in C^\infty(\mathcal{M}_\eta, \mathbb{R}), \forall f_{\eta'} \in C^\infty(\mathcal{M}_{\eta'}, \mathbb{R}), \left( f_\eta \sim f_{\eta'} \Leftrightarrow \widehat{f_\eta} \sim \widehat{f_{\eta'}} \right).$ $\square$

Finally, we obtain the advertised holomorphic representation for a choice of Kähler structure on the symplectic manifolds $\mathcal{M}_\eta$. Requiring the factorizing maps to be holomorphic is enough to ensure that the holomorphic subspaces of the prequantum Hilbert spaces $\mathcal{H}_\eta$ set up above will correctly decompose over the already arranged tensor product factorizations, as can be shown by proving the corresponding factorizing properties of the orthogonal projections on these (closed) vector subspaces.

**Definition 3.13** A factorizing system of Kähler manifolds is a factorizing system of phase spaces $(\mathcal{L}, \mathcal{M}, \varphi)^\times$ [15, def. 2.12] such that:

1. for all $\eta \in \mathcal{L}$, $\mathcal{M}_\eta$ is equipped with a complex structure $J_\eta$ such that $(\mathcal{M}_\eta, \Omega_\eta, J_\eta)$ is a Kähler manifold (def. A.5);

2. for all $\eta \prec \eta' \in \mathcal{L}$, $\mathcal{M}_{\eta' \to \eta}$ is equipped with a complex structure $J_{\eta' \to \eta}$ such that $(\mathcal{M}_{\eta' \to \eta}, \Omega_{\eta' \to \eta}, J_{\eta' \to \eta})$ is a Kähler manifold;

3. for all $\eta \prec \eta' \in \mathcal{L}$, $\varphi_{\eta' \to \eta}$ is holomorphic, and for all $\eta \prec \eta' \prec \eta'' \in \mathcal{L}$, $\varphi_{\eta'' \to \eta' \to \eta}$ is holomorphic.

**Proposition 3.14** We consider the same objects as in prop. 3.11, but we now moreover assume that $(\mathcal{L}, \mathcal{M}, \varphi)^\times$ is a factorizing system of Kähler manifolds. We define:

1. for $\eta \in \mathcal{L}$, $\mathcal{H}^{\text{Holo}}_\eta := \mathcal{H}^{\text{preQ}}_\eta \cap \text{Holo}\left(\mathcal{M}_\eta \to \mathcal{B}_\eta\right)$ (prop. A.6);

2. for $\eta \prec \eta' \in \mathcal{L}$, $\mathcal{H}^{\text{Holo}}_{\eta' \to \eta} := \mathcal{H}^{\text{preQ}}_{\eta' \to \eta} \cap \text{Holo}\left(\mathcal{M}_{\eta' \to \eta} \to \mathcal{B}_{\eta' \to \eta}\right)$ and for $\eta \in \mathcal{L}$, $\mathcal{H}^{\text{Holo}}_{\eta \to \eta} := \mathcal{H}^{\text{preQ}}_{\eta \to \eta} = \mathbb{C}$.

Then, for all $\eta \preccurlyeq \eta'$, $\Phi^{\text{preQ}}_{\eta' \to \eta} \langle \mathcal{H}^{\text{Holo}}_{\eta'} \rangle = \mathcal{H}^{\text{Holo}}_{\eta' \to \eta} \otimes \mathcal{H}^{\text{Holo}}_\eta$ and for all $\eta \preccurlyeq \eta' \preccurlyeq \eta''$, $\Phi^{\text{preQ}}_{\eta'' \to \eta' \to \eta} \langle \mathcal{H}^{\text{Holo}}_{\eta'' \to \eta} \rangle = \mathcal{H}^{\text{Holo}}_{\eta'' \to \eta'} \otimes \mathcal{H}^{\text{Holo}}_{\eta' \to \eta}$. Hence, defining:

3. for $\eta \preccurlyeq \eta' \in \mathcal{L}$, $\Phi^{\text{Holo}}_{\eta' \to \eta} := \Phi^{\text{preQ}}_{\eta' \to \eta} \big|_{\mathcal{H}^{\text{Holo}}_{\eta'} \to \mathcal{H}^{\text{Holo}}_{\eta' \to \eta} \otimes \mathcal{H}^{\text{Holo}}_\eta}$;

4. and for $\eta \preccurlyeq \eta' \preccurlyeq \eta'' \in \mathcal{L}$, $\Phi^{\text{Holo}}_{\eta'' \to \eta' \to \eta} := \Phi^{\text{preQ}}_{\eta'' \to \eta' \to \eta} \big|_{\mathcal{H}^{\text{Holo}}_{\eta'' \to \eta} \to \mathcal{H}^{\text{Holo}}_{\eta'' \to \eta'} \otimes \mathcal{H}^{\text{Holo}}_{\eta' \to \eta}}$;

$(\mathcal{L}, \mathcal{H}^{\text{Holo}}, \Phi^{\text{Holo}})^{\otimes}$ is a projective system of quantum state spaces.

**Proof** First, for every $\eta \in \mathcal{L}$, we can define a complex structure on $\mathcal{B}_\eta$ in the following way: for $z \in \mathcal{B}_\eta$, with $x = \Pi_{\mathcal{B}_\eta}(z)$, we have $T_z(\mathcal{B}_\eta) = \text{Hor}_z(\mathcal{B}_\eta, \nabla_\eta) \oplus T_z\left(\Pi_{\mathcal{B}_\eta}^{-1}(z)\right)$, where $\text{Hor}_z(\mathcal{B}_\eta, \nabla_\eta)$ can be identified $T_x(\mathcal{M}_\eta)$, and thus equipped with the lift of the complex structure $J_{\eta,x}$, while on $T_z\left(\Pi_{\mathcal{B}_\eta}^{-1}(z)\right)$, the multiplication by $i$ provide a natural complex structure. With this, a cross-section of $\mathcal{B}_\eta$ is holomorphic if and only if it is holomorphic as a map $\mathcal{M}_\eta \to \mathcal{B}_\eta$.

Similarily, for every $\eta \preccurlyeq \eta' \in \mathcal{L}$, we have a complex structure on $\mathcal{B}_{\eta' \to \eta}$. Eq. (3.8.1), together with defs. 3.6.1, 3.6.3 and the holomorphicity of $\varphi_{\eta' \to \eta}$, ensures that $\zeta_{\eta' \to \eta}$ is holomorphic as a map $\mathcal{B}_{\eta' \to \eta} \times \mathcal{B}_\eta \to \mathcal{B}_{\eta'}$.

For $\eta \in \mathcal{L}$, $\Phi_{\eta \to \eta}^{\text{preQ}}$ is a trivial identification, hence the desired result holds. Thus, we consider $\eta \prec \eta' \in \mathcal{L}$. We first want to prove that $\Phi_{\eta' \to \eta}^{\text{preQ}} \circ \Pi_{\eta'}^{\text{Holo}} \circ \Phi_{\eta' \to \eta}^{\text{preQ},-1} = \Pi_{\eta' \to \eta}^{\text{Holo}} \otimes \Pi_\eta^{\text{Holo}}$, where $\Pi_{\eta'}^{\text{Holo}} : \mathcal{H}_{\eta'}^{\text{preQ}} \to \mathcal{H}_{\eta'}^{\text{Holo}}$ is the orthogonal projection on the closed vector subspace $\mathcal{H}_{\eta'}^{\text{Holo}}$ in $\mathcal{H}_{\eta'}^{\text{preQ}}$, and $\Pi_\eta^{\text{Holo}}$, $\Pi_{\eta' \to \eta}^{\text{Holo}}$ are defined analogously.

Let $t \in \mathcal{H}_{\eta' \to \eta}^{\text{preQ}}$, $s \in \mathcal{H}_\eta^{\text{preQ}}$ and define $\bar{t} = \Pi_{\eta' \to \eta}^{\text{Holo}} t$ and $\bar{s} = \Pi_\eta^{\text{Holo}} s$. By definition of $\Phi_{\eta' \to \eta}^{\text{preQ}}$, we have:

$$\forall y, x \in \mathcal{M}_{\eta' \to \eta} \times \mathcal{M}_\eta, \ \Phi_{\eta' \to \eta}^{\text{preQ},-1}\left(\bar{t} \otimes \bar{s}\right) \circ \varphi_{\eta' \to \eta}^{-1}(y, x) = \zeta_{\eta' \to \eta}(\bar{t}(y), \bar{s}(x)).$$

But, as a composition of holomorphic maps, $(y, x) \mapsto \zeta_{\eta' \to \eta}(\bar{t}(y), \bar{s}(x))$ is holomorphic. Hence, $\varphi_{\eta' \to \eta}$ being holomorphic, $\Phi_{\eta' \to \eta}^{\text{preQ},-1}\left(\bar{t} \otimes \bar{s}\right) \in \mathcal{H}_{\eta'}^{\text{Holo}}$.

Let $s' \in \mathcal{H}_{\eta'}^{\text{Holo}}$. Using the volume-preserving property of $\varphi_{\eta' \to \eta}$, we compute:

$$\left\langle s', \Pi_{\eta'}^{\text{Holo}} \circ \Phi_{\eta' \to \eta}^{\text{preQ},-1}(t \otimes s) \right\rangle_{\mathcal{H}_{\eta'}} = \left\langle s', \widetilde{\zeta}_{\eta' \to \eta}(t, s) \right\rangle_{\mathcal{H}_{\eta'}}$$

$$= \int_{\mathcal{M}_{\eta' \to \eta}} d\mu_{\eta' \to \eta}(y) \int_{\mathcal{M}_\eta} d\mu_\eta(x) \ \left\langle s' \circ \varphi_{\eta' \to \eta}^{-1}(y, x), \zeta_{\eta' \to \eta}(t(y), s(x)) \right\rangle_{\mathcal{B}_{\eta'}}.$$

For $y, x \in \mathcal{M}_{\eta' \to \eta} \times \mathcal{M}_\eta$ we define $u^y(x) \in \mathcal{B}_\eta^x$ such that:

$$\forall u' \in \mathcal{B}_\eta^x, \ \left\langle s' \circ \varphi_{\eta' \to \eta}^{-1}(y, x), \zeta_{\eta' \to \eta}(t(y), u') \right\rangle_{\mathcal{B}_{\eta'}} = \left\langle u^y(x), u' \right\rangle_{\mathcal{B}_\eta},$$

$u^y(x)$ is well-defined, since the left-hand side is a $\mathbb{C}$-linear function of $u'$ (from def. 3.6.3).

Since $s' \in \mathcal{H}_{\eta'}^{\text{Holo}}$, the map $x \mapsto s' \circ \varphi_{\eta' \to \eta}^{-1}(y, x)$ is holomorphic, and for any (local) anti-holomorphic cross-section $u'$ of $\mathcal{B}_\eta$, the map:

$$x \mapsto \left\langle \zeta_{\eta' \to \eta}(t(y), u'), s' \circ \varphi_{\eta' \to \eta}^{-1}(y, x) \right\rangle_{\mathcal{B}_{\eta'}},$$

is holomorphic (for the connection is a $\mathcal{U}(1)$-connection, so the parallel transport preserve the scalar product). Therefore, the cross-section $u^y$ is holomorphic. Moreover, using def. 3.6.2:

$$\forall x \in \mathcal{M}_\eta, \ |u^y(x)| \leqslant \left|s' \circ \varphi_{\eta' \to \eta}^{-1}(y, x)\right| |t(y)|,$$

thus, by Fubini theorem, for almost every $y$ in $\left(\mathcal{M}_{\eta' \to \eta}, d\mu_{\eta' \to \eta}\right)$, $u^y \in \mathcal{H}_\eta^{\text{Holo}}$.

Hence, for almost every $y \in \mathcal{M}_{\eta' \to \eta}$:

$$\int_{\mathcal{M}_\eta} d\mu_\eta(x) \ \left\langle s' \circ \varphi^{-1}_{\eta' \to \eta}(y, x), \zeta_{\eta' \to \eta}(t(y), s(x)) \right\rangle_{\mathcal{B}_{\eta'}} = \int_{\mathcal{M}_\eta} d\mu_\eta(x) \ \left\langle u^y(x), s(x) \right\rangle_{\mathcal{B}_\eta}$$

$$= \langle u^y, s \rangle_{\mathcal{H}_\eta} = \langle u^y, \Pi^{\text{Holo}}_\eta s \rangle_{\mathcal{H}_\eta} = \int_{\mathcal{M}_\eta} d\mu_\eta(x) \ \left\langle s' \circ \varphi^{-1}_{\eta' \to \eta}(y, x), \zeta_{\eta' \to \eta}(t(y), \overline{s}(x)) \right\rangle_{\mathcal{B}_{\eta'}}.$$

And we can prove in a similar way that, for almost every $x \in \mathcal{M}_{\eta' \to \eta}$:

$$\int_{\mathcal{M}_{\eta' \to \eta}} d\mu_{\eta' \to \eta}(y) \ \left\langle s' \circ \varphi^{-1}_{\eta' \to \eta}(y, x), \zeta_{\eta' \to \eta}(t(y), \overline{s}(x)) \right\rangle_{\mathcal{B}_{\eta'}} =$$

$$= \int_{\mathcal{M}_{\eta' \to \eta}} d\mu_{\eta' \to \eta}(y) \ \left\langle s' \circ \varphi^{-1}_{\eta' \to \eta}(y, x), \zeta_{\eta' \to \eta}(\overline{t}(y), \overline{s}(x)) \right\rangle_{\mathcal{B}_{\eta'}}.$$

Therefore, we arrive at:

$$\left\langle s', \Pi^{\text{Holo}}_{\eta'} \circ \Phi^{\text{preQ},-1}_{\eta' \to \eta}(t \otimes s) \right\rangle_{\mathcal{H}_{\eta'}} = \left\langle s', \Phi^{\text{preQ},-1}_{\eta' \to \eta}(\overline{t} \otimes \overline{s}) \right\rangle_{\mathcal{H}_{\eta'}}.$$

Since this holds for all $s' \in \mathcal{H}^{\text{Holo}}_{\eta'}$ and we have already proved that $\Phi^{\text{preQ},-1}_{\eta' \to \eta}(\overline{t} \otimes \overline{s}) \in \mathcal{H}^{\text{Holo}}_{\eta'}$, we have:

$$\Pi^{\text{Holo}}_{\eta'} \circ \Phi^{\text{preQ},-1}_{\eta' \to \eta}(t \otimes s) = \Phi^{\text{preQ},-1}_{\eta' \to \eta}(\overline{t} \otimes \overline{s}),$$

which gives us the announced result:

$$\Phi^{\text{preQ}}_{\eta' \to \eta} \circ \Pi^{\text{Holo}}_{\eta'} \circ \Phi^{\text{preQ},-1}_{\eta' \to \eta} = \Pi^{\text{Holo}}_{\eta' \to \eta} \otimes \Pi^{\text{Holo}}_\eta.$$

Hence, $\Phi^{\text{preQ}}_{\eta' \to \eta} \left\langle \mathcal{H}^{\text{Holo}}_{\eta'} \right\rangle = \Phi^{\text{preQ}}_{\eta' \to \eta} \circ \Pi^{\text{Holo}}_{\eta'} \circ \Phi^{\text{preQ},-1}_{\eta' \to \eta} \left\langle \mathcal{H}^{\text{preQ}}_{\eta' \to \eta} \otimes \mathcal{H}^{\text{preQ}}_\eta \right\rangle = \Pi^{\text{Holo}}_{\eta' \to \eta} \otimes \Pi^{\text{Holo}}_\eta \left\langle \mathcal{H}^{\text{preQ}}_{\eta' \to \eta} \otimes \mathcal{H}^{\text{preQ}}_\eta \right\rangle = \mathcal{H}^{\text{Holo}}_{\eta' \to \eta} \otimes \mathcal{H}^{\text{Holo}}_\eta$. And the relation involving $\Phi^{\text{preQ}}_{\eta'' \to \eta' \to \eta}$ can be proved in a similar way. □

Note that in an holomorphic representation, the evaluation of the holomorphic wavefunction at a given point in phase space is a bounded linear form (via an argument similar to the proof of prop. A.6), hence is dual to a vector in the Hilbert space: this defines the coherent state centered around this classical point. Now, if we choose an element in the projective family of symplectic manifolds we started from, ie. a projective family of points $(x_\eta)_{\eta \in \mathcal{L}}$, we can form a projective family of quantum states, by considering, in each $\mathcal{H}^{\text{Holo}}_\eta$, the coherent state centered around $x_\eta$. This family of states will moreover be of the form considered in theorem 2.9, so we can apply this result to get a corresponding inductive limit Hilbert space and characterize its range in the quantum projective state space (for example, the Fock representation we will consider in [16, prop. 3.17] can be obtained in this manner).

# 4 Outlook

While we have discussed extensively how the classical structures presented in [15, section 2] can be converted into their quantum analogues, we have not yet formalized how to infer from the strategy exposed in [15, section 3] a program for dealing with constraints at the quantum level. Nevertheless, an example will be considered in [16, subsection 3.2], suggesting how such a program would look like.

Our main motivation for studying this projective approach to quantum field theory being its application to quantum gravity, we will in forthcoming work construct projective quantum state spaces closely related to the Hilbert spaces currently used in Loop Quantum Gravity (LQG) and Loop Quantum Cosmology (LQC). An important goal here is to obtain states in the full theory that are almost symmetric both in configuration and momentum variables, and that we could identify with the states of the reduced theory. This problem is closely related to the search for good coherent states and this is where extending the space of states can make a difference. Also, symmetry reducing a theory is mathematically the same as imposing second class constraints and will thus involve the construction of an appropriate regularization strategy for their quantum implementation.

Other interesting directions for further work include the development of quantization prescriptions going beyond the framework laid out in section 3. The ultimate goal would be to have a general procedure, building upon geometric quantization, to quantize any projective system of classical phase spaces, as soon as we give us a consistent family of polarizations thereon.

In particular, it should be possible to relax the requirement of having a global factorization. Recall that in our discussion of the local factorization result [15, prop. 2.10], we had identified two different kinds of obstructions that could prevent it from holding globally. In both cases, we can sketch a route for proceeding to quantization nevertheless. The first kind of obstruction is realized when $\mathcal{M}_{\eta'}$ cannot be written as a Cartesian product, but at least can be seen as an open subset of a bigger manifold $\widetilde{\mathcal{M}}_{\eta'} := \mathcal{M}_{\eta' \to \eta} \times \mathcal{M}_\eta$. This suggests to deal with this situation by a slight generalization of def. 2.1, allowing $\mathcal{H}_{\eta'}$ to be a closed vector subspace in a bigger Hilbert space $\widetilde{\mathcal{H}}_{\eta'} := \mathcal{H}_{\eta' \to \eta} \otimes \mathcal{H}_\eta$. Then, the density matrices over $\mathcal{H}_{\eta'}$ could be seen as density matrices over $\widetilde{\mathcal{H}}_{\eta'}$, with support restricted to $\mathcal{H}_{\eta'}$. Thus, it would still be possible define a map $\text{Tr}_{\eta' \to \eta} : \mathcal{S}_{\eta'} \to \mathcal{S}_\eta$ by first embedding $\mathcal{S}_{\eta'}$ in $\widetilde{\mathcal{S}}_{\eta'}$ and then tracing over $\mathcal{H}_{\eta' \to \eta}$. While such a map could no longer be seen as a partial trace over a tensor factor in $\mathcal{H}_{\eta' \to \eta}$, it should still retain the properties that we really need for the formalism to make sense (in particular, appropriate compatibility with the evaluation of expectation values). The other obstacle for a global factorization is illustrated by taking $\mathcal{M}_{\eta'}$ as a covering space of $\mathcal{M}_\eta$: in this case we still have the option of writing $\mathcal{M}_{\eta'} \simeq \mathcal{M}_{\eta' \to \eta} \times \mathcal{M}_\eta$ with a discrete space $\mathcal{M}_{\eta' \to \eta}$, but we have to accept that the identification will not be everywhere smooth: there will be cuts, and the disposition of these cuts will, when going over to the quantum theory, be imprinted in the precise interpretation of the observables (we will provide an example of this procedure when investigating LQC).

## Acknowledgements

This work has been financially supported by the Université François Rabelais, Tours, France.

This research project has been supported by funds to Emerging Field Project "Quantum Geometry" from the FAU Erlangen-Nuernberg within its Emerging Fields Initiative.

# A Appendix: Geometric quantization

The aim of this appendix is to import a few definitions and properties from geometric quantization, that are needed in particular for section 3. We try here to give a short self-contained introduction, leading rapidly to the definition of the holomorphic representation on a Kähler manifold [27, sections 8.4 & 9.2], and of the position representation arising from a choice of configuration variables on a symplectic manifold [27, sections 4.5 & 9.3]. Accordingly, we skip advanced aspects, including underlying insights and technical subtleties.

In this appendix all manifolds are assumed to be smooth, *finite dimensional* manifolds and all maps between them are assumed to be smooth.

## A.1 Prequantization

**Definition A.1** An hermitian line bundle is a vector bundle $(\mathcal{B}, \Pi_\mathcal{B}, \mathcal{M})$ associated to a $\mathcal{U}(1)$-principal bundle on a smooth manifold $\mathcal{M}$ via the standard action of $\mathcal{U}(1)$ on $\mathbb{C}$. Since the $\mathbb{C}$-linear structure and the Hermitian product $\langle \cdot , \cdot \rangle : z, z' \mapsto z^{*}z'$ on $\mathbb{C}$ is preserved under the action of $\mathcal{U}(1)$, each fiber of $\mathcal{B}$ can be equipped with a natural hermitian structure.

Any connection in the $\mathcal{U}(1)$-principal bundle defines a covariant derivative $\nabla$ on $\mathcal{B}$, and we can associate to its curvature a $\text{Lie}(\mathcal{U}(1)) \approx \mathbb{R}$-valued 2-form $D\nabla$ on $\mathcal{M}$, such that for any cross-section $s$ of $\mathcal{B}$ and any vector fields $X, Y$ on $\mathcal{M}$:

$$[\nabla_X, \nabla_Y](s) = \nabla_{[X,Y]}(s) + i D\nabla(X, Y) s. \tag{A.1.1}$$

**Definition A.2** Let $\mathcal{M}$ be a symplectic manifold (with symplectic structure $\Omega$). A prequantum bundle $(\mathcal{B}, \nabla)$ on $\mathcal{M}$ is an hermitian line bundle $\mathcal{B}$, with base $\mathcal{M}$, equipped with a connection, with corresponding covariant derivative $\nabla$ and such that:

$$D\nabla = -\Omega.$$

On $\mathcal{M}$, we define the symplectic volume form $\omega := \frac{1}{p!} \Omega \wedge \ldots \wedge \Omega = \frac{1}{p!} \Omega^{\wedge p}$ (where $p := \dim \mathcal{M}/2$) and the corresponding measure $\mu_\omega$.

**Definition A.3** Considering the same objects as in def. A.2, we define the prequantum Hilbert space $\mathcal{H}_{\text{preQ}}$ as the space $L_2(\mathcal{M} \to \mathcal{B}, d\mu_\omega)$ of (equivalence classes up to almost-everywhere equality of) cross-sections of $\mathcal{B}$ whose norm, defined using the hermitian structure on $\mathcal{B}$, is square-integrable with respect to $\mu_\omega$.

For $f \in C^\infty(\mathcal{M}, \mathbb{C})$, we define the prequantization $\widehat{f}$ of $f$ as a (densely defined) operator on $\mathcal{H}_{\text{preQ}}$ by:

$$\forall s \in \mathcal{D}_f \subset \mathcal{H}_{\text{preQ}}, \ \widehat{f}s := fs + i \nabla_{X_f} s,$$

where $\nabla_{X_f} := \nabla_{X_{\text{Re}(f)}} + i \nabla_{X_{\text{Im}(f)}}$.

**Proposition A.4** Let $f, g \in C^\infty(\mathcal{M}, \mathbb{C})$. Then:
$$\left[\widehat{f}, \widehat{g}\right] = i\, \widehat{\{f, g\}}.$$

and:
$$\forall s, s' \in \mathcal{D}_f, \left\langle \widehat{f^*}(s), s' \right\rangle = \left\langle s, \widehat{f}(s') \right\rangle.$$

**Proof** Let $s \in \mathcal{D}_{f,g}$ (defining the common domain $\mathcal{D}_{f,g}$ such that both $\widehat{f}\widehat{g}s$ and $\widehat{g}\widehat{f}s$ are well-defined; since $\mathcal{D}_{f,g}$ contains at least the compactly supported smooth cross-sections, it is dense in $\mathcal{H}_{\text{preQ}}$). Using eq. (A.1.1) we have:

$$\left[\widehat{f}, \widehat{g}\right](s) = -\left[\nabla_{X_f}, \nabla_{X_g}\right](s) + i\, d_{X_f}(g)\, s - i\, d_{X_g}(f)\, s$$

$$= -\nabla_{[X_f, X_g]}(s) + i\,\Omega\left(X_f, X_g\right)s + i\,\Omega\left(X_g, X_f\right)s - i\,\Omega\left(X_f, X_g\right)s$$

$$= -\nabla_{X_{\{f,g\}}}(s) + i\,\{f, g\}\, s = i\,\widehat{\{f, g\}}(s).$$

Let $s, s' \in \mathcal{D}_f$. We have for all $X$ vector field on $\mathcal{M}$:
$$\forall x \in \mathcal{M},\, d_X \langle s, s' \rangle(x) = \langle \nabla_X(s)(x), s'(x) \rangle + \langle s(x), \nabla_X(s')(x) \rangle, \tag{A.4.1}$$
for $\nabla$ comes from a $\mathcal{U}(1)$-connection. Hence, we get:

$$\left\langle \widehat{f^*}(s), s' \right\rangle = \int d\mu_\omega(x)\, \langle i\,\nabla_{X_{f^*}}(s)(x), s'(x)\rangle + \int d\mu_\omega(x)\, \langle f^* s(x), s'(x)\rangle$$

$$= -\int d\mu_\omega(x)\, i\, d_{X_f}\langle s, s'\rangle(x) + \int d\mu_\omega(x)\, \langle s(x), i\,\nabla_{X_f}(s')(x)\rangle + \int d\mu_\omega(x)\, \langle s(x), f\, s'(x)\rangle$$

$$= \int i\, \langle s, s'\rangle\, (\mathfrak{L}_{X_f}\omega) + \int d\mu_\omega(x)\, \left\langle s(x), \widehat{f}(s')(x)\right\rangle$$

(using Stokes theorem [17, theorem 10.23]; $\mathcal{M}$ is assumed to be without boundary, or $\mathcal{D}_f$ is required to ensure suitable fall-off conditions)

$$= \int d\mu_\omega(x)\, \left\langle s(x), \widehat{f}(s')(x)\right\rangle = \left\langle s, \widehat{f}(s')\right\rangle$$

(for $X_f$ generates symplectomorphisms, thus preserving the symplectic volume form $\omega$).
$\square$

The prequantization of $\mathcal{M}$ leads to a faithful representation of the full Poisson-algebra $C^\infty(\mathcal{M}, \mathbb{C})$. However, this representation is typically much too big (as is to be expected from the Groenewold-Van-Hove theorem [9] and generalizations thereof [8]), so the next step will be to implement additional prescriptions yielding a physically admissible Hilbert space (at the cost of restricting which observables can be quantized).

## A.2 Holomorphic representation

To discuss holomorphic quantization we need to equip $\mathcal{M}$ with an almost complex structure $J$ (def. A.5.1), which is required to be integrable (def. A.5.2, ensuring the existence of local holomorphic coordinates, thus making $J$ into a complex structure for $\mathcal{M}$) and compatible with the symplectic structure $\Omega$ (def. A.5.3). An additional positivity requirement (def. A.5.4) allows to define from $\Omega$ and $J$ a Riemannian metric on $\mathcal{M}$ (the so-called Kähler metric) and makes $\mathcal{M}$ into a Kähler manifold [14, section IX.4].

**Definition A.5** A Kähler manifold $(\mathcal{M}, \Omega, J)$ is a symplectic manifold $(\mathcal{M}, \Omega)$ equipped with a smooth field $J$ satisfying:

1. $\forall x \in \mathcal{M}$, $J_x$ is an endomorphism of $T_x(\mathcal{M})$ such that $J_x^2 = -\mathrm{id}_{T_x(\mathcal{M})}$;
2. $\forall X, Y \in \mathcal{T}^\infty(\mathcal{M})$, $[JX, JY] - J[X, JY] - J[JX, Y] - [X, Y] = 0$ (where $\mathcal{T}^\infty(\mathcal{M})$ is the space of smooth vector fields on $\mathcal{M}$);
3. $\forall x \in \mathcal{M}$, $\forall v, w \in T_x(\mathcal{M})$, $\Omega_x(J_x v, J_x w) = \Omega(v, w)$;
4. $\forall x \in \mathcal{M}$, $\forall v \neq 0 \in T_x(\mathcal{M})$, $\Omega_x(v, J_x v) > 0$.

**Proposition A.6** Let $\mathcal{M}$ be a Kähler manifold and $(\mathcal{B}, \nabla)$ be a prequantum bundle on $\mathcal{M}$. We define the holomorphic quantization $\mathcal{H}_{\text{Holo}}$ of $\mathcal{M}$ to be $\mathcal{H}_{\text{Holo}} := \mathcal{H}_{\text{preQ}} \cap \text{Holo}(\mathcal{M} \to \mathcal{B})$, where $\text{Holo}(\mathcal{M} \to \mathcal{B})$ is the space of holomorphic cross-sections of $\mathcal{B}$:

$$\text{Holo}(\mathcal{M} \to \mathcal{B}) := \{s \in C^\infty(\mathcal{M} \to \mathcal{B}) \mid \forall x \in \mathcal{M}, \forall v \in T_x(\mathcal{M}), \nabla_{Jv} s = i \nabla_v s\}.$$

$\mathcal{H}_{\text{Holo}}$ is a closed vector subspace of $\mathcal{H}_{\text{preQ}}$, hence is itself an Hilbert space.

**Proof** Let $(s_\alpha)_{\alpha \in A}$ be a net in $\mathcal{H}_{\text{Holo}}$ that converges (for the norm $\|\cdot\|_{\text{preQ}}$) to $s \in \mathcal{H}_{\text{preQ}}$.

Let $x \in \mathcal{M}$. There exist a neighborhood $U$ of $x$ in $\mathcal{M}$, holomorphic coordinates $z^1, \ldots, z^p$ ($2p := \dim \mathcal{M}$) on $U$ and a real valued function $K(z, z^*)$ on $U$ such that [27, section 5.4]:

$$\forall x' \in U, \Omega_{x'} = i \frac{\partial^2 K}{\partial z^j \, \partial z^{l,*}}(x') \, dz^j \wedge dz^{l,*},$$

and from A.5.4 $\dfrac{\partial^2 K}{\partial z^j \, \partial z^{l,*}}$ has to be a positive definite hermitian matrix at every point in $U$. The symplectic measure is then given over $U$ by:

$$\mu_\omega := \beta \, \mu_{\mathbb{C}}^{(p)} = 2^p \det\left(\frac{\partial^2 K}{\partial z \, \partial z^*}\right) \mu_{\mathbb{C}}^{(p)} \text{ (where } \mu_{\mathbb{C}}^{(p)} \text{ is the standard measure on } \mathbb{C}^p\text{)}.$$

We choose $t \in \Pi_{\mathcal{B}}^{-1} \langle x \rangle$ (the fiber of $\mathcal{B}$ above $x$), with $|t| = e^{-K(x)/2}$, and we define the cross-section $r$ of $\mathcal{B}|_U$ by:

$$r(x) = t \quad \& \quad \forall x' \in U, \forall v \in T_{x'}(\mathcal{M}), \nabla_v r(x') = -\left(\frac{\partial K}{\partial z^j}(x') \left[dz^j\right]_{x'}(v)\right) r(x').$$

We can check using eq. (A.1.1) that this characterizes a well-defined cross-section of $\mathcal{B}|_U$ and we have moreover:

$$\forall x' \in U, \forall v \in T_{x'}(\mathcal{M}), \nabla_{Jv} r(x') = i \nabla_v r(x'),$$

$$\text{and} \quad [d \langle r, r \rangle]_{x'}(v) = -dK_{x'}(v) \langle r, r \rangle (x'),$$

so $r$ is an holomorphic cross-section of $\mathcal{B}|_U$ and $\forall x' \in U, |r(x')| = e^{-K(x')/2}$.

Next, for all $\alpha \in A$ we can define $f_\alpha$ as the holomorphic function $f_\alpha : U \to \mathbb{C}$ such that $\forall x' \in U, s_\alpha(x') = f_\alpha(x') r(x')$. Similarly, we define $f : U \to \mathbb{C}$ such that $\forall x' \in U, s(x') = f(x') r(x')$.

Let $\epsilon > 0$ and let $U_1$ be a closed ball (with respect to the coordinates $z^1, \ldots, z^p$) of center $x$ and radius $r > 0$ such that $U_1 \subset U$ and $\forall x' \in U_1, \beta(x') e^{-K(x')} > \epsilon$. Let $U_2$ be the closed ball of center $x$ and radius $r/2$. For all $x' \in U_2$ we call $U_2^{x'}$ the closed ball of center $x'$ and radius $r/2$. Hence, $U_2^{x'} \subset U_1$. For $g$ an holomorphic function on $U_1 \to \mathbb{C}$, we have:

$$\forall x' \in U_2, g(x') = \frac{8^p}{r^{2p}} \left[ \prod_{k=1}^p \int_0^{r/2} q^k \, dq^k \right] g(x')$$

$$= \frac{4^p}{\pi^p \, r^{2p}} \left[ \prod_{k=1}^p \int_0^{r/2} dq^k \int_0^{2\pi} q^k \, d\theta^k \right] g\left(x' + \left(q^1 e^{i\theta^1}, \ldots, q^p e^{i\theta^p}\right)\right),$$

hence:

$$\forall x' \in U_2, |g(x')|^2 \leqslant \left(\frac{4}{\pi r^2}\right)^{2p} \left[ \int_{U_2^{x'}} d\mu_{\mathbb{C}}^{(p)}(z) \; |g(z)| \right]^2$$

$$\leqslant \left(\frac{4}{\pi r^2}\right)^{2p} \frac{\pi^p r^{2p}}{4^p} \int_{U_2^{x'}} d\mu_{\mathbb{C}}^{(p)}(z) \; |g(z)|^2 \text{ (by convexity of } x \mapsto x^2)$$

$$\leqslant \frac{4}{\pi r^2} \int_{U_1} d\mu_{\mathbb{C}}^{(p)}(z) \; |g(z)|^2.$$

Therefore, for all $\alpha, \alpha' \in A$:

$$\forall x' \in U_2, |f_\alpha(x') - f_{\alpha'}(x')|^2 \leqslant \frac{4}{\pi r^2} \int_{U_1} d\mu_{\mathbb{C}}^{(p)}(z) \; |f_\alpha(z) - f_{\alpha'}(z)|^2$$

$$\leqslant \frac{1}{\epsilon} \frac{4}{\pi r^2} \int_{U_1} d\mu_\omega(z) \; |f_\alpha(z) - f_{\alpha'}(z)|^2 \; |r|^2$$

$$\leqslant \frac{4}{\epsilon \pi r^2} \|s_\alpha - s_{\alpha'}\|_{\text{preQ}}^2,$$

hence the net $\left(f_\alpha|_{U_2}\right)_{\alpha \in A}$ converges uniformly to a function $f' : U_2 \to \mathbb{C}$. Cauchy's integral formula implies that $f'$ is holomorphic on the interior of $U_2$. On the other hand, the net $\left(f_\alpha|_{U_2}\right)_{\alpha \in A}$ converges in $L_2$-norm to $f|_{U_2}$ (for $\forall \alpha \in A, \left\|f_\alpha|_{U_2} - f|_{U_2}\right\|_2 \leqslant \frac{1}{\sqrt{\epsilon}} \|s_\alpha - s\|_{\text{preQ}}$), hence $f' = f|_{U_2}$ ($\mu_\omega$-almost-everywhere). Therefore, $s \in \mathcal{H}_{\text{Holo}}$. □

Since we restrict the quantum Hilbert space to $\mathcal{H}_{\text{Holo}}$, we should also restrict the admissible observables to be the ones that stabilize $\mathcal{H}_{\text{Holo}}$ (note that we do not discuss here whether the intersection of $\mathcal{H}_{\text{Holo}}$ with the dense domain of such an observable will also be dense in $\mathcal{H}_{\text{Holo}}$; this is a non-trivial question, for the usual tools based on bump functions are not available in the holomorphic class).

**Proposition A.7** We consider the same objects as in prop. A.6. We define:

$$\mathcal{O}_{\text{Holo},\mathbb{C}} := \{f \in C^\infty(\mathcal{M}, \mathbb{C}) \mid \forall Y \in \mathcal{T}^\infty(\mathcal{M}), \exists Z \in \mathcal{T}^\infty(\mathcal{M}) \ / \ [Y + iJY, X_f] = Z + iJZ\},$$

where $\mathcal{T}^\infty(\mathcal{M})$ is the space of smooth vector fields on $\mathcal{M}$.

Then, for all $f \in \mathcal{O}_{\text{Holo},\mathbb{C}}$, $\widehat{f}$ stabilizes $\mathcal{H}_{\text{Holo}}$.

**Proof** Let $f \in \mathcal{O}_{\text{Holo},\mathbb{C}}$ and $s \in \mathcal{D}_f \cap \mathcal{H}_{\text{Holo}}$. Let $x \in \mathcal{M}$ and $v \in T_x(\mathcal{M})$. Then, there exists $Y \in \mathcal{T}^\infty(\mathcal{M})$ such that $Y_x = v$ (we can construct such a vector field using local smooth coordinates around $x$ and an appropriate bump function). Since $f \in \mathcal{O}_{\text{Holo},\mathbb{C}}$, there also exists $Z \in \mathcal{T}^\infty(\mathcal{M})$ such that $[Y + iJY, X_f] = Z + iJZ$. Hence:

$$\nabla_{Jv,x} \widehat{f} s = \left(\nabla_{JY} \widehat{f} s\right)(x)$$

$$= \left(f\left(\nabla_{JY} s\right)\right)(x) + \left(\Omega(X_f, JY) s\right)(x) + i\left(\nabla_{JY} \nabla_{X_f} s\right)(x)$$

$$= \left(f\left(\nabla_{JY} s\right)\right)(x) - i\left(\nabla_{[X_f, JY]} s\right)(x) + i\left(\nabla_{X_f} \nabla_{JY} s\right)(x) \quad \text{(using eq. (A.1.1))}$$

$$= \left(f\left(\nabla_{JY} s\right)\right)(x) + \left(\nabla_{Z+iJZ} s\right)(x) + \left(\nabla_{[X_f, Y]} s\right)(x) + i\left(\nabla_{X_f} \nabla_{JY} s\right)(x)$$

$$= i\left(f\left(\nabla_Y s\right)\right)(x) + \left(\nabla_{[X_f, Y]} s\right)(x) - \left(\nabla_{X_f} \nabla_Y s\right)(x) \quad \text{(using } s \in \mathcal{H}_{\text{Holo}})$$

$$= i \nabla_Y \left(\widehat{f} s\right)(x) = i \nabla_{v,x} \widehat{f} s,$$

therefore $\widehat{f} s \in \mathcal{H}_{\text{Holo}}$. $\square$

**Proposition A.8** We consider the same objects as in prop. A.7. Let $r$ be a nowhere-vanishing cross-section of $\mathcal{B}$ such that $r \in \mathcal{H}_{\text{Holo}}$ and let $\mu_r$ be the measure on $\mathcal{M}$ defined by $\mu_r = \langle r, r \rangle \mu_\omega$. Then, the map:

$$\Phi_r \ : \ L_2(\mathcal{M}, d\mu_r) \cap \text{Holo}(\mathcal{M}) \ \to \ \mathcal{H}_{\text{Holo}}$$
$$\psi \mapsto \psi r \qquad ,$$

is an Hilbert space isomorphism.

If $f \in \mathcal{O}_{\text{Holo},\mathbb{C}}$ and $\psi \in \Phi_r^{-1}\langle \mathcal{D}_f \rangle$, we have:

$$\widehat{f}^r \psi := \left(\Phi_r^{-1} \widehat{f} \Phi_r\right) \psi = f \psi + i (d_{X_f} \psi) + \frac{i}{2} X_f^r \psi, \tag{A.8.1}$$

where $X_f^r$ is defined by $2 \nabla_{X_f} r = X_f^r r$.

**Proof** Let $s \in \mathcal{H}_{\text{Holo}}$. Since $r$ is a nowhere-vanishing holomorphic cross-section there exists a unique smooth function $\psi : \mathcal{M} \to \mathbb{C}$ such that $s = \psi r$. Moreover, for all $x \in \mathcal{M}$ and all $v \in T_x(\mathcal{M})$:

$$i\,(d_v\psi)\,r = i\,\nabla_v s - i\,\psi\,\nabla_v r = \nabla_{Jv} s - \psi\,\nabla_{Jv} r = (d_{Jv}\psi)\,r\,,$$

hence $\psi \in \mathrm{Holo}(\mathcal{M})$. Moreover:

$$\|s\|^2_{\mathrm{Holo}} = \int_{\mathcal{M}} d\mu_\omega(x)\,\langle \psi(x)\,r(x),\,\psi(x)\,r(x)\rangle = \int_{\mathcal{M}} d\mu_r(x)\,\psi^*(x)\,\psi(x) = \|\psi\|^2_{2,\mu_r}.$$

Therefore, $\psi \in L_2(\mathcal{M},\,d\mu_r)$. But since we have $\Phi_r(\psi) = s$ and $\|s\|_{\mathrm{Holo}} = \|\psi\|_{2,\mu_r}$, $\Phi_r$ is an Hilbert space isomorphism.

Eq. (A.8.1) can be checked from the definition of $\widehat{(\cdot)}$ (def. A.3). $\square$

## A.3 Position representation

We now turn to the position representation. We describe a choice of configuration variables as a map $\gamma$ from the phase space into the configuration space. The typical example occurs when $\mathcal{M}$ is given as a cotangent bundle (with its canonical symplectic structure) in which case $\gamma$ is simply the projection on the base manifold.

**Definition A.9** Let $\mathcal{M}$ be a symplectic manifold. A configuration space for $\mathcal{M}$ is a manifold $\mathcal{C}$ and a surjective map $\gamma : \mathcal{M} \to \mathcal{C}$ such that:
1. $\forall x \in \mathcal{M},\ \mathrm{Im}(T_x\gamma) = T_{\gamma(x)}(\mathcal{C})$;
2. $\forall x \in \mathcal{M},\ \mathrm{Ker}(T_x\gamma) = \big(\mathrm{Ker}(T_x\gamma)\big)^\perp := \{v \in T_x(\mathcal{M}) \mid \forall w \in \mathrm{Ker}(T_x\gamma),\ \Omega_{\mathcal{M},x}(v,w) = 0\}$.
3. $\forall y \in \mathcal{C},\ \gamma^{-1}\langle\{y\}\rangle$ is connected.

**Definition A.10** Let $\mathcal{M}$ be a symplectic manifold, $(\mathcal{C}, \gamma)$ be a configuration space for $\mathcal{M}$ and $(\mathcal{B}, \nabla)$ be a prequantum bundle on $\mathcal{M}$. A configuration quantum bundle on $\mathcal{C}$ is an hermitian line bundle $\mathcal{B}_\mathcal{C}$, with base $\mathcal{C}$, and a smooth map $\Gamma : \mathcal{B} \to \mathcal{B}_\mathcal{C}$ such that:
1. $\forall z \in \mathcal{B},\ \Pi_{\mathcal{B}_\mathcal{C}} \circ \Gamma(z) = \gamma \circ \Pi_{\mathcal{B}}(z)$ (where $\Pi_{\mathcal{B}}$ and $\Pi_{\mathcal{B}_\mathcal{C}}$ are the bundle projections);
2. $\forall z \in \mathcal{B},\ \forall \lambda \in \mathbb{C},\ \Gamma(\lambda z) = \lambda\,\Gamma(z)$;
3. $\forall z \in \mathcal{B},\ |\Gamma(z)| = |z|$;
4. $\forall z \in \mathcal{B},\ \mathrm{Ker}\,T_z\Gamma \subset \mathrm{Hor}_z(\mathcal{B}, \nabla)$ (where $\mathrm{Hor}_z(\mathcal{B}, \nabla)$ is defined as the $\nabla$-horizontal subspace of $T_z(\mathcal{B})$).

**Proposition A.11** Let $\mathcal{M}$ be a symplectic manifold, $(\mathcal{C}, \gamma)$ be a configuration space for $\mathcal{M}$ and $(\mathcal{B}, \nabla)$ be a prequantum bundle on $\mathcal{M}$. If, for all $y \in \mathcal{C}$, $\gamma^{-1}\langle\{y\}\rangle$ is simply-connected, then there exists a configuration quantum bundle $(\mathcal{B}_\mathcal{C}, \Gamma)$ on $\mathcal{C}$.

**Proof** *Definition of $\mathcal{B}_\mathcal{C}$.* Let $y \in \mathcal{C}$ and let $x \in \gamma^{-1}\langle y\rangle$. Since the derivative of $\gamma$ is surjective at any point in $\mathcal{M}$ (def. A.9.1), we have, by the rank theorem [17, theorem 5.13], $T_x\left(\gamma^{-1}\langle y\rangle\right) = \mathrm{Ker}\,T_x\gamma$. So, using def. A.9.2, $T_x\left(\gamma^{-1}\langle y\rangle\right) = T_x\left(\gamma^{-1}\langle y\rangle\right)^\perp$, hence $\Omega_x|_{T_x(\gamma^{-1}\langle y\rangle)} = 0$. Therefore, if we call $\left(\mathcal{B}_y, \nabla_y\right)$ the restriction of $(\mathcal{B}, \nabla)$ over $\gamma^{-1}\langle y\rangle$, the connection $\nabla_y$ is flat.

Therefore, $z \mapsto \mathrm{Hor}_z(\mathcal{B}, \nabla) \cap \mathrm{Ker}\,[T_z(\gamma \circ \Pi_\mathcal{B})]$ is a smooth involutive tangent distribution on $\mathcal{B}$, so by the global Frobenius theorem [17, theorem 14.13], it defines a foliation of $\mathcal{B}$. Moreover, if $\Lambda$ is a leaf of this foliation, there exists $y \in \mathcal{C}$ such that $\Pi_\mathcal{B}\langle\Lambda\rangle = \gamma^{-1}\langle y\rangle$ and $\Pi_\mathcal{B}|_{\Lambda \to \gamma^{-1}\langle y\rangle}$ is a diffeomorphism. Indeed, the leaf $\Lambda$ being connected by definition, $\gamma \circ \Pi_\mathcal{B}$ is constant on $\Lambda$, so there exists $y \in \mathcal{C}$ such that $\Pi_\mathcal{B}\langle\Lambda\rangle \subset \gamma^{-1}\langle y\rangle$, ie. $\Lambda \subset \mathcal{B}_y$, and, $\gamma^{-1}\langle y\rangle$ being simply-connected, $\Lambda$ is just a global horizontal cross-section of $(\mathcal{B}_y, \nabla_y)$ [14, corollary II.9.2].

We define $\mathcal{B}_\mathcal{C}$ as the set of all leaves and $\Gamma : \mathcal{B} \to \mathcal{B}_\mathcal{C}$ as the quotient map. Since $\gamma \circ \Pi_\mathcal{B}$ is constant on a leaf, we can define a map $\Pi_{\mathcal{B}_\mathcal{C}}$ on $\mathcal{B}_\mathcal{C}$ such that $\gamma \circ \Pi_\mathcal{B} = \Pi_{\mathcal{B}_\mathcal{C}} \circ \Gamma$. Moreover, for any leaf $\Lambda$ and any $\lambda \in \mathbb{C}$, $\lambda.\Lambda$ is also a leaf, therefore, we can define an action of $\mathbb{C}$ on $\mathcal{B}_\mathcal{C}$ such that $\forall z \in \mathcal{B}, \forall \lambda \in \mathbb{C}, \Gamma(\lambda z) = \lambda \Gamma(z)$. And since $\nabla$ is a $\mathcal{U}(1)$ connection, the norm $|\cdot|$ on $\mathcal{B}$ is constant on each leaf, so we also can define a norm on $\mathcal{B}_\mathcal{C}$ such that $\forall z \in \mathcal{B}, |\Gamma(z)| = |z|$.

*Local description of the quotient.* Let $x \in \mathcal{M}$ and let $y = \gamma(x)$. Let $U_1$ be an open neighborhood of $x$ in $\mathcal{M}$ and $\varphi_1 : U_1 \times \mathbb{C} \to \mathcal{B}$ a local trivialization of the bundle $\mathcal{B}$. Since $T\gamma$ is surjective at any point in $\mathcal{M}$, there exist, by the rank theorem, open neighborhoods $V_2$ of $y$ in $\mathcal{C}$, $W_2$ of $0$ in $\mathbb{R}^p$ ($p := \dim \mathcal{M}/2 = \dim \mathcal{C}$), and $U_2$ of $x$ in $U_1$, and a diffeomorphism $\varphi_2 : V_2 \times W_2 \to U_2$ such that $\gamma \circ \varphi_2$ is the first projection map $V_2 \times W_2 \to V_2$.

Next, by definition of a foliation, there exist a neighborhood $T_2$ of $\varphi_1(x, 0)$ in $\varphi_1\langle U_2 \times \mathbb{C}\rangle$, neighborhoods $Q_2$ and $R_2$ of $0$ in $\mathbb{R}^{p+2}$ and $\mathbb{R}^p$ respectively and a diffeomorphism $\psi_1 : Q_2 \times R_2 \to T_2$ such that for any $u \in \mathcal{B}_\mathcal{C}$ there exists a (possibly empty) countable subset $Q_u \subset Q_2$ with:

$$\Gamma^{-1}\langle u\rangle \cap T_2 = \psi_1\langle Q_u \times R_2\rangle\,. \tag{A.11.1}$$

Let $V_3$, $W_3$ and $S_3$ be neighborhoods of $y$ in $V_2$, $0$ in $W_2$ and $0$ in $\mathbb{C}$ respectively, such that $\varphi_1\langle\varphi_2\langle V_3 \times W_3\rangle \times S_3\rangle \subset T_2$, and define:

$$\psi_2 : V_3 \times S_3 \to Q_2$$
$$v, \lambda \mapsto \pi_{Q_2} \circ \psi_1^{-1} \circ \varphi_1(\varphi_2(v, 0), \lambda)\,,$$

where $\pi_{Q_2} : Q_2 \times R_2 \to Q_2$ is the first projection map. Since we have:

$$[T_{(0,0)}\psi_1]\langle\{0\} \times T_0(R_2)\rangle = \mathrm{Hor}_{\varphi_1(x,0)}(\mathcal{B}, \nabla) \cap \mathrm{Ker}\,[T_{\varphi_1(x,0)}(\gamma \circ \Pi_\mathcal{B})]$$

and $[T_{(x,0)}\varphi_1]\langle[T_{y,0}\,d\varphi_2]\langle T_y(V_3) \times \{0\}\rangle \times \mathbb{C}\rangle \cap \mathrm{Hor}_{\varphi_1(x,0)}(\mathcal{B}, \nabla) \cap \mathrm{Ker}\,[T_{\varphi_1(x,0)}(\gamma \circ \Pi_\mathcal{B})] = \{0\}$,

$[T_{y,0}\psi_2]$ is surjective, thus invertible, so by the inverse function theorem [17, theorem 5.11], we can narrow $W_3$ and $S_3$ so that there exists a neighborhood $Q_3$ of $0$ in $Q_2$ with $\psi_2$ inducing a diffeomorphism $V_3 \times S_3 \to Q_3$.

Now, we define $T_3 := \psi_1\langle Q_3 \times R_2\rangle \cap \varphi_1\langle\varphi_2\langle V_3 \times W_3\rangle \times \mathbb{C}\rangle$ and:

$$\varphi_3 : T_3 \to V_3 \times S_3 \times W_3$$
$$z \mapsto \psi_2^{-1} \circ \pi_{Q_3} \circ \psi_1^{-1}(z), \pi_{W_3} \circ \varphi_2^{-1} \circ \Pi_\mathcal{B}(z)\,.$$

Precomposing $\varphi_3$ by $\varphi_1 \circ (\varphi_2 \times \mathrm{id}_\mathbb{C})$, we can check that $[T_{\varphi_1(x,0)}\varphi_3]$ is injective, thus invertible, so using again the inverse function theorem, there exist neighborhoods $T$, $V$, $W$ and $S$ of $\varphi_1(x, 0)$ in $\mathcal{B}$, $y$ in $\mathcal{C}$, $0$ in $\mathbb{R}^p$ and $0$ in $\mathbb{C}$ respectively, and a diffeomorphism $\varphi : V \times W \times S \to T$ satisfying, for all $v, \lambda \in V \times S$:

$$\exists u \in \mathcal{B}_\mathcal{C} \,/\, \Gamma^{-1}\langle u \rangle \cap \operatorname{Im} \varphi = \varphi \langle \{v\} \times W \times \{\lambda\} \rangle \quad \& \quad \Pi_{\mathcal{B}_\mathcal{C}}(u) = v \tag{A.11.2}$$

(using eq. (A.11.1) and the injectivity of the restriction of $\Pi_\mathcal{B}$ to the leaf $\Gamma^{-1}\langle u \rangle$, together with $\varphi \langle \{v\} \times W \times \{\lambda\} \rangle = \psi_1 \langle \{\psi_2(v,\lambda)\} \times R_2 \rangle \cap \operatorname{Im} \varphi$) and for all $v, w, \lambda \in V \times W \times S$:

$$\forall \mu \in \mathbb{C} \,/\, \mu\lambda \in S, \quad \varphi(v, w, \mu\lambda) = \mu\, \varphi(v, w, \lambda) \tag{A.11.3}$$

(we can first check this for $w = 0$, and then use the previous point, since for all $u \in \mathcal{B}_\mathcal{C}$ and all $\mu \in \mathbb{C}$, $\Gamma^{-1}\langle \mu.u \rangle = \mu \,.\, \Gamma^{-1}\langle u \rangle$).

Finally, using eq. (A.11.3), we can extend $S$ to be all $\mathbb{C}$ while still satisfying eqs. (A.11.2) and (A.11.3).

*Compatibility of the local descriptions.* Let $x_0, x_1 \in \mathcal{M}$ such that $\gamma(x_0) = \gamma(x_1) =: y$. There exists a path $\kappa : [0, 1] \to \gamma^{-1}\langle y \rangle$ such that $\gamma(0) = x_0$ and $\gamma(1) = x_1$ (simple connectedness implies path connectedness).

Using the preceding point, for all $t \in [0, 1]$ there exist open neighborhoods $V_t$ of $y$ in $\mathcal{C}$, $W_t$ of $0$ in $\mathbb{R}^p$ and $U_t$ of $\kappa(t)$ in $\mathcal{M}$, and a diffeomorphism $\varphi_t : V_t \times W_t \times \mathbb{C} \to \Pi_\mathcal{B}^{-1}\langle U_t \rangle$ satisfying eq. (A.11.2) and eq. (A.11.3). For any $t \in [0, 1]$, we call $\pi_t$ the projection map $\pi_t : V_t \times W_t \times \mathbb{C} \to V_t \times \mathbb{C}$ and we define the smooth map $\Gamma_t := \pi_t \circ \varphi_t^{-1} : \Pi_\mathcal{B}^{-1}\langle U_t \rangle \to V_t \times \mathbb{C}$.

Next, there exist $t_1, \ldots, t_{N-1}$ ($1 \leqslant N < \infty$) such that $(U_{t_i})_{0 \leqslant i \leqslant N}$ is an open cover of $\kappa\langle [0, 1] \rangle$ (where we set $t_0 = 0$ and $t_N = 1$), and $\forall i \leqslant N - 1$, $\kappa^{-1}\langle U_{t_i} \cap U_{t_{i+1}} \rangle \neq \varnothing$.

We define $V = \bigcap_{0 \leqslant i \leqslant N-1} \gamma\langle U_{t_i} \cap U_{t_{i+1}} \rangle$. $\gamma$ is an open map (for $T\gamma$ is surjective at any point), therefore $V$ is an open subset of $\mathcal{C}$, and for any $i \leqslant N - 1$, there exists $t \in [0, 1]$ such that $\kappa(t) \in U_{t_i} \cap U_{t_{i+1}}$, hence $y = \gamma \circ \kappa(t) \in U_{t_i} \cap U_{t_{i+1}}$. Thus, $V$ is an open neighborhood of $y$ in $\mathcal{C}$.

Let $i \leqslant N-1$. The maps $\Gamma_{t_i}|_{\Pi_\mathcal{B}^{-1}\langle U_{t_i} \cap U_{t_{i+1}} \cap \gamma^{-1}\langle V \rangle \rangle \to V \times \mathbb{C}}$ and $\Gamma_{t_{i+1}}|_{\Pi_\mathcal{B}^{-1}\langle U_{t_i} \cap U_{t_{i+1}} \cap \gamma^{-1}\langle V \rangle \rangle \to V \times \mathbb{C}}$ are smooth, surjective, their derivatives are surjective at each point and they are constant on each other level sets (using eq. (A.11.2)), therefore the rank theorem implies [17, prop. 5.21] the existence of a diffeomorphism $\Phi_i : V \times \mathbb{C} \to V \times \mathbb{C}$ such that:

$$\forall x \in \Pi_\mathcal{B}^{-1}\langle U_{t_i} \cap U_{t_{i+1}} \cap \gamma^{-1}\langle V \rangle \rangle, \ \Gamma_{t_i}(x) = \Phi_i \circ \Gamma_{t_{i+1}}(x).$$

Thus, eq. (A.11.2) leads to:

$$\forall v \in V, \forall \lambda \in \mathbb{C}, \exists u \in \mathcal{B}_\mathcal{C} \,/$$

$$\Gamma^{-1}\langle u \rangle \cap \operatorname{Im} \varphi_{t_i} = \varphi_{t_i} \langle \{v\} \times W_{t_i} \times \{\lambda\} \rangle$$

$$\& \quad \Gamma^{-1}\langle u \rangle \cap \operatorname{Im} \varphi_{t_{i+1}} = \varphi_{t_{i+1}} \circ \widetilde{\Phi}_i^{-1} \langle \{v\} \times W_{t_{i+1}} \times \{\lambda\} \rangle,$$

where $\widetilde{\Phi}_i$ is defined naturally from $\Phi_i$ as a map $\widetilde{\Phi}_i : V \times W_{t_{i+1}} \times \mathbb{C} \to V \times W_{t_{i+1}} \times \mathbb{C}$.

Defining $\Phi := \Phi_0 \circ \ldots \circ \Phi_{N-1} : V \times \mathbb{C} \to V \times \mathbb{C}$ and $\widetilde{\Phi} : V \times W_{t_N} \times \mathbb{C} \to V \times W_{t_N} \times \mathbb{C}$, we have:

$$\forall v \in V, \forall \lambda \in \mathbb{C}, \forall w_0 \in W_0, \forall w_1 \in W_1, \Gamma \circ \varphi_0(v, w_0, \lambda) = \Gamma \circ \varphi_1 \circ \widetilde{\Phi}^{-1}(v, w_1, \lambda).$$

This way we have proved that for any $x_0, x_1 \in \mathcal{M}$ such that $\gamma(x_0) = \gamma(x_1) =: y$, there exist open neighborhoods $V$ of $y$ in $\mathcal{C}$, $W_0$ and $W_1$ of $0$ in $\mathbb{R}^p$, $\widetilde{U}_0$ of $x_0$ in $\mathcal{M}$ and $\widetilde{U}_1$ of $x_1$ in $\mathcal{M}$, diffeomorphisms $\widetilde{\varphi}_0 : V \times W_0 \times \mathbb{C} \to \Pi_{\mathcal{B}}^{-1}\langle \widetilde{U}_0 \rangle$ and $\widetilde{\varphi}_1 : V \times W_1 \times \mathbb{C} \to \Pi_{\mathcal{B}}^{-1}\langle \widetilde{U}_1 \rangle$, and an injective map $\psi : V \times \mathbb{C} \to \mathcal{B}_e$, such that:

$$\forall v \in V, \forall \lambda \in \mathbb{C}, \forall w \in W_{0/1}, \psi(v, \lambda) = \Gamma \circ \widetilde{\varphi}_{0/1}(v, w, \lambda)$$

$$\forall v \in V, \forall \lambda \in \mathbb{C}, \Pi_{\mathcal{B}_e} \circ \psi(v, \lambda) = v$$

and $\quad \forall v, w \in V \times W_{0/1}, \forall \lambda \in \mathbb{C}, \widetilde{\varphi}_{0/1}(v, w, \lambda \cdot) = \lambda\, \widetilde{\varphi}_{0/1}(v, w, \cdot)$.

*Topological, differentiable and bundle structures on $\mathcal{B}_e$.* We equip $\mathcal{B}_e$ with the final topology induced by $\Gamma$ (so that $U \subset \mathcal{B}_e$ is open iff $\Gamma^{-1}\langle U \rangle$ is open in $\mathcal{B}$). The previous point, together with $\gamma \circ \Pi_{\mathcal{B}} = \Pi_{\mathcal{B}_e} \circ \Gamma$, ensures that $\Gamma$ is an open map for this topology (because the preimage of the image of an open subset of $\mathcal{B}$ is an open subset of $\mathcal{B}$), and that we can use the local descriptions of the quotient to define a bundle structure on $\mathcal{B}_e$, with respect to which $\Gamma$ will be a smooth surjective map with surjective derivative at each point. We can check that this structure is then compatible with the projection $\Pi_{\mathcal{B}_e}$ and the action of $\mathbb{C}$ on $\mathcal{B}_e$ defined above. □

Since the cross-sections of $\mathcal{B}$ that are $\nabla$-horizontal over the level sets of $\gamma$ are typically non-normalizable, we need to introduce a measure on $\mathcal{C}$. In general, there is however no preferred choice for this measure, hence we will associate to any smooth measure [7, section 11.4] on $\mathcal{C}$ a corresponding Hilbert space and we will restore the independence with respect to the choice of measure by providing identifications between these different Hilbert spaces.

**Definition A.12** Let $\mathcal{C}$ be a smooth manifold. A smooth measure $\mu$ on $\mathcal{C}$ is a Borel measure on $\mathcal{C}$ such that, for any smooth coordinate chart $\varphi : U \to \mathbb{R}^p$ ($p := \dim \mathcal{C}$) on an open subset $U$ of $\mathcal{C}$, there exists a smooth, nowhere vanishing, strictly positive function $\alpha_\varphi : U \to \mathbb{R}$ satisfying:

$$\mu|_U = \alpha_\varphi \left[ \varphi^* \mu_{\mathbb{R}}^{(p)} \right],$$

where $\mu_{\mathbb{R}}^{(p)}$ is the Lebesgue measure on $\mathbb{R}^p$.

In particular, the measure $\mu_\omega$ associated to a nowhere vanishing volume form $\omega$ on $\mathcal{C}$ is a smooth measure on $\mathcal{C}$.

For any smooth vector field $X$ on $\mathcal{C}$ we define its divergence with respect to $\mu$ as the smooth function $\text{div}_\mu X$ on $\mathcal{C}$ satisfying:

$$\mathfrak{L}_X \mu = (\text{div}_\mu X)\, \mu\,.$$

Finally, for any two smooth measures $\mu, \mu'$ on $\mathcal{C}$ there exists a unique strictly positive smooth function $\alpha$ on $\mathcal{C}$ such that $\mu' = \alpha \mu$.

**Definition A.13** We consider the same objects as in def. A.10. Let $\mu$ be a smooth measure on $\mathcal{C}$. The position representation with measure $\mu$ is the Hilbert space $\mathcal{H}_{\text{Pos}}^\mu := L_2(\mathcal{C} \to \mathcal{B}_e, d\mu)$ of cross-sections of $\mathcal{B}_e$ with square-integrable norm with respect to $\mu$.

As underlined above, trimming the prequantum representation down to a physically pertinent size comes at the price of restricting the algebra of observables that can be quantized. In the position representation, this quantization condition requires that the Hamiltonian flow of an admissible observable should send level sets of $\gamma$ onto level sets of $\gamma$. In the case of a cotangent bundle, the quantizable functions are therefore the ones that depend at most linearly on the momentum variables.

**Proposition A.14** We consider the same objects as in def. A.13. We define:

$$\mathcal{O}_{\text{Pos}} := \left\{ f \in C^\infty(\mathcal{M}, \mathbb{R}) \mid \exists \overline{X}_f \in \mathcal{T}^\infty(\mathcal{C}), \forall x \in \mathcal{M}, T_x\gamma\left(X_{f,x}\right) = \overline{X}_{f,\gamma(x)} \right\},$$

where $\mathcal{T}^\infty(\mathcal{C})$ is the space of smooth vector fields on $\mathcal{C}$, and:

$$\mathcal{O}_{\text{Pos},\mathbb{C}} := \{ f \in C^\infty(\mathcal{M}, \mathbb{C}) \mid \operatorname{Re} f, \operatorname{Im} f \in \mathcal{O}_{\text{Pos}} \}.$$

Then, for $f \in \mathcal{O}_{\text{Pos}}$, we can define the quantization $\widehat{f}^\mu$ of $f$ as a densely defined operator on $\mathcal{H}^\mu_{\text{Pos}}$ by:

$$\forall s \in \mathcal{D}_f^\mu, \forall y \in \mathcal{C}, \left(\widehat{f}^\mu s\right)(y) := f(x)s(y) + i\Gamma\left(\nabla_{X_{f,x}}\widetilde{s}(x)\right) + \frac{i}{2}\left(\operatorname{div}_\mu \overline{X}_f\right)(y)s(y),$$

where $x$ is any point in $\gamma^{-1}\langle y\rangle$, $\widetilde{s}$ is the cross-section of $\mathcal{B}$ such that $\Gamma \circ \widetilde{s} = s \circ \gamma$, and $\operatorname{div}_\mu \overline{X}_f \in C^\infty(\mathcal{C}, \mathbb{R})$ is such that $\mathfrak{L}_{\overline{X}_f}\mu = \left(\operatorname{div}_\mu \overline{X}_f\right)\mu$. For $f \in \mathcal{O}_{\text{Pos},\mathbb{C}}$, we define $\widehat{f}^\mu := \widehat{\operatorname{Re}(f)}^\mu + \widehat{\operatorname{Im}(f)}^\mu$.

Moreover, we have for all $f, g \in \mathcal{O}_{\text{Pos},\mathbb{C}}$:

$$\{f, g\}_\mathcal{M} \in \mathcal{O}_{\text{Pos},\mathbb{C}}, \quad \left[\widehat{f}^\mu, \widehat{g}^\mu\right] = i\widehat{\{f, g\}}^\mu,$$

and $\quad \forall s, s' \in \mathcal{D}_f^\mu, \left\langle s', \widehat{f}^\mu(s) \right\rangle = \left\langle \widehat{f^*}^\mu(s'), s \right\rangle.$

**Proof** Let $f \in \mathcal{O}_{\text{Pos}}$ and $s \in \mathcal{D}_f^\mu$. The cross-section $\widetilde{s}$ such that $\Gamma \circ \widetilde{s} = s \circ \gamma$ is well-defined, since for any $y \in \mathcal{C}$, $x \in \gamma^{-1}\langle y\rangle$ and $w \in \Pi_{\mathcal{B}_\mathcal{C}}^{-1}\langle y\rangle$, there is a unique $z \in \Pi_\mathcal{B}^{-1}\langle x\rangle$ such that $\Gamma(z) = w$ (this follows from def. A.10.2). We now want to prove that $f(x)s(y) + i\Gamma\left(\nabla_{X_{f,x}}\widetilde{s}(x)\right)$ does not depend on the choice of $x$ in $\gamma^{-1}\langle y\rangle$.

Let $V$ be any smooth vector field on $\mathcal{M}$ such that $\forall x \in \mathcal{M}, V_x \in \operatorname{Ker} T_x\gamma$. We have $\nabla_V \widetilde{s} = 0$ (since $\forall x \in \mathcal{M}, d_V\widetilde{s}(x) \in \operatorname{Ker} T_{s(x)}\Gamma \subset \operatorname{Hor}_{s(x)}(\mathcal{B}, \nabla)$ using def. A.10.4), therefore:

$$\forall x \in \mathcal{M}, \nabla_{V_x}\nabla_{X_{f,x}}\widetilde{s}(x) = \left[\nabla_{V_x}, \nabla_{X_{f,x}}\right]\widetilde{s}(x)$$

$$= \nabla_{[V_x, X_{f,x}]}\widetilde{s}(x) - i\Omega_{\mathcal{M},x}\left(V_x, X_{f,x}\right)\widetilde{s}(x),$$

and $T\gamma\left([V, X_f]\right) = 0$ (using $\forall x \in \mathcal{M}, V_x \in \operatorname{Ker} T_x\gamma$ and $T_x\gamma\left(X_{f,x}\right) = \overline{X}_{f,\gamma(x)}$), hence $\forall x \in \mathcal{M}, \nabla_{V_x}\nabla_{X_{f,x}}\widetilde{s}(x) = i\left(d_{V_x}f\right)\widetilde{s}(x)$. Therefore:

$$\forall x \in \mathcal{M}, \nabla_{V_x}\left[f(x)\widetilde{s}(x) + i\nabla_{X_{f,x}}\widetilde{s}(x)\right] = 0. \tag{A.14.1}$$

Let $z \in \mathcal{B}$ and let $V^{\text{HOR}}$ be the $\nabla$-horizontal lift on $\mathcal{B}$ of the vector field $V$ on $\mathcal{M}$. Using $T\gamma(V) = 0$ together with def. A.10.1, we have $[T_{\Gamma(z)}\Pi_{\mathcal{B}_\mathcal{C}}] \circ [T_z\Gamma](V_z^{\text{HOR}}) = 0$, so there exists $u \in \mathbb{C}$ such that $T_z\Gamma\left(V_z^{\text{HOR}}\right) = [T_1(\cdot \Gamma(z))](u) = T_z\Gamma \circ [T_1(\cdot z)](u)$ (where we used def. A.10.2 to get the second equality). Thus, using def. A.10.4, $V_z^{\text{HOR}} - [T_1(\cdot z)](u) \in \text{Hor}_z(\mathcal{B}, \nabla)$, therefore $u = 0$, and $T_z\Gamma(V_z^{\text{HOR}}) = 0$.

Hence, eq. (A.14.1) becomes:

$$\left[T\left(x \mapsto \Gamma\left[f(x)\widetilde{s}(x) + i\nabla_{X_{f,x}}\widetilde{s}(x)\right]\right)\right](V) = 0,$$

so $\Gamma[f\widetilde{s} + i\nabla_{X_f}\widetilde{s}] = f(s \circ \gamma) + i\Gamma(\nabla_{X_f}\widetilde{s})$ is constant on the level sets of $\gamma$. This ensures that $\widehat{f}^\mu s$ is well-defined as a cross-section of $\mathcal{B}_\mathcal{C}$.

Let $f, g \in \mathcal{O}_{\text{Pos}}$ (since $\widehat{\cdot}^\mu$ is $\mathbb{C}$-linear, and $[\cdot, \cdot]$, $\{\cdot, \cdot\}$ are $\mathbb{C}$-bilinear, it is enough to consider $\mathbb{R}$-valued functions to prove the commutator relations). Using the characterization of $\mathcal{O}_{\text{Pos}}$, we have:

$$\forall x \in \mathcal{M}, \ T_x\gamma\left(X_{\{f,g\},x}\right) = T_x\gamma\left([X_f, X_g]_x\right) = \left[\overline{X}_f, \overline{X}_g\right]_{\gamma(x)},$$

hence $\{f, g\} \in \mathcal{O}_{\text{Pos}}$ with $\overline{X}_{\{f,g\}} = \left[\overline{X}_f, \overline{X}_g\right]$.

Let $s \in \mathcal{D}^\mu_{f,g}$ (where, as in the proof of prop. A.4, the common domain $\mathcal{D}^\mu_{f,g}$, such that both $\widehat{f}^\mu \widehat{g}^\mu s$ and $\widehat{g}^\mu \widehat{f}^\mu s$ are well-defined, is dense in $\mathcal{H}^\mu_{\text{Pos}}$). Like in the proof of prop. A.4, we have:

$$\left[f + i\nabla_{X_f}, g + i\nabla_{X_g}\right]\widetilde{s} = i\{f, g\}\widetilde{s} - \nabla_{X_{\{f,g\}}}\widetilde{s}.$$

On the other hand, we can rewrite the definition of $\widehat{f}^\mu$ as:

$$\widetilde{\left(\widehat{f}^\mu s\right)} = \left(f + i\nabla_{X_f} + \frac{i}{2}(\text{div}_\mu \overline{X}_f) \circ \gamma\right)\widetilde{s} \tag{A.14.2}$$

thus:

$$\left(\widetilde{\left[\widehat{f}^\mu, \widehat{g}^\mu\right] s}\right) = \left(i\{f, g\} - \nabla_{X_{\{f,g\}}} + \frac{1}{2}\left(d_{X_g}(\text{div}_\mu \overline{X}_f) \circ \gamma - d_{X_f}(\text{div}_\mu \overline{X}_g) \circ \gamma\right)\right)\widetilde{s}.$$

Next, we have:

$$d_{X_g}(\text{div}_\mu \overline{X}_f) \circ \gamma - d_{X_f}(\text{div}_\mu \overline{X}_g) \circ \gamma = \left(d_{\overline{X}_g}(\text{div}_\mu \overline{X}_f) - d_{\overline{X}_f}(\text{div}_\mu \overline{X}_g)\right) \circ \gamma,$$

and $\mathcal{L}_{\overline{X}_g}\left(\mathcal{L}_{\overline{X}_f}\mu\right) = \left(d_{\overline{X}_g}(\text{div}_\mu \overline{X}_f)\right)\mu + (\text{div}_\mu \overline{X}_g)(\text{div}_\mu \overline{X}_f)\mu$, therefore, using $\overline{X}_{\{f,g\}} = \left[\overline{X}_f, \overline{X}_g\right]$:

$$d_{X_g}(\text{div}_\mu \overline{X}_f) \circ \gamma - d_{X_f}(\text{div}_\mu \overline{X}_g) \circ \gamma = -\left(\text{div}_\mu \overline{X}_{\{f,g\}}\right) \circ \gamma.$$

Hence, using eq. (A.14.2) for $\widehat{\{f, g\}}^\mu$:

$$\left[\widehat{f}^\mu, \widehat{g}^\mu\right] s = i\widehat{\{f, g\}}^\mu s.$$

Lastly, let $f \in \mathcal{O}_{\text{Pos},\mathbb{C}}$ and $s, s' \in \mathcal{D}^\mu_f$. Using def. A.10.3, we have $\forall x \in \mathcal{M}, \langle \widetilde{s}'(x), \widetilde{s}(x)\rangle = \langle s' \circ \gamma(x), s \circ \gamma(x)\rangle$ and combining eq. (A.14.2) with eq. (A.4.1):

$$\forall x \in \mathcal{M}, \ \left\langle \widetilde{s}', \widetilde{\widehat{f}^\mu s}\right\rangle(x) = i\left(\text{div}_\mu \overline{X}_f\right) \circ \gamma(x) \langle \widetilde{s}', \widetilde{s}\rangle(x) + i d_{X_{f,x}}\langle \widetilde{s}', \widetilde{s}\rangle(x) + \left\langle \widetilde{\widehat{f^*}^\mu s'}, \widetilde{s}\right\rangle(x),$$

therefore:

$$\forall y \in \mathcal{C}, \ \left\langle s', \widehat{f}^\mu s\right\rangle(y) = i\left(\text{div}_\mu \overline{X}_f\right)(y) \langle s', s\rangle(y) + i d_{\overline{X}_{f,y}}\langle s', s\rangle(y) + \left\langle \widehat{f^*}^\mu s', s\right\rangle(y).$$

Now, using Stokes theorem [2, theorem 7.7] ($\mathcal{C}$ is assumed to be without boundary, or $\mathcal{D}_f^\mu$ is required to ensure suitable fall-off conditions) and the definition of $\overline{X}_f$, we have:

$$\int_\mathcal{C} d\mu(y) \ \left(\text{div}_\mu \overline{X}_f\right)(y) \langle s', s\rangle(y) + d_{\overline{X}_{f,y}} \langle s', s\rangle(y) = \int_\mathcal{C} \mathfrak{L}_X \left[\langle s', s\rangle \, d\mu\right] = 0 \,,$$

thus $\left\langle s', \widehat{f}^\mu s \right\rangle = \left\langle \widehat{f^*}^\mu s', s \right\rangle$. □

**Proposition A.15** We consider the same objects as in def. A.10. Let $\mu$ and $\mu'$ be two smooth measures on $\mathcal{C}$ and let $\mathcal{H}_{\text{Pos}}^\mu$ and $\mathcal{H}_{\text{Pos}}^{\mu'}$ be the corresponding position representations. Then there exists a Hilbert space isomorphism $\Phi_{\mu \to \mu'} : \mathcal{H}_{\text{Pos}}^\mu \to \mathcal{H}_{\text{Pos}}^{\mu'}$ such that:

$$\forall f \in \mathcal{O}_{\text{Pos},\mathcal{C}}, \ \widehat{f}^{\mu'} = \Phi_{\mu \to \mu'} \circ \widehat{f}^\mu \circ \Phi_{\mu \to \mu'}^{-1} \,. \tag{A.15.1}$$

Moreover, we can define these maps in such a way that for any three smooth measures $\mu$, $\mu'$, and $\mu''$, on $\mathcal{C}$, $\Phi_{\mu \to \mu''} = \Phi_{\mu' \to \mu''} \circ \Phi_{\mu \to \mu'}$. Thus, this family of maps provides a position representation $\mathcal{H}_{\text{Pos}}$, that can be consistently identified with $\mathcal{H}_{\text{Pos}}^\mu$ for any $\mu$.

**Proof** Let $\mu$ and $\mu'$ be two smooth measures on $\mathcal{C}$. Then, there exists a unique $\alpha \in C^\infty(\mathcal{C}, \mathbb{R}_+^*)$ such that $\mu' = \alpha \mu$ (def. A.12). We define $\Phi_{\mu \to \mu'}$ by:

$$\Phi_{\mu \to \mu'} : \mathcal{H}_{\text{Pos}}^\mu \to \mathcal{H}_{\text{Pos}}^{\mu'}$$
$$s \mapsto \frac{1}{\sqrt{\alpha}} s \,.$$

The factor $\frac{1}{\sqrt{\alpha}}$ ensures that $\Phi_{\mu \to \mu'}$ is a unitary map and we can check that for any three positive volume forms $\mu$, $\mu'$, and $\mu''$, $\Phi_{\mu \to \mu''} = \Phi_{\mu' \to \mu''} \circ \Phi_{\mu \to \mu'}$. In particular, $\Phi_{\mu \to \mu'}$ is then invertible, hence it is a Hilbert space isomorphism.

Lastly, eq. (A.15.1) follows from:

$$\forall f \in \mathcal{O}_{\text{Pos}}, \ 2\sqrt{\alpha} \left( d_{\overline{X}_f} \frac{1}{\sqrt{\alpha}} \right) + \left(\text{div}_{\mu'} \overline{X}_f\right) = \left(\text{div}_\mu \overline{X}_f\right) \,.$$

□

# B References